\pdfoutput=1
\documentclass[aps,pre,preprint,amsmath,amssymb,superscriptaddress]{revtex4-2}
\usepackage{fancyhdr}
\usepackage{fancybox}
\usepackage{color}
\usepackage{graphicx}% Include figure files
\usepackage{dcolumn}% Align table columns on decimal point
\usepackage{bm}% bold math
\usepackage{hyperref}% add hypertext capabilities
\hypersetup{
	colorlinks=true,
	linkcolor=blue,
	filecolor=black,      
	urlcolor=red,
	citecolor=red,
}
\usepackage{amssymb}
\usepackage{mathrsfs}
\usepackage{color}
\usepackage{soul}
\usepackage{subfigure}
\usepackage{soul}
\usepackage{CJK}
\begin{document}
\preprint{APS/PRF}
\begin{CJK*}{GB}{} % Use default fonts from CJK (see below)
\title{Lamb dilatation and its hydrodynamic viscous flux in near-wall incompressible flows}

\author{Tao Chen}
\email[First author:]{1601111553@pku.edu.cn}
%\thanks{}
\affiliation{
	Department of Mechanics and Aerospace Engineering, Southern University of Science and Technology, Shenzhen 518055, China
}
\author{Tianshu Liu}
\email[Corresponding author:]{tianshu.liu@wmich.edu}
\affiliation{Department of Mechanical and Aerospace Engineering, Western Michigan University, Kalamazoo, Michigan, 49008, USA}

\date{\today}

\begin{abstract}
In this paper, we introduce a new physical concept referred to as the wall-normal Lamb dilatation flux (WNLDF), which is defined as the wall-normal derivative of the Lamb dilatation (namely, the divergence of the Lamb vector) multiplied by the dynamic viscosity for incompressible viscous flows. 
It is proved that the boundary Lamb dilatation flux (BLDF, namely the wall-normal Lamb dilatation flux at the wall) is determined by the boundary enstrophy flux (BEF) and the surface curvature-induced contribution.
As the first step to explore this new concept, the present study only considers flow past a stationary no-slip flat wall without curvature-fluid dynamic coupling effects.
It is found that the temporal-spatial evolution rate of the WNLDF is contributed by four source terms, which can be explicitly expressed using the fundamental surface quantities including skin friction (or surface vorticity) and surface pressure. 
Therefore, the instantaneous near-wall flow structures are directly related to the WNLDF in both laminar and turbulent flows. As an example, for the turbulent channel flow at $Re_{\tau}=180$, the intensification of the temporal-spatial evolution rate of the WNLDF is caused by the strong wall-normal velocity event (SWNVE) associated with quasi-streamwise vortices and high intermittency of the viscous sublayer. In addition, near the SWNVE, this evolution rate is mainly contributed by the coupling between skin friction and the boundary enstrophy gradient, as well as the coupling between the skin friction divergence and the boundary enstrophy.
The exact results presented in this paper could provide new physical insights into complex near-wall flows.
\end{abstract}
%\keywords{}

\maketitle
\end{CJK*}
\section{Introduction}
The cross product of the vorticity $\bm{\omega}$ and the velocity $\bm{u}$ defines the Lamb vector $\bm{L}=\bm{\omega}\times\bm{u}$ which characterizes the essential nonlinearity of the convective fluid acceleration in the Navier-Stokes (NS) equation.
Since the Lamb vector is associated with both the vorticity and velocity fields, its spatial structures are usually highly localized and more compact when compared to those of velocity or vorticity alone~\cite{WuJZ2006book,YangYT2007}.
Since the pioneering works of Prandtl~\cite{Prandtl1918} in developing the finite wing theory, the physical connotations and significance of the Lamb vector were extensively explored and interpreted from different perspectives. The relevant topics include modern aerodynamic vortex force theory~\cite{Lighthill1962,Karman1935,Marongiu2013AIAA,WuJZ2018PAS,LiuTianshu2021}, turbulence modeling~\cite{Wu1999POF}, Lamb surfaces~\cite{Sposito1997}, the initial formation of the Lamb vector and its strong influence on global flow~\cite{YangYT2007}, applications in vortex dynamics and fluid hydrodynamics~\cite{Batchelor1967,Saffman1992,WuJZ2006book,WangShizhao2019} and relation between the Lamb vector and the depression of turbulence nonlinearity~\cite{Tsinober1990,Shtilman1992}.
For compressible viscous flows, the shearing and dilatational processes are coupled and linked by virtue of the Lamb vector and its relevant derivatives. The curl of the Lamb vector $\bm{\nabla}\times\bm{L}$ drives the shearing process of coherent vortical structures by affecting the advection, stretching and tilting processes of vorticity lines, while its divergence $\vartheta_{L}\equiv\bm{\nabla}\cdot\bm{L}$ (referred to as the Lamb dilatation in the present paper by analogy with the velocity dilatation $\vartheta\equiv\bm{\nabla}\cdot\bm{u}$) acts as an acoustic source term in Lighthill's wave equation~\cite{Lighthill1952,Howe1975} during the compression and expansion processes.

However, for either incompressible or compressible viscous flows, there are only very limited theoretical studies on the Lamb dilatation.
In addition to serving as a robust indicative quantity for near-wall coherent structures~\cite{Adrian2007,Liuchaoqun2014,XuAM2015,Yang2016,Jamenez2018,LeeCB2019},
it is believed that the Lamb dilatation provides a rigorous foundation for the analysis of momentum and energy redistribution and therefore new insight into development and evolution of coherent structures in near-wall flows.
Relevant progresses on the Lamb dilatation are briefly discussed as follows.

Hamman {\it et al.}~\cite{Hamman2008} derived the transport equation for the Lamb dilatation and explored its mathematical and physical properties.
They found that the Lamb dilatation had the capacity to characterize the temporal-spatial evolution of high- and low-momentum flow regions, which was directly related to the turbulent mixing mechanism.
They further demonstrated that the self-sustaining near-wall ejection and sweep motions could be well captured by the source-sink distribution of the Lamb dilatation in turbulent channel flows. In addition, they pointed out that
decreasing the area over which regions of positive and negative Lamb vector divergence interact could lead to pressure drag reduction.
Xu {\it et al.}~\cite{ChenXuLu2010s} carried out a large-eddy simulation of the compressible flow past a wavy cylinder.
The two-layer structures of the shear layer separated from the cylinder were found to be well captured using the instantaneous Lamb dilatation. Further evaluations showed that the evolution of the turbulent shear layer was directly related to the momentum exchange between the strong straining and vortical motions that basically corresponded to positive and negative Lamb dilatation, respectively. Compared to flow past a circular cylinder, the interaction area of positive and negative Lamb dilatation was obviously decreased by the wavy surface, which was consistent with pressure drag reduction mechanism proposed by Hamman {\it et al.}~\cite{Hamman2008}
Dynamical roles of the Lamb dilatation were also demonstrated and visualized in compressible flow past an airfoil~\cite{ChenXuLu2010} and in a hypersonic turbulent boundary layer~\cite{TongFulin2017,XuDehao2021JFM}.
Marmanis~\cite{Marmanis1998} and Sridhar~\cite{Sridhar1998} showed interesting correspondences between hydrodynamic and electromagnetic quantities, and independently proposed a set of hydrodynamic Maxwell equations. By analogy with the electric charge density, the Lamb dilatation was referred to as the hydrodynamic charge density.
Rousseaux~\cite{Rousseaux2007} performed experimental measurements of both the Lamb vector and the Lamb dilatation 
by employing two-dimensional particle image velocimetry (PIV). Compared to those pressure-minimum-based vortex identification criteria, the Lamb dilatation takes into account both the influences from both pressure and kinetic energy, which is shown as an useful quantity in detecting and characterizing coherent flow structures.
Kollmann~\cite{Kollmann2006} analyzed the Lamb vector lines in a swirling jet flow and found that the set of critical points of the 
Lamb vector field contained stable and unstable manifolds associated with the high shear regions, which was clearly visualized by the Lamb dilatation.

The aforementioned studies on the Lamb dilatation mainly focus on its dynamical roles in bulk flow regions and its indicative nature to coherent structures.
However, to the authors' knowledge, near-wall behaviours of the Lamb dilatation and its wall-normal hydrodynamic viscous flux were not yet investigated, particularly from theoretical perspectives. In addition, their relations with the fundamental surface quantities (such as skin friction $\bm{\tau}$ and surface pressure $p_{\partial B}$) were also not explored, as well as the temporal-spatial evolution rate of the wall-normal Lamb dilatation flux (WNLDF). Therefore, establishing these less known but important relations naturally becomes the main objective of the present study. Before the formal development of theory, a brief summary of previous theoretical progresses on relations between near-wall flows and surface quantities is given in order to facilitate the readers to understand our work, which also provides a good starting point for the present study.

In fact, skin friction $\bm{\tau}$ and surface pressure $p_{\partial B}$ have been considered as the footprints of near-wall incompressible viscous flows~\cite{Bewley2004,Liu2019PAS,ChenTao2021POF}.
By applying the Taylor-series expansion to the NS equations on a stationary wall, the near-wall flow variables
in a small vicinity of the wall
are uniquely determined by $\bm{\tau}$, $p_{\partial B}$ and their relevant temporal-spatial derivatives at the wall~\cite{Bewley2004,Liu2018AIA,Liu2019PAS,ChenTao2021POF}.
However, skin friction $\bm{\tau}$ and surface pressure $p_{\partial B}$ are not independent but are intrinsically coupled 
via the boundary enstrophy flux (BEF) $f_{\Omega}\equiv\mu[\partial\Omega/\partial n]_{\partial B}$~\cite{WuJZ2006book,Liu2016MST,Liu2018AIA,Liu2019PAS,WuJZ2018PAS}, where $\mu$ is the dynamic viscosity, $\Omega=\omega^2/2$ is the fluid enstrophy and $\partial B$ denotes the physical quantities on the wall. 
A concise but exact on-wall relation ($\bm{\tau}-p_{\partial B}$ relation) between $\bm{\tau}$, $p_{\partial B}$ and $f_{\Omega}$ was derived, interpreted and generalized by Liu {\it et al.}~\cite{Liu2016MST} and Chen {\it et al.}~\cite{ChenTao2019POF}
It is indicated that the BEF can be generated through the viscous coupling between the skin friction and surface pressure gradient during the vorticity-wall interaction. Effect of the surface curvature on the BEF was also discussed in detail as well as relations between skin friction and other surface physical quantities~\cite{ChenTao2019POF,Miozzi2019}. It is noted that an integrated BEF is proposed by Sudharsan {\it et al.}~\cite{Sudharsan2022JFM} as a physically-intuitive vorticity-based criterion to characterize leading-edge-type dynamic stall onset.

Global skin friction field is difficult to measure in fluid experiments~\cite{Costantini2021,Orlu2020,Liu2016MST,Liu2019PAS}
while the surface pressure field is relatively easier to be measured using pressure-sensitive paint (PSP)~\cite{Liu2021Book}.
Interestingly, by modeling the BEF properly, a global skin friction field can be
extracted from the measured surface pressure field by solving the Euler-Lagrangian equation in an unified 
variational framework. Further exploration in this direction indicates that the variation of local skin friction topology 
is closely related to local pressure pertubation in complex separated and attached flows~\cite{Liu2018AIA,ChenTao2019POF,ChenTao2021POF}.
In a single-phase turbulent channel flow at the frictional Reynolds number $Re_\tau=180$, Chen {\it et al.}~\cite{ChenTao2021POF} found that the strong wall-normal velocity events (SWNVEs) induced by the near-wall quasi-streamwise vortices were strongly correlated with high-magnitude BEF regions, which accounted for the high intermittent turbulence feature of the viscous sublayer.
A most recent statistical study performed by Guerrero {\it et al.}~\cite{Guerrero2020,Guerrero2022JFM} in a turbulent pipe flow also revealed that the rare back flow events were signatures of near-wall self-sustaining processes, which were closely related to the nonlinear enstrophy stretching and intensification mechanisms.
Relevant turbulence statistics on the skin friction, surface pressure, wall heat flux and their fluctuating components were also analyzed and reported~\cite{PanC2018,ChinRC2018,Cardesa2019,Cheng2020,DongSiwei2020,Gibeau2021,YuMing2021,TongFulin2022,YuMing2022}.
The importance of near-wall viscous stress work to wall heat flux was discussed and highlighted by Zhang and Xia~\cite{ZhangPeng2020} and Sun {\it et al.}~\cite{SunD2021}
Relation between the temporal-spatial evolution rate of the wall-normal enstrophy flux and those surface quantities was also discussed in both laminar and turbulent flows by Chen {\it et al.}~\cite{ChenTao2021POFb} in enlightment of previous works for the boundary vorticity flux (BVF)~\cite{Andreopoulos1996,WuJZ1996,WuJZWuJM1998}. Near the SWNVEs, the evolution
of the wall-normal enstrophy flux was found to be dominated by the wall-normal diffusion of the vortex stretching term.

Most recently, Chen and Liu~\cite{ChenTao2022AIPa}
generalized the near-wall Taylor-series expansion solution to the compressible Navier-Stokes-Fourier (NSF) system. As a direct application, by assuming constant viscosities, different physical mechanisms that were responsible for initial formation of the Lamb vector in the viscous sublayer were clearly elucidated, extending the work of Yang {\it et al.}~\cite{YangYT2007} It was found that the tangential Lamb
vector was contributed by the terms related to both the boundary vorticity divergence and the skin friction divergence. Subsequently, the temporal-spatial evolution of the Lamb vector and its magnitude in the viscous sublayers was investigated in a laminar Hiemenz flow (an attachment line flow) and a turbulent channel flow~\cite{ChenTao2022AIPb}. The significance of the skin friction divergence was highlighted by Liu {\it et al.}~\cite{LiuT2021wedge}, and Chen and Liu~\cite{ChenTao2022AIPb} in complex separation and attached flow regions.

This paper is organized as follows. In Section~\ref{NS}, the NS equations and its near-wall
Taylor-series expansion solution are briefly discussed. Then, in Section~\ref{Lamb vector and Lamb dilatation},
the Lamb vector and the Lamb dilatation are naturally introduced, followed by their physical interpretations.
Next, in Section~\ref{transport_eq}, we present a new version of the transport equation for the Lamb dilatation
based on the Lamb vector transport equation obtained by Wu {\it et al.}~\cite{Wu1999POF} The equivalence between different versions of the transport equation for the
Lamb dilatation are elucidated. In Section~\ref{BLDFPH}, the boundary Lamb dilatation flux (BLDF) is introduced as a new concept, which is found to provide a rational foundation for pressure drag reduction mechanism. For a stationary flat wall, we derive the exact relation between the temporal-spatial evolution rate of the wall-normal enstrophy flux at the wall and fundamental surface physical quantities in Section~\ref{TSEW}.
In Section~\ref{Numerical_method}, necessary ingredients for the numerical methods used are briefly discussed. Then, in Section~\ref{analysis}, by using these new theoretical results, two simulations are carefully done to explore the underlying physics in instantaneous separation and attachment line flows. Conclusions are made in Section~\ref{Conclusions}. Appendix~\ref{Appendix1} documents some necessary technical details of derivations.

\section{Navier-Stokes equations, Lamb vector and Lamb dilatation}\label{Navier-Stokes equations and Lamb vector}
\subsection{Navier-Stokes equations}\label{NS}
The dynamics of an incompressible viscous Newtonian fluid can be well described by the Navier-Stokes (NS) equations:
\begin{subequations}
	\begin{eqnarray}\label{momentum_eq}
		\frac{\partial\bm{u}}{\partial t}+\bm{u}\bm{\cdot}\bm{\nabla}\bm{u}=-\bm{\nabla}\left(\frac{p}{\rho}\right)+\nu\nabla^2\bm{u},
	\end{eqnarray}
	\begin{eqnarray}\label{continuity_eq}
		\bm{\nabla}\bm{\cdot}\bm{u}=0,
	\end{eqnarray}
\end{subequations}
where $\rho$ is the constant fluid density, $\bm{u}$ is the velocity and $p$ is the pressure. $\nu=\mu/\rho$ is the kinematic viscosity with $\mu$ being the dynamic viscosity. The body force potential has been combined into the pressure. Therefore, the body force does not exist explicitly in the following discussion.

The decomposition of the velocity gradient tensor $\bm{\nabla}\bm{u}$ gives the strain rate tensor $\bm{S}$ and the rotation tensor $\bm{A}$ as
\begin{eqnarray}
	\bm{S}=\frac{1}{2}\left(\bm{\nabla}\bm{u}^{T}+\bm{\nabla}\bm{u}\right),~~\bm{A}=\frac{1}{2}\left(\bm{\nabla}\bm{u}^{T}-\bm{\nabla}\bm{u}\right).
\end{eqnarray}
The rotation tensor $\bm{A}$ satisfies the relation $2\bm{A}\bm{\cdot}\bm{v}=\bm{\omega}\times\bm{v}$ for any vector $\bm{v}$
with $\bm{\omega}=\bm{\nabla}\times\bm{u}$ being the vorticity. The fluid enstrophy is defined as $\Omega\equiv\omega^2/2$.

In this paper, the no-slip and no-penetration boundary condition $\bm{u}_{\partial B}=\bm{0}$ is considered on a general stationary curved boundary surface $\partial B$ with the normal vector $\bm{n}$ directing from the wall to the fluid. A physical quantity with the subscript $\partial B$ denotes its value at the wall and $\bm{\nabla}_{\partial B}$ represents the tangential surface gradient operator. 
The surface Laplacian operator is denoted as $\nabla_{\partial B}^{2}\equiv\bm{\nabla}_{\partial B}\bm{\cdot}\bm{\nabla}_{\partial B}$.
The surface curvature tensor is denoted as $\bm{K}=-\bm{\nabla}_{\partial B}\bm{n}$, whose trace $tr(\bm{K})$ is twice the mean surface curvature $H$, namely, $tr(\bm{K})=-\bm{\nabla}_{\partial B}\bm{\cdot}\bm{n}=2H=\kappa_1+\kappa_2$, where $\kappa_1$ and $\kappa_2$ represent the two principal curvatures of the surface. Obviously, $\bm{K}$ vanishes for a flat surface. Combined with proper boundary conditions, Eqs.~\eqref{momentum_eq} and~\eqref{continuity_eq} give a complete description for incompressible viscous flows.

Interestingly, in a small vicinity of the solid wall $\partial B$, applying the near-wall Taylor-series expansion
to Eqs.~\eqref{momentum_eq} and~\eqref{continuity_eq} yields an explicit asymptotic solution of the NS system which uniquely relates the near-wall velocity $\bm{u}$ and pressure $p$ to the fundamental surface quantities (including skin friction $\bm{\tau}=\mu\bm{\omega}_{\partial B}\times\bm{n}$, surface pressure $p_{\partial B}$ and surface curvature tensor $\bm{K}$) as well as their temporal-spatial derivatives on the surface~\cite{Bewley2004,Liu2018AIA,ChenTao2021POF,ChenTao2022AIPa}. Namely,
\begin{eqnarray}
	\bm{u}=f_1\left(y;\bm{\tau},\bm{\nabla}_{\partial B}p_{\partial B},\bm{\nabla}_{\partial B}\bm{\cdot}\bm{\tau},\nabla_{\partial B}^2p_{\partial B},tr(\bm{K}),\cdots\right),
\end{eqnarray}
\begin{eqnarray}
	{p}=f_2\left(y;p_{\partial B},\bm{\nabla}_{\partial B}\bm{\cdot}\bm{\tau},\nabla_{\partial B}^2p_{\partial B},tr(\bm{K}),\cdots\right),
\end{eqnarray}
where $y$ is the wall-normal coordinate measured from the surface in the fluid side.
However, the physical and geometrical variables ($\bm{\tau}$, $p_{\partial B}$ and $\bm{K}$) are generally
not independent but are coupled through the boundary enstrophy flux (BEF) $f_{\Omega}\equiv\mu[\partial\Omega/\partial n]_{\partial B}$, satisfying the exact relation~\cite{Liu2016MST,ChenTao2019POF}
\begin{eqnarray}\label{BEF}
	\bm{\tau}\bm{\cdot}\bm{\nabla}_{\partial B}p_{\partial B}+\mu^2\bm{\omega}_{\partial B}\bm{\cdot}\bm{K}\bm{\cdot}\bm{\omega}_{\partial B}
	=\mu f_{\Omega}.
\end{eqnarray}
The first term in Eq.~\eqref{BEF} represents the non-linear coupling between the skin friction and the surface pressure gradient. The second term in Eq.~\eqref{BEF} represents the interaction between the boundary vorticity and the surface curvature via the dynamic viscosity. The sum of these two contributions is physically balanced by the wall-normal diffusion of the enstrophy.
Generally, the BEF is dominated by the $\bm{\tau}-p_{\partial B}$ coupling term when the surface curvature contribution is not significant. Since the enstrophy is usually created at the wall and then diffuses into the fluid, the BEF usually remains negative in most attached flows while its sign could be positive near the separation and attachment lines or isolated critical points~\cite{Liu2018AIA,ChenTao2021POF}.
Of particular interest are small-Reynolds-number viscous flows where the surface curvature contribution cannot be neglected in the total BEF. The sign of the surface curvature contribution is obviously dependent on the surface type~\cite{ChenTao2019POF}. Proper modeling the BEF could realize the transformation between the footprints (skin friction and surface pressure) in an unified variational framework~\cite{Liu2018AIA,ChenTao2019POF}.

\subsection{Lamb vector and Lamb dilatation}\label{Lamb vector and Lamb dilatation}
The convective acceleration in Eq.~\eqref{momentum_eq} can be decomposed as 
\begin{eqnarray}
	\bm{u}\bm{\cdot}\bm{\nabla}\bm{u}=\bm{\omega}\times\bm{u}+\bm{\nabla}\left(\frac{1}{2}u^2\right)\equiv\bm{L}+\bm{\nabla}K,
\end{eqnarray}
where $K=u^2/2$ is the kinetic energy per unit mass. The cross product of the velocity and vorticity defines the Lamb vector $\bm{L}\equiv\bm{\omega}\times\bm{u}$, which represents the critical non-linear features of the viscous flows, while the kinetic energy potential gradient can be absorbed into the pressure gradient to form the stagnation enthalpy $h_0=p/\rho+u^2/2$ or the Bernoulli function~\cite{Hamman2008,Wu1999POF}. Using the incompressibility condition Eq.~\eqref{continuity_eq} and the identity $\bm{\nabla}\times\bm{\omega}=\bm{\nabla}\times(\bm{\nabla}\times\bm{u})=-\nabla^2\bm{u}$, we recast Eq.~\eqref{momentum_eq} as
\begin{eqnarray}\label{NS2}
	\frac{\partial\bm{u}}{\partial t}+\bm{L}=-\bm{\nabla}h_0-\nu\bm{\nabla}\times\bm{\omega}.
\end{eqnarray}

The Lamb dilatation is defined as the divergence of the Lamb vector, namely $\vartheta_L\equiv\bm{\nabla}\bm{\cdot}\bm{L}$. Taking divergences of both sides of Eq.~\eqref{NS2} yields a Poisson equation for the stagnation enthalpy $h_0$ where the Lamb dilatation $\vartheta_{L}$ (with a minus sign) acts as a source term, i.e.,
$\nabla^2h_0=-\vartheta_{L}$.
Physically, a region with positive Lamb dilatation ($\vartheta_{L}>0$) implies a concentrated stagnation enthalpy, while a region with negative Lamb dilatation ($\vartheta_{L}<0$) implies a depleted stagnation enthalpy.
Usually, straining motions tend to correspond to regions with positive Lamb dilatation, while vortical motions tend to correspond to regions with negative Lamb dilatation. Interaction between positive and negative regions of the Lamb dilatation will effect a time rate of change of momentum and thereby causes a distinct change in the overall flow dynamics~\cite{Hamman2008}. In this sense, the spatial structure of the Lamb dilatation is directly related to the energy redistribution and momentum transfer between the flow regions with positive and negative Lamb dilatation. 

\section{Transport equation for Lamb dilatation}\label{transport_eq}
In this section, we provide two versions of the transport equations for the Lamb dilatation $\vartheta_{L}$. The first version comes from our derivation by taking divergence of the transport equation for the Lamb vector $\bm{L}$ previously obtained by Wu {\it et al.}~\cite{Wu1999POF}, while the second version is established by Hamman {\it et al.}~\cite{Hamman2008} Mathematically, we will prove the equivalence of these two different versions. The former seems more compact and concise compared to the latter, which will be used in the following discussion.

\subsection{Derivation and physical interpretation}
By taking the curl of both sides of Eq.~\eqref{NS2}, the vorticity transport equation can be derived as
\begin{eqnarray}\label{vorticity_transport}
	\frac{\partial\bm{\omega}}{\partial t}=-\bm{\nabla}\times\bm{L}+\nu\nabla^2\bm{\omega}
	=-\bm{u}\bm{\cdot}\bm{\nabla}\bm{\omega}+\bm{\omega}\bm{\cdot}\bm{\nabla}\bm{u}+\nu\nabla^2\bm{\omega},
\end{eqnarray}
where the convection, stretching and tilting effects on vorticity lines originate from the curl of the Lamb vector $\bm{\nabla}\times\bm{L}$.

Taking a cross product of the vorticity $\bm{\omega}$ with Eq.~\eqref{NS2} and that of the velocity $\bm{u}$ with Eq.~\eqref{vorticity_transport}, Wu {\it et al.}~\cite{Wu1999POF} first obtained the transport equation for the Lamb vector,
\begin{eqnarray}\label{Lamb_vector}
	\frac{\partial\bm{L}}{\partial t}+\bm{u}\bm{\cdot}\bm{\nabla}\bm{L}
	=-\bm{L}\bm{\cdot}\bm{\nabla}\bm{u}
	+\bm{\nabla}h_{0}\times\bm{\omega}
	+\nu\nabla^2\bm{L}+\bm{q},
\end{eqnarray}
where the left hand side of Eq.~\eqref{Lamb_vector} represents the material derivative of the Lamb vector. In the right hand side of Eq.~\eqref{Lamb_vector}, the first term  denotes the inviscid stretching and tilting effects of the $\bm{L}$-lines. The second term stands for the interaction between the gradient of the stagnation enthalpy $\bm{\nabla}h_0$ and the vorticity $\bm{\omega}$. Note that $\bm{\nabla}h_{0}\times\bm{\omega}=-2\bm{A}\bm{\cdot}\bm{\nabla}h_0=-2\lVert\bm{\nabla}h_0\rVert\bm{A}\bm{\cdot}\bm{n}_{0}$, where $n_0\equiv\bm{\nabla}h_0/\lVert\bm{\nabla}h_0\rVert$ is the unit normal vector of the isosurface of the stagnation enthalpy. Therefore, this term is determined by the magnitude of the stagnation enthalpy gradient and the increment of the unit normal vector $\bm{n}_0$ due to the rotational action exerted by the tensor $\bm{A}$. The third term represents the viscous diffusion of the Lamb vector. The last term $\bm{q}$ can be written as the divergence of a second-order tensor, namely,
\begin{eqnarray}\label{qvis}
	\bm{q}\equiv-2\nu\frac{\partial\bm{\omega}}{\partial x_j}\times\frac{\partial\bm{u}}{\partial x_j}=\nu\bm{\nabla}\bm{\cdot}
	\left(4\bm{S}\bm{\cdot}\bm{A}+\bm{\omega\omega}\right).
\end{eqnarray}
Therefore, $\bm{q}$ can be interpreted as a viscous source. The detailed proof of Eq.~\eqref{qvis} can be found in the appendix of our recent paper~\cite{ChenTao2022AIPb}.

Using the following identities
\begin{eqnarray}
	\bm{\nabla}\bm{\cdot}\left(\bm{u}\bm{\cdot}\bm{\nabla}\bm{L}\right)=\bm{\nabla}\bm{L}\bm{:}\bm{\nabla}\bm{u}^{T}+\bm{u}\bm{\cdot}\bm{\nabla}\vartheta_{L},
\end{eqnarray}
\begin{eqnarray}
	\bm{\nabla}\bm{\cdot}\left(\bm{L}\bm{\cdot}\bm{\nabla}\bm{u}\right)=\bm{\nabla}\bm{L}\bm{:}\bm{\nabla}\bm{u}^{T}+\bm{L}\bm{\cdot}\bm{\nabla}\left(\bm{\nabla}\bm{\cdot}\bm{u}\right)=\bm{\nabla}\bm{L}\bm{:}\bm{\nabla}\bm{u}^{T},
\end{eqnarray}
\begin{eqnarray}
	\bm{\nabla}\bm{\cdot}\left(\bm{\nabla}h_0\times\bm{\omega}\right)=-\bm{\nabla}h_0\bm{\cdot}\left(\bm{\nabla}\times\bm{\omega}\right),
\end{eqnarray}
and taking the divergences of both sides of Eq.~\eqref{Lamb_vector}, we obtain the transport equation for the Lamb dilatation, i.e.,
\begin{eqnarray}\label{transport_Lamb_dilatation}
	\frac{\partial\vartheta_L}{\partial t}
	+\bm{u}\bm{\cdot}\bm{\nabla}\vartheta_{L}
	=-2\bm{\nabla}\bm{L}\bm{:}\bm{\nabla}\bm{u}^T
	-\bm{\nabla}h_0\bm{\cdot}\left(\bm{\nabla}\times\bm{\omega}\right)+\nu\nabla^2\vartheta_{L}+\bm{\nabla}\bm{\cdot}\bm{q},
\end{eqnarray}
where the left hand side of Eq.~\eqref{transport_Lamb_dilatation} represents the material derivative of the Lamb dilatation. In the right hand side of Eq.~\eqref{transport_Lamb_dilatation}, the first term denotes the double contraction of the Lamb vector gradient and the transposed velocity gradient. The second term represents the coupling between the curl of the vorticity and the gradient of the stagnation enthalpy. The other two terms denote the viscous diffusion of the Lamb dilatation and the divergence of the viscous source term $\bm{q}$, respectively.

\subsection{Equivalence between different formulations}
It should be noted that Hamman {\it et al.}~\cite{Hamman2008} also reported another version for the transport equation of the Lamb dilatation, which read
\begin{eqnarray}\label{HKK}
	\frac{\partial\vartheta_L}{\partial t}
	+\bm{u}\bm{\cdot}\bm{\nabla}\vartheta_{L}
	&=&-2\bm{\nabla}\bm{L}\bm{:}\bm{\nabla}\bm{u}^T
	-\bm{\nabla}h_0\bm{\cdot}\left(\bm{\nabla}\times\bm{\omega}\right)\nonumber\\
	&&-\nu\lVert\bm{\nabla}\times\bm{\omega}\rVert^2+\nu\bm{u}\bm{\cdot}\nabla^2\left(\bm{\nabla}\times\bm{\omega}\right)
	-2\nu\bm{\omega}\bm{\cdot}\nabla^2\bm{\omega}.
\end{eqnarray}

Here, we further prove the equivalence of Eqs.~\eqref{transport_Lamb_dilatation} and~\eqref{HKK}. First, we see that direct vector field analysis gives the following identity
\begin{eqnarray}\label{id1}
	\nu\nabla^2\bm{\omega}\times\bm{u}+\nu\bm{\omega}\times\nabla^2\bm{u}=\nu\nabla^2\bm{L}+\bm{q}.
\end{eqnarray}
Then, by taking divergences of both sides of Eq.~\eqref{id1}, we have
\begin{eqnarray}\label{id2}
	\nu\bm{\nabla}\bm{\cdot}(\nabla^2\bm{\omega}\times\bm{u})+\nu\bm{\nabla}\bm{\cdot}(\bm{\omega}\times\nabla^2\bm{u})=\nu\nabla^2\vartheta_{L}+\bm{\nabla}\bm{\cdot}\bm{q}.
\end{eqnarray}
The first and the second terms in the left hand side of Eq.~\eqref{id2} can be evaluated as
\begin{eqnarray}\label{id3}
	\nu\bm{\nabla}\bm{\cdot}(\nabla^2\bm{\omega}\times\bm{u})&=&\nu\bm{u}\bm{\cdot}\bm{\nabla}\times\nabla^2\bm{\omega}-\nu\nabla^2\bm{\omega}\bm{\cdot}\bm{\nabla}\times\bm{u}\nonumber\\
	&=&\nu\bm{u}\bm{\cdot}\nabla^2(\bm{\nabla}\times\bm{\omega})-\nu\bm{\omega}\bm{\cdot}\nabla^2\bm{\omega},
\end{eqnarray}
\begin{eqnarray}\label{id4}
	\nu\bm{\nabla}\bm{\cdot}(\bm{\omega}\times\nabla^2\bm{u})
	&=&\nu\nabla^2\bm{u}\bm{\cdot}\bm{\nabla}\times\bm{\omega}-\nu\bm{\omega}\bm{\cdot}\bm{\nabla}\times\nabla^2\bm{u}\nonumber\\
	&=&-\nu\lVert\bm{\nabla}\times\bm{\omega}\rVert^2-\nu\bm{\omega}\bm{\cdot}\nabla^2\bm{\omega}.
\end{eqnarray}
Substituting Eq.~\eqref{id3} and~\eqref{id4} into Eq.~\eqref{id2} gives the following result:
\begin{eqnarray}\label{id5}
	-\nu\lVert\bm{\nabla}\times\bm{\omega}\rVert^2+\nu\bm{u}\bm{\cdot}\nabla^2\left(\bm{\nabla}\times\bm{\omega}\right)
	-2\nu\bm{\omega}\bm{\cdot}\nabla^2\bm{\omega}=\nu\nabla^2\vartheta_{L}+\bm{\nabla}\bm{\cdot}\bm{q}.
\end{eqnarray}
From Eq.~\eqref{id5}, the equivalence between Eqs.~\eqref{transport_Lamb_dilatation} and~\eqref{HKK} is proved.

On a general stationary curved wall, the terms $-\nabla h_0\bm{\cdot}\left(\nabla\times\bm{\omega}\right)$
and $-\nu\lVert\bm{\nabla}\times\bm{\omega}\rVert^2$ cancel out with each other because a natural reduction of the NS equations to the boundary gives a on-wall force balance equation, namely,  $\rho^{-1}\left[\nabla p\right]_{\partial B}=\left[\nabla h_0\right]_{\partial B}=-\nu\left[\bm{\nabla}\times\bm{\omega}\right]_{\partial B}=\rho^{-1}\bm{\nabla}_{\partial B}p_{\partial B}+\rho^{-1}(\bm{\nabla}_{\partial B}\bm{\cdot}\bm{\tau})\bm{n}$. Although the total fluid acceleration on the wall is zero, the acceleration caused by either the pressure gradient or the viscous force is determined by the surface pressure gradient and the skin friction divergence. Additionally, applying the enstrophy transport equation on the wall, we have $\partial\Omega_{\partial B}/\partial t=\left[\nu\bm{\omega}\bm{\cdot}\nabla^2\bm{\omega}\right]_{\partial B}$. Combining these two results, Eq.~\eqref{HKK} reduces to $\partial[\vartheta_{L}]_{\partial B}/\partial t=-2\partial\Omega_{\partial B}/\partial t$ at the wall. This identity obviously holds by taking time derivatives of both sides of Eq.~\eqref{BLD}.

\section{Boundary Lamb dilatation flux and it physical interpretation}\label{BLDFPH}
\subsection{Boundary Lamb dilatation flux}
Due to the velocity boundary condition $\bm{u}_{\partial B}=\bm{0}$,
the Lamb vector $\bm{L}$ vanishes at the stationary wall and the boundary vorticity vector must be parallel to the tangent plane of the wall. However, once the vorticity is generated by virtue of the viscosity $\mu$ and surface pressure gradient $\bm{\nabla}_{\partial B}p_{\partial B}$ at the solid boundary and diffuses into the interior of the fluid, the spatial vortical structures rapidly become three-dimensionalized within a very limited distance away from the wall. Therefore, compared to the near-wall velocity and vorticity fields, a more compact and localized distribution of the Lamb vector will be observed slightly away from the wall. Indeed, the Lamb vector as the main aerodynamic force constituent is concentrated in a very thin layer (within about $10\%$ boundary-layer thickness) inside the boundary layer attached to the surface~\cite{WuJZ2006book,LiuTianshu2021}. This Lamb-vector maximum is the major
contributor to the entire vortex force acting on an airfoil~\cite{WuJZ2006book}.

The sources and sinks of the Lamb vector can be well described by the Lamb dilatation $\vartheta_L$~\cite{Hamman2008}.
The Lamb dilatation can be decomposed as the sum of a flexion product $\bm{u}\bm{\cdot}\bm{\nabla}\times\bm{\omega}$ and a non-positive part associated with the enstrophy $\Omega$, namely,
\begin{eqnarray}\label{vartheta_wall}
	\vartheta_L=\bm{\nabla}\bm{\cdot}\bm{L}=\bm{u}\bm{\cdot}\bm{\nabla}\times\bm{\omega}-2\Omega.
\end{eqnarray}
Interestingly, different from the Lamb vector itself, the Lamb dilatation at the wall (i.e., $\left[\vartheta_{L}\right]_{\partial B}$) generally does not vanish. From Eq.~\eqref{vartheta_wall}, we have
\begin{eqnarray}\label{BLD}
	\left[\vartheta_{L}\right]_{\partial B}=-2\Omega_{\partial B},
\end{eqnarray}
which implies that the non-positive Lamb dilatation at the wall is solely determined by the boundary enstrophy and thus the on-wall dissipation rate. A local maximum point of the boundary enstrophy must be a local minimum point of the boundary Lamb dilatation.

Like the BEF in Eq.~\eqref{BEF}, the wall-normal variation of the Lamb dilatation can be inferred from its boundary flux.
For incompressible viscous flows, by applying the Gauss theorem, the volume integral of the viscous diffusion term $\mu\nabla^2\vartheta_L$ over a closed volume $B$ yields a surface integral of its viscous flux $\mu\bm{\hat n}\bm{\cdot}\bm{\nabla}\vartheta_{L}$ over the boundary surface $\partial B$, i.e.,
\begin{eqnarray}
	\int_{B}\mu\nabla^2\vartheta_L dV=\oint_{\partial B}\mu\bm{\hat n}\bm{\cdot}\bm{\nabla}\vartheta_{L}dS,
\end{eqnarray}
where $\hat{\bm{n}}$ represents the outward unit normal vector of the boundary surface. 
By analogy with the concept of BEF, we can introduce the boundary Lamb dilatation flux (BLDF) as
\begin{eqnarray}
	f_{\vartheta_L}\equiv\mu\bm{ n}\bm{\cdot}\left[\bm{\nabla}\vartheta_{L}\right]_{\partial B},
\end{eqnarray}
which measures the diffusion rate of the Lamb dilatation through the boundary with the unit normal vector $\bm{n}=-\hat{\bm{n}}$ pointing from the boundary to the fluid.

For a general stationary curved surface, by performing wall-normal derivative to both sides of Eq.~\eqref{vartheta_wall} and applying the resulting equation on the wall, we obtain
\begin{eqnarray}\label{fL}
	f_{\vartheta_L}=-3\mu^{-1}\bm{\tau}\bm{\cdot}\bm{\nabla}_{\partial B}p_{\partial B}
	-2\mu\bm{\omega}_{\partial B}\bm{\cdot}\bm{K}\bm{\cdot}\bm{\omega}_{\partial B},
\end{eqnarray}
Substituting Eq.~\eqref{BEF} into Eq.~\eqref{fL} gives the exact relation between $f_{\vartheta_L}$ and $f_{\Omega}$: 
\begin{eqnarray}\label{fthetaL}
	f_{\vartheta_L}=-3f_{\Omega}+\mu\bm{\omega}_{\partial B}\bm{\cdot}\bm{K}\bm{\cdot}\bm{\omega}_{\partial B}.
\end{eqnarray}

As the first step of this theoretical study, we only consider flow past a stationary flat wall ($\bm{K}=\bm{0}$) in the present paper.
Indeed, theoretical results for a stationary flat wall are relatively simple in mathematics, although the detailed derivation is not trivial as shown in Appendix~\ref{Appendix1}.
Theoretical formulae for a stationary curved wall (and an arbitrarily moving and deforming wall) are much more complex due to the presence
of the coupling terms among different kinematic, dynamic and geometric quantities, which will be further explored and reported in a separate paper. For example, skin friction on an deforming wall will include not only the surface-vorticity-induced contribution, but also the contribution from the additional vorticity caused by local angular velocity of the wall. These complex coupling mechanisms will be directly related to high-fidelity numerical topics, which cannot be thoroughly addressed in a single research.

Therefore, for a stationary flat wall, Eqs.~\eqref{fL} and~\eqref{fthetaL} reduce to a concise relation between the BLDF and BEF:
\begin{eqnarray}\label{ff}
	f_{\vartheta_L}=-3f_{\Omega}=-3\mu^{-1}\bm{\tau}\bm{\cdot}\bm{\nabla}_{\partial B}p_{\partial B}.
\end{eqnarray}

Three comments are made here. {\it Firstly}, it is now clear that both the BLDF and BEF are determined by the nonlinear coupling mechanism between the skin friction $\bm{\tau}$ and the surface pressure gradient $\bm{\nabla}_{\partial B}p_{\partial B}$. The wall-normal diffusion of the Lamb dilatation is closely related to the boundary enstrophy generation and fundamental surface physical quantities.

{\it Secondly}, a negative BEF region ($f_\Omega<0$) must correspond to a positive BLDF region ($f_{\vartheta_{L}}>0$) for any instantaneous field in wall-bounded viscous flows, indicating a wall-normal increase of the Lamb dilatation inside a small vicinity of the wall.
In fact, in a region of this kind, we have $[\partial\vartheta_{L}/\partial n]_{\partial B}=\mu^{-1}f_{\vartheta_{L}}>0$,
which implies that $\vartheta_{L}$ increases monotonically with the wall-normal coordinate $y$ from its wall value $-2\Omega_{\partial B}$ in a small vicinity of the wall. According to the physical interpretation of the Lamb dilatation in Section~\ref{Lamb vector and Lamb dilatation}, these exists a trend to attenuate the vorticity bearing motion while to enhance the straining motion in negative BEF regions with the increasing wall-normal distance. In addition, the BEF remains negative in the sense of ensemble average~\cite{Liu2018AIA,ChenTao2021POF}, which implies that the averaged BLDF should be positive, implying an increase of the Lamb dilatation along the wall-normal direction in a small vicinity of the wall.
This analysis is consistent with the statistical results in a turbulent channel flow of $Re_{\tau}=590$~\cite{Hamman2008}: the mean Lamb dilatation increases from its negative minimum value at the wall till its positive maximum value is achieved at $y^+\approx12$ and the mean Bernoulli function (stagnation enthalpy) transitions from a subharmonic (local energy depletion) to superharmonic (local energy concentration) state.

{\it Thirdly}, characteristic points in the BEF field that usually indicate local flow separation or attachment must also be those of the BLDF field.
The BLDF provides a new physically-intuitive interpretation to the BEF because the Lamb dilatation has been shown as an effective structural indicator to coherent fluid motions and self-sustaining temporal-spatial interactions in the near-wall region~\cite{Hamman2008,Rousseaux2007,ChenXuLu2010}.

\subsection{BLDF and aerodynamic force}
As a direct application of Eq.~\eqref{fL}, a new total aerodynamic force expression is presented based on the Lamb dilatation and the corresponding 
physical mechanism of pressure drag reduction is clearly elucidated.

Considering a two-dimensional body $B$ bounded by a closed surface $\partial B$, the curvature contribution vanishes because the vorticity is perpendicular to the plane.
Under the continuity assumption, Eq.~\eqref{BLD} implies that there always exists a small vicinity of the wall such that $\vartheta_{L}=-\lVert\vartheta_{L}\rVert\leq0$.
Therefore, if $\vartheta_{L}\neq0$ holds for the region considered (extreme conditions are not included at present), Eq.~\eqref{fL} can be formally rewritten as 
\begin{eqnarray}\label{dpds}
	\frac{dp}{ds}=-\frac{f_{\vartheta_{L}}}{3\left[\sqrt{\lVert\vartheta_{L}\rVert}\right]_{\partial B}}
	=\frac{2}{3}\mu\left[\frac{\partial\sqrt{\lVert\vartheta_{L}\rVert}}{\partial n}\right]_{\partial B},
\end{eqnarray}
where $s$ is the arc length measured from a reference location $\bm{x}_0$ to the other location $\bm{x}$ on $\partial B$. Integrating Eq.~\eqref{dpds} from $\bm{x}_{0}=\bm{x}(0)$ to $\bm{x}=\bm{x}(s)$ gives the surface pressure distribution,
\begin{eqnarray}\label{surface_pressure}
	p_{\partial B}(\bm{x}(s))-p({\bm{x}_0})=\int_{0}^{s}\frac{2}{3}\mu\left[\frac{\partial\sqrt{\lVert\vartheta_{L}\rVert}}{\partial n}\right]_{\partial B}(s^\prime)ds^\prime,
\end{eqnarray}
which is directly related to the wall-normal viscous diffusion of the square root of the Lamb dilatation along the skin friction line. Note that $\oint_{\partial B}\bm{n}ds=\bm{0}$ and the skin friction $\bm{\tau}=\mu\bm{\omega}_{\partial B}\times\bm{n}$, using Eq.~\eqref{surface_pressure} to eliminate the surface pressure, the total aerodynamic force can be evaluated as
\begin{eqnarray}\label{total_force}
	\bm{F}&=&\oint_{\partial B}(-p_{\partial B}\bm{n}+\bm{\tau})ds\nonumber\\
	&=&-\frac{2}{3}\mu\oint_{\partial B}\left[\int_{0}^{s}\left[\frac{\partial\sqrt{\lVert\vartheta_{L}\rVert}}{\partial n}\right]_{\partial B}(s^\prime)ds^\prime\right]\bm{n}(s)ds+\mu\oint_{\partial B}\bm{\omega}_{\partial B}(s)\times\bm{n}(s)ds.
\end{eqnarray}
Physical interpretation of Eq.~\eqref{total_force} is given as follows.
On the one hand, it explicitly reveals the critical role of viscosity in generating the total force.
Lift and drag must coexist as a result of the incompressible viscous flow over an airfoil.
On the other hand, the lift force is mainly contributed by the first term due to the surface pressure integral while both the two terms can contribute to the drag force (including the pressure drag and skin friction drag). Therefore, balancing the positive and negative regions of the Lamb dilatation in the near-wall region gives the possibility to decrease the wall-normal variation of the near-wall Lamb dilatation and the magnitude of the wall-normal derivative of the square root of the Lamb dilatation, thereby reducing the pressure drag while the lift is sustained. Compared to the pressure drag reduction theory proposed by Hamman {\it et al.}~\cite{Hamman2008}, the present analysis provides a more natural and essential way to elucidate the physical mechanisms of pressure drag reduction and lift enhancement. It is noted that Liu {\it et al.}~\cite{Liu2017AIAA} gave a force formula similar to Eq.~\eqref{total_force} based on the BEF. Therefore, both the BEF and BLDF are useful physical concepts in interpreting the origin of the aerodynamic force.

\section{Temporal-spatial evolution of wall-normal Lamb dilatation flux}\label{TSEW}
In this section, we will discuss the temporal-spatial evolution rate of the wall-normal Lamb dilatation flux (WNLDF) $F_{\vartheta_L}\equiv\mu\partial\vartheta_{L}/\partial n$ on a solid wall $\partial B$. 
Applying $\mu\partial/\partial{n}$ to both sides of Eq.~\eqref{transport_Lamb_dilatation} and reducing it to the wall, we have
\begin{subequations}\label{LF}
	\begin{eqnarray}\label{CR}
		\left[\mathcal{L}F_{\vartheta_L}\right]_{\partial B}=Q_1+Q_2+Q_3+Q_4\equiv Q,
	\end{eqnarray}
	where the commonly used temporal-spatial evolution operator is $\mathcal{L}\equiv\partial/\partial t-\nu\nabla^2$ and the left hand side of Eq.~\eqref{CR} is the temporal-spatial evolution rate of the WNLDF on the wall.
	The source terms $Q_1$, $Q_2$, $Q_3$ and $Q_4$ are respectively given by
	\begin{eqnarray}\label{CRa}
		Q_1=-\left[\mu\frac{\partial}{\partial n}\left(\bm{u}\bm{\cdot}\bm{\nabla}\vartheta_{L}\right)\right]_{\partial B},
	\end{eqnarray}
	\begin{eqnarray}\label{CRb}
		Q_2=-2\left[\mu\frac{\partial}{\partial n}\left(\bm{\nabla L}\bm{:}\bm{\nabla u}^T\right)\right]_{\partial B},
	\end{eqnarray}
	\begin{eqnarray}\label{CRc}
		Q_3=-\left[\mu\frac{\partial}{\partial n}\left(\bm{\nabla}h_0\bm{\cdot}\bm{\nabla}\times\bm{\omega}\right)\right]_{\partial B},
	\end{eqnarray}
	\begin{eqnarray}\label{CRd}
		Q_4=\left[\mu\frac{\partial}{\partial n}\left(\bm{\nabla}\bm{\cdot}\bm{q}\right)\right]_{\partial B}.
	\end{eqnarray}
\end{subequations}
Eq.~\eqref{CR} indicates that the temporal-spatial evolution rate of the WNLDF at the wall is contributed by the superposition of different wall-normal viscous diffusion effects at the wall. Obviously, we have $\left[F_{\vartheta_L}\right]_{\partial B}=f_{\vartheta_L}$. 
Different source terms can be uniquely determined by using the fundamental surface physical quantities (skin friction $\bm{\tau}$, surface pressure $p_{\partial B}$, surface vorticity $\bm{\omega}_{\partial B}$, surface enstrophy $\Omega_{\partial B}$, etc.) as well as their temporal and tangential spatial derivatives on the wall.
After some calculations, explicit expressions of Eqs.~\eqref{CRa}--~\eqref{CRd} are given as
\begin{subequations}\label{LF2}
	\begin{eqnarray}\label{aa}
		Q_1=2\bm{\tau}\bm{\cdot}\bm{\nabla}_{\partial B}\Omega_{\partial B},
	\end{eqnarray}
	\begin{eqnarray}\label{bb}
		Q_2=\underbrace{4\bm{\tau}\bm{\cdot}\bm{\nabla}_{\partial B}\Omega_{\partial B}}_{Q_{21}}
		\underbrace{-4\left(\bm{\nabla}_{\partial B}\bm{\cdot}\bm{\tau}\right)\Omega_{\partial B}}_{Q_{22}},
	\end{eqnarray}
	\begin{eqnarray}\label{cc}
		Q_3&=&\underbrace{-2\left(\bm{\nabla}_{\partial B}\bm{\cdot}\bm{\tau}\right)\Omega_{\partial B}}_{Q_{31}}
		\underbrace{-\frac{2}{\rho}\bm{\nabla}_{\partial B}\left(\bm{\nabla}_{\partial B}\bm{\cdot}\bm{\tau}\right)\bm{\cdot}\bm{\nabla}_{\partial B}p_{\partial B}}_{Q_{32}}\nonumber\\
		& &
		\underbrace{+\frac{2}{\rho}\left(\bm{\nabla}_{\partial B}\bm{\cdot}\bm{\tau}\right)
			(\nabla_{\partial B}^{2}p_{\partial B})}_{Q_{33}}\underbrace{+\frac{1}{\mu}\frac{\partial\bm{\tau}}{\partial t}\bm{\cdot}\bm{\nabla}_{\partial B}p_{\partial B}}_{Q_{34}},
	\end{eqnarray}
	\begin{eqnarray}\label{dd}
		Q_4&=&
		\underbrace{\frac{6}{\rho}\bm{\nabla}_{\partial B}\bm{\tau}\bm{:}\bm{\nabla}_{\partial B}\bm{\nabla}_{\partial B}p_{\partial B}}_{Q_{41}}
		\underbrace{+\frac{6}{\mu}\frac{\partial\bm{\tau}}{\partial t}\bm{\cdot}\bm{\nabla}_{\partial B}p_{\partial B}}_{Q_{42}}
		\underbrace{+\frac{2}{\rho}(\bm{\nabla}_{\partial B}\bm{\cdot}\bm{\tau})({\nabla}_{\partial B}^{2}p_{\partial B})}_{Q_{43}}\nonumber\\
		& &\underbrace{-\frac{4}{\rho}\nabla_{\partial B}^{2}\bm{\tau}\bm{\cdot}\bm{\nabla}_{\partial B}p_{\partial B}}_{Q_{44}}
		\underbrace{-\frac{2}{\rho}\bm{\nabla}_{\partial B}(\bm{\nabla}_{\partial B}\bm{\cdot}\bm{\tau})\bm{\cdot}\bm{\nabla}_{\partial B}p_{\partial B}}_{Q_{45}}
		\nonumber\\
		& &\underbrace{+\frac{2}{\mu}\bm{\tau}\bm{\cdot}\mathcal{L}_{\partial B}\left(\bm{\nabla}_{\partial B}p_{\partial B}\right)}_{Q_{46}}
		\underbrace{+4\bm{\tau}\bm{\cdot}\bm{\nabla}_{\partial B}\Omega_{\partial B}}_{Q_{47}}
		\underbrace{-4\left(\bm{\nabla}_{\partial B}\bm{\cdot}\bm{\tau}\right)\Omega_{\partial B}}_{Q_{48}}.
	\end{eqnarray}
\end{subequations}
where ${\mathcal{L}}_{\partial B}\equiv\partial/\partial{t}-\nu\nabla_{\partial B}^{2}$ represents the temporal-spatial evolution operator on the surface.
When the surface physical quantities can be determined, different contributions to the temporal-spatial evolution rate of the WNLDF at the wall can be quantitatively computed by using Eqs.~(\ref{aa})--~(\ref{dd}).
The dominant terms should depend on the specific flow type, which could be determined by experiments and numerical simulations.

Some derivation details are included in Appendix~\ref{Appendix1}, where $\bm{\sigma}$ represents the boundary vorticity flux (BVF).
The concept of the BVF was first introduced by Lighthill~\cite{Lighthill1963} to explain the vorticity creation rate at the solid boundary in two-dimensional case and was generalized by Panton~\cite{Panton1984} to three-dimensional case. A general explicit decomposition of the BVF was first obtained by Wu and Wu~\cite{WuJZ1996,WuJZWuJM1998} by recasting the NS equations to the solid wall with the no-slip boundary condition.
For a stationary flat wall ($\bm{\omega}_{\partial B}\bm{\cdot}\bm{n}=\bm{0}$ and $\bm{K}=\bm{0}$), according to Lighthill-Panton-Wu's definition, $\bm{\sigma}$ can be expressed as
\begin{eqnarray}\label{BVF}
	\bm{\sigma}&&\equiv\mu\bm{n}\bm{\cdot}[\bm{\nabla\omega}]_{\partial B}\nonumber\\
	&&=-\mu\left[\bm{n}\times(\bm{\nabla}\times\bm{\omega})\right]_{\partial B}+\mu\left[(\bm{n}\times\bm{\nabla})\times\bm{\omega}\right]_{\partial B}\nonumber\\
	&&=\bm{n}\times\bm{\nabla}_{\partial B}p_{\partial B}-\mu\left(\bm{\nabla}_{\partial B}\bm{\cdot}\bm{\omega}_{\partial B}\right)\bm{n}.
\end{eqnarray}

It should be mentioned that Lyman~\cite{Lyman1990} also proposed a kinematically equivalent definition of the BVF.
There has some controversy over these two definitions and the role of viscosity in vorticity creation process.
Based on Lyman's definition of the BVF, Morton~\cite{Morton1984} and Terrington {\it et al.}~\cite{Terrington2021} insisted that the vorticity (or circulation) creation on a solid wall was an inviscid process, independent of the viscosity. The role of viscosity was to drive the vorticity diffusion process
immediately after vorticity generation due to the boundary acceleration
and surface pressure gradient. However, Wu and Wu~\cite{WuJZWuJM1998} argued
that both the viscosity and no-slip boundary condition were essential to the vorticity creation process because
the interface vortex sheet on a free-slip solid boundary did not represent the
rotation of fluid elements and therefore did not represent a layer of vorticity on the boundary.
In fact, the continuity of the  acceleration at the boundary is a natural result derived from that of the velocity (namely, the no-slip boundary condition), which is obviously not consistent with Morton's inviscid interpretation. In other words, the viscous flow with the viscosity $\mu\rightarrow 0$ is essentially different with the purely inviscid flow with $\mu=0$. From modern aerodynamic perspective, as recently claimed by Wu {\it et al.}~\cite{WuJZ2018PAS} and Liu~\cite{LiuTianshu2021}, the generation of the circulation and lift is a viscous-flow phenomenon. The Kutta-Joukowski inviscid circulation theory for lift, including the Kutta condition for determining the circulation, is only rationalized in the viscous-flow framework.
In this paper, Lighthill's definition of BVF is applied.

\section{Numerical method}\label{Numerical_method}
After the above theoretical derivations and physical interpretations, we will simulate a two-dimensional (2D) laminar lid-driven cavity flow and a three-dimensional (3D) turbulent channel flow, in order to demonstrate the application of these exact relations preliminarily.
A brief summary of the main contents and innovative points is given as follows.

For 2D lid-driven cavity flow, the lattice Boltzmann method (LBM)~\cite{ChenDoolen1998} is applied by solving the following equation:
\begin{eqnarray}
	f(\bm{x}+\bm{\xi}\Delta{t},\bm{\xi},t+\Delta{t})=f(\bm{x},\bm{\xi},t)+\frac{f^{eq}-f}{\tau_{LBM}}
	+\left(1-\frac{1}{2\tau_{LBM}}\right)\frac{\bm{b}\bm{\cdot}\bm{c}}{RT}f^{eq}\Delta{t},
\end{eqnarray}
where $f$ represents the transformed distribution function, $f^{eq}$ is the Maxwellian equilibrium,
$\bm{x}$ is the spatial location, $t$ is the time, $\bm{\xi}$ is the particle velocity.
$\bm{c}=\bm{\xi}-\bm{u}$ is the thermal fluctuating velocity where $\bm{u}$ is the hydrodynamic velocity.
$R$ is the gas constant and $T$ is the temperature. The dimensionless relaxation time $\tau_{LBM}=\tau/\Delta{t}+1/2$ where $\tau=\nu/RT$, $\nu$ is the kinematic viscosity and $\Delta t$ is the time step. $\bm{b}$ is the body force per unit mass.

The density $\rho$, the momentum $\rho\bm{u}$ and the viscous stress tensor $\bm{\Pi}$ are updated by using
\begin{eqnarray}\label{hydroQ}
	&&\rho=\int fd\bm{\xi},\rho\bm{u}=\int\bm{\xi}fd\bm{\xi}+\frac{\Delta t}{2}\rho\bm{b},\nonumber\\
	&&\bm{\Pi}=-\left(1-\frac{1}{2\tau_{LBM}}\right)\int\bm{c}\bm{c}\left(f-f^{eq}\right)d\bm{\xi}.
\end{eqnarray}
By utilizing the Gauss-Hermite quadrature~\cite{Shan2006}, the integrals over the continuous particle velocity space can be 
numerically calculated. In order to eliminate the compressibility errors inherited in the LBM,
the density $\rho$ is partitioned into a background density $\rho_0$ plus a fluctuating part $\delta{\rho}$: $\rho=\rho_0+\delta{\rho}$
and the Maxwellian equilibrium is truncated to $O(Ma^2)$ in the Hermite expansion~\cite{HeLuo1997}. In addition, a modified
on-wall bounce back scheme is used where the first lattice node is directly located at the wall, so that the momentum balance can be ensured in the near-wall region. The code has been validated by comparing with the published
benchmark data sets~\cite{XuHe2003,AbdelMigid2017}.

Simulation of the 3D turbulent channel flow is performed by a multi-relaxation-time lattice Boltzmann method (MRT-LBM). The code was originally developed and validated by Wang {\it et al.}~\cite{WangLP2016}, which was recently compared with the most accurate online DNS data accepted by the turbulence community at the moment~\cite{ChenTao2021POF}.
A modified on-wall bounce back is used in the simulation, which guarantees not only the statistical robustness but also the momentum balance in the near-wall region.
The streaming and collision steps are implemented in the particle velocity space and the transformed momentum space, respectively. Some parameters can be flexibly tuned to enhance the numerical instability because they do not influence the hydrodynamics at the NS order according to the Chapman-Enskog analysis.
Instead of using the finite difference scheme, the strain rate tensor $S_{ij}=(\partial u_i/\partial x_j+\partial u_j/\partial x_i)/2$ is directly obtained from the non-equilibrium moments $m^{neq}$ by summing over the contributions from different particle directions. Then, the skin friction $\bm{\tau}$ at the wall is obtained using $\tau_{x}=2\mu[S_{yx}]_{y=0}$ and $\tau_{z}=2\mu[S_{yz}]_{y=0}$. The boundary vorticity $\bm{\omega}_{\partial B}$ is then computed using the on-wall orthogonality relation $\bm{\omega}_{\partial B}=\mu^{-1}\bm{n}\times\bm{\tau}$.

Evaluating the source terms ($Q_1$, $Q_2$, $Q_3$ and $Q_4$) based on their original definitions involves the calculation of the high-order wall-normal velocity derivatives at the wall and thus requires more available data points in the near-wall region or in the ghost cells out of the computational domain.
By the use of Eqs.~\eqref{aa}--~\eqref{dd}, these source terms can be computed from the information of the surface quantities as well as their temporal and tangential spatial derivatives at the wall. In this way,
evaluating the high-order wall-normal velocity derivatives based on their original definition is technically circumvented.
\section{Numerical simulation and analysis}\label{analysis}
\subsection{2D lid-driven square cavity flow}\label{cavity_flow_sec}
In this subsection, we simulate a 2D steady lid-driven square cavity flow
at the Reynolds number $Re\equiv UL/\nu=1000$, where $U$ is the lid speed, $L$ is the cavity height 
and $\nu$ is the kinematic viscosity. $256\times256$ uniform meshes are used in the simulation.
The Mach number $Ma\equiv U/\sqrt{RT}$ is equal to $0.173$ such that the Knudsen number
$Kn\equiv\sqrt{RT}\tau/L=Ma/Re$ is $0.000173$, where $\tau$ is the dimensional relaxation time. This sufficiently low Knudsen number implies that the flow is entirely at the continuum regime 
which can be described by the NS equations accurately.
The variations of the velocity and pressure along the horizontal and vertical centerlines have been extensively used as the canonical benchmarks for numerical tests. However, to the authors' knowledge, both the BLDF and the temporal-spatial evolution rate of the WNLDF at the bottom wall are not studied yet.
\begin{figure}[h!]
	\centering
	\includegraphics[width=0.8\columnwidth,trim={0.1cm 2.9cm 0.1cm 2.8cm},clip]{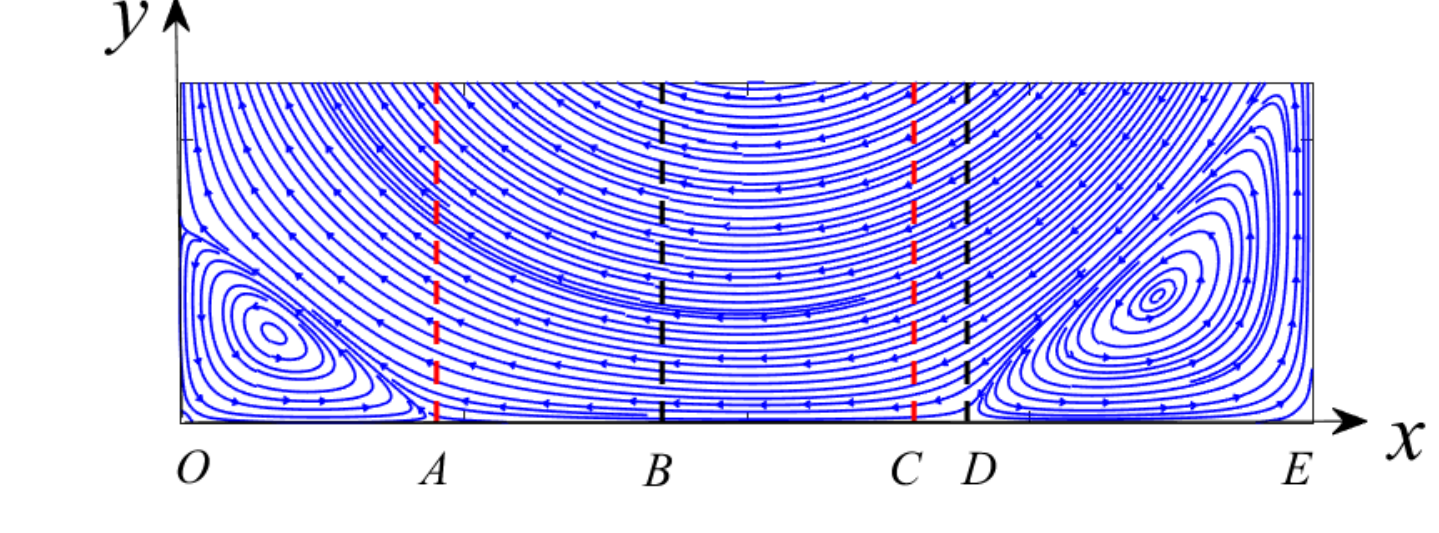}
	\caption{Near-wall streamlines in a 2D lid-driven cavity flow at $Re=1000$. The origin of the coordinate system $(x^*,y^*)=(x,y)/L$ locates at the point $O(0,0)$ and $E(1,0)$ represents the right bottom corner point. $A(0.2266,0)$ is the separation point and $D(0.6953,0)$ is the attachment point on the bottom wall. $B(0.4258,0)$ and $C(0.6484,0)$ denote the local minimum and maximum surface pressure points, respectively.} 
	\label{cavity_streamlines}
\end{figure}

\begin{figure}[h]
	\centering
	\subfigure[$\lVert\bm{\tau}^{*}\rVert$]{
		\begin{minipage}[t]{0.5\linewidth}
			\centering
			\includegraphics[width=1.0\columnwidth,trim={0.1cm 0.14cm 0.1cm 0.6cm},clip]{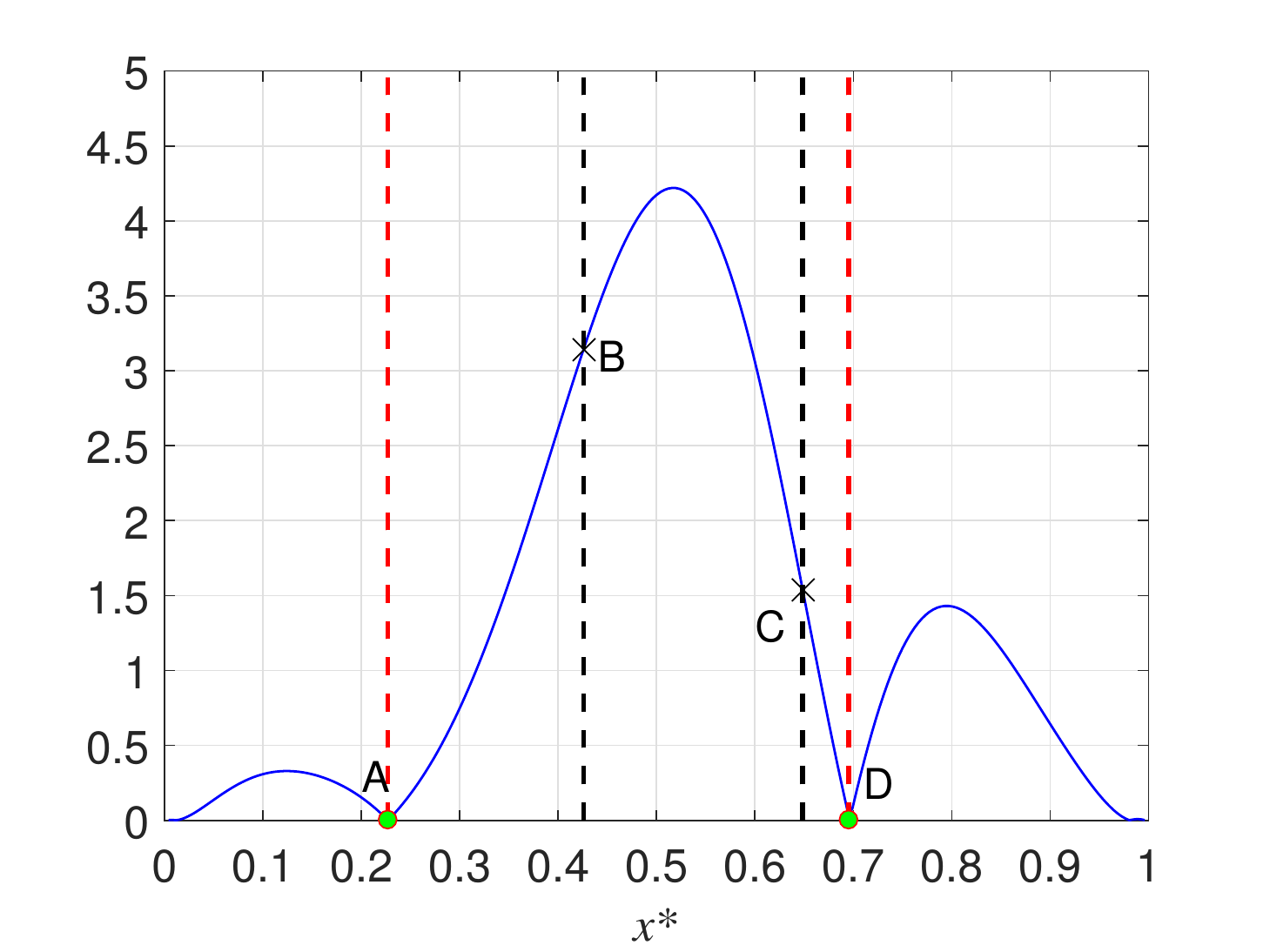}
			\label{cavity_sqrtBLD}
		\end{minipage}%
	}%
	\subfigure[$f_{\vartheta_{L}}^{*}$]{
		\begin{minipage}[t]{0.5\linewidth}
			\centering
			\includegraphics[width=1.0\columnwidth,trim={0.1cm 0.14cm 0.1cm 0.6cm},clip]{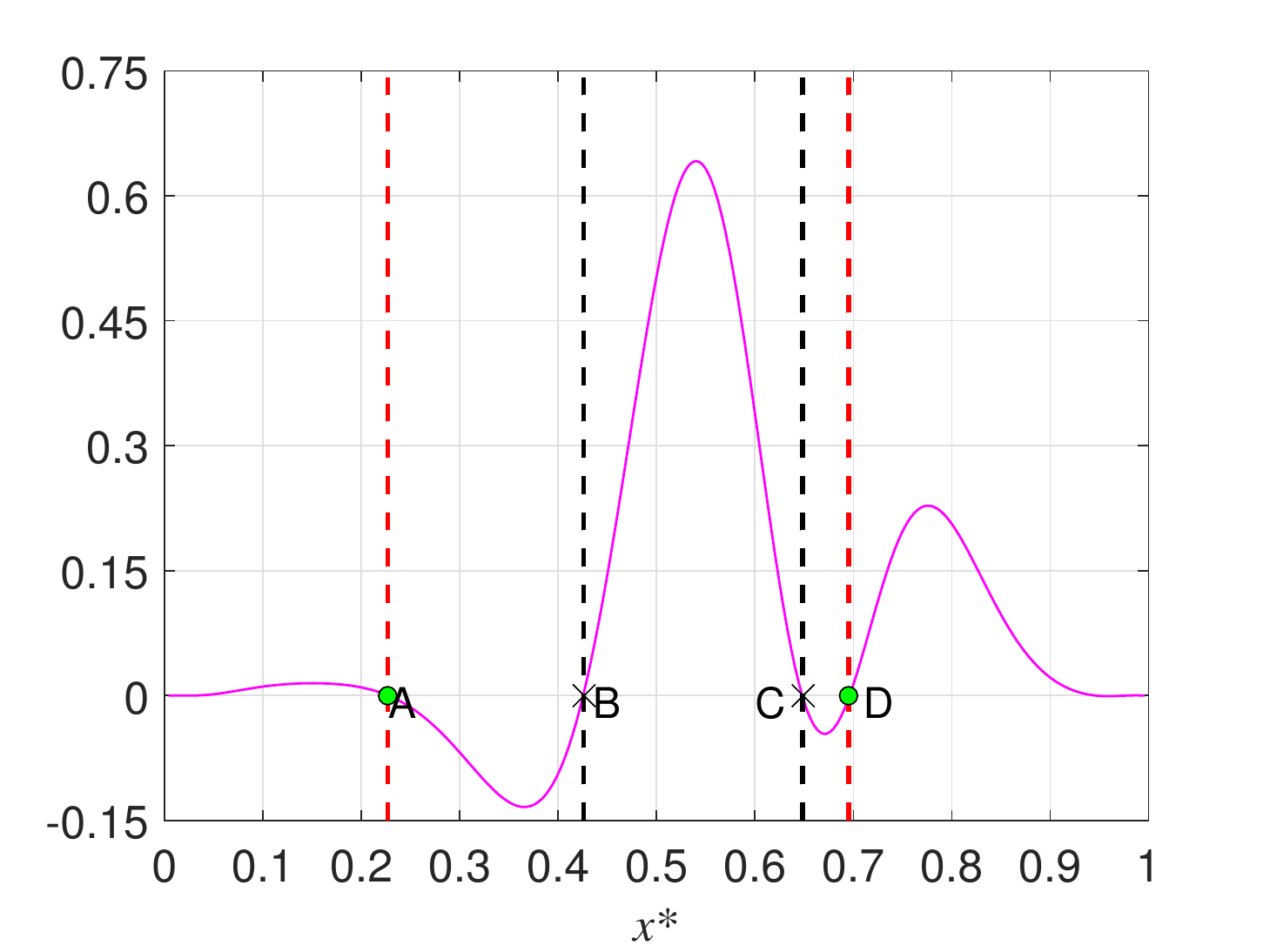}
			\label{fvarthetaL}
		\end{minipage}%
	}%
	\caption{Normalized surface quantities on the bottom wall in 2D lid-driven cavity at $Re=1000$. ({\it a}) skin friction magnitude $\lVert\bm{\tau}^{*}\rVert$ and ({\it b}) the boundary Lamb dilatation flux (BLDF) $f_{\vartheta_{L}}^{*}\equiv f_{\vartheta_{L}}/(\rho U^3/L^2)$.} 
	\label{xx0}
\end{figure}

Fig.~\ref{cavity_streamlines} shows the streamlines near the bottom wall. Due to interaction between the near-wall viscous flow and the bottom wall, distinct topological features can be observed at the bottom wall. The separation point $A$ and the attachment point $D$ correspond the two zero skin friction points, while $B$ and $C$ denote the local pressure minimum and maximum points, respectively. Eq.~\eqref{BLD} implies that $\left[\sqrt{\lVert\vartheta_{L}^{*}\rVert}\right]_{\partial B}=\lVert\bm{\tau}^{*}\rVert=\lVert\bm{\omega}_{\partial B}^{*}\rVert$. Therefore, the zero points of the boundary Lamb dilatation are consistent with the zero points of skin friction field (or surface vorticity field), as shown in Fig.~\ref{cavity_sqrtBLD}.
Three local maximum points of the skin friction magnitude can also be observed in $OA$, $BC$ and $DE$, respectively.

In Fig.~\ref{fvarthetaL}, the two zero skin friction points ($A$ and $D$) and the two pressure extreme points ($B$ and $C$) are just the four zero-crossing points of the BLDF. From Eq.~\eqref{ff}, skin friction $\tau_x$ and surface pressure gradient $\partial p_{\partial B}/\partial x$ are coupled through the BLDF $f_{\vartheta_{L}}$, namely, $\tau_x\cdot\partial p_{\partial B}/\partial x=-\mu f_{\vartheta_{L}}/3$.
Therefore, $f_{\vartheta_{L}}=0$ implies that $\tau_x=0$ or $\partial p_{\partial B}/\partial x=0$. The solutions of $\tau_x=0$ correspond to the two zero skin friction points ($A$ and $D$), while those of $\partial p_{\partial B}/\partial x=0$ correspond to the two pressure extreme points ($B$ and $C$). Positive BLDF is found in the main contact region between the primary vortex and the wall ($BC$), as well as the regions below the two small corner vortices. In constrast, the signs of the BLDF become negative in the regions $AB$ and $CD$. Positive (negative) BLDF indicates that the enhancement (attenuation) of the straining motion and the attenuation (enhancement) of the vortical motion along the wall-normal direction in a small vicinity of the wall.

\begin{figure}[h]
	\centering
	\subfigure[]{
		\begin{minipage}[t]{0.5\linewidth}
			\centering
			\includegraphics[width=1.0\columnwidth,trim={0.1cm 0.14cm 0.1cm 0.6cm},clip]{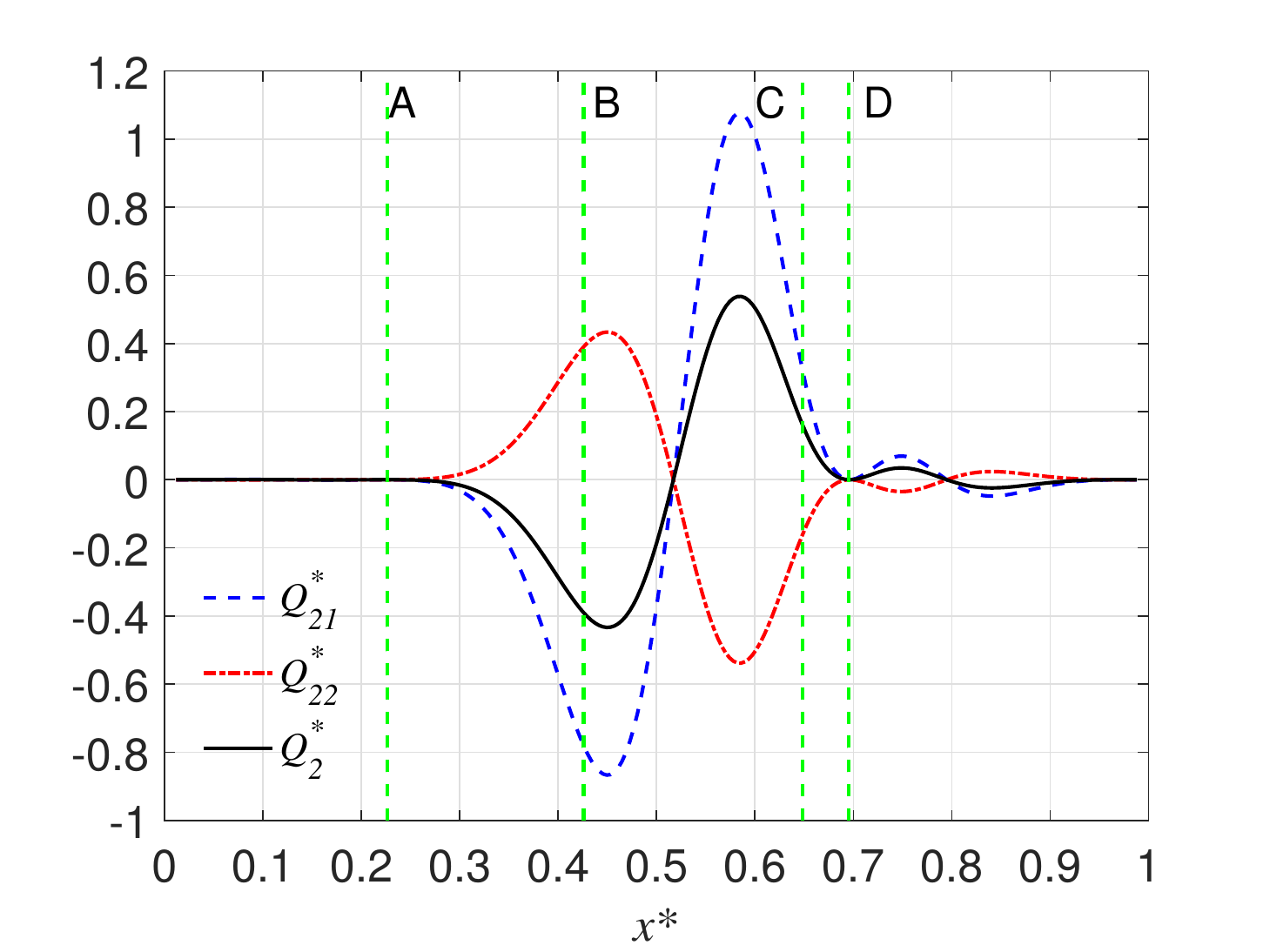}
			\label{cavity_Q2}
		\end{minipage}%
	}%
	\subfigure[]{
		\begin{minipage}[t]{0.5\linewidth}
			\centering
			\includegraphics[width=1.0\columnwidth,trim={0.1cm 0.14cm 0.1cm 0.6cm},clip]{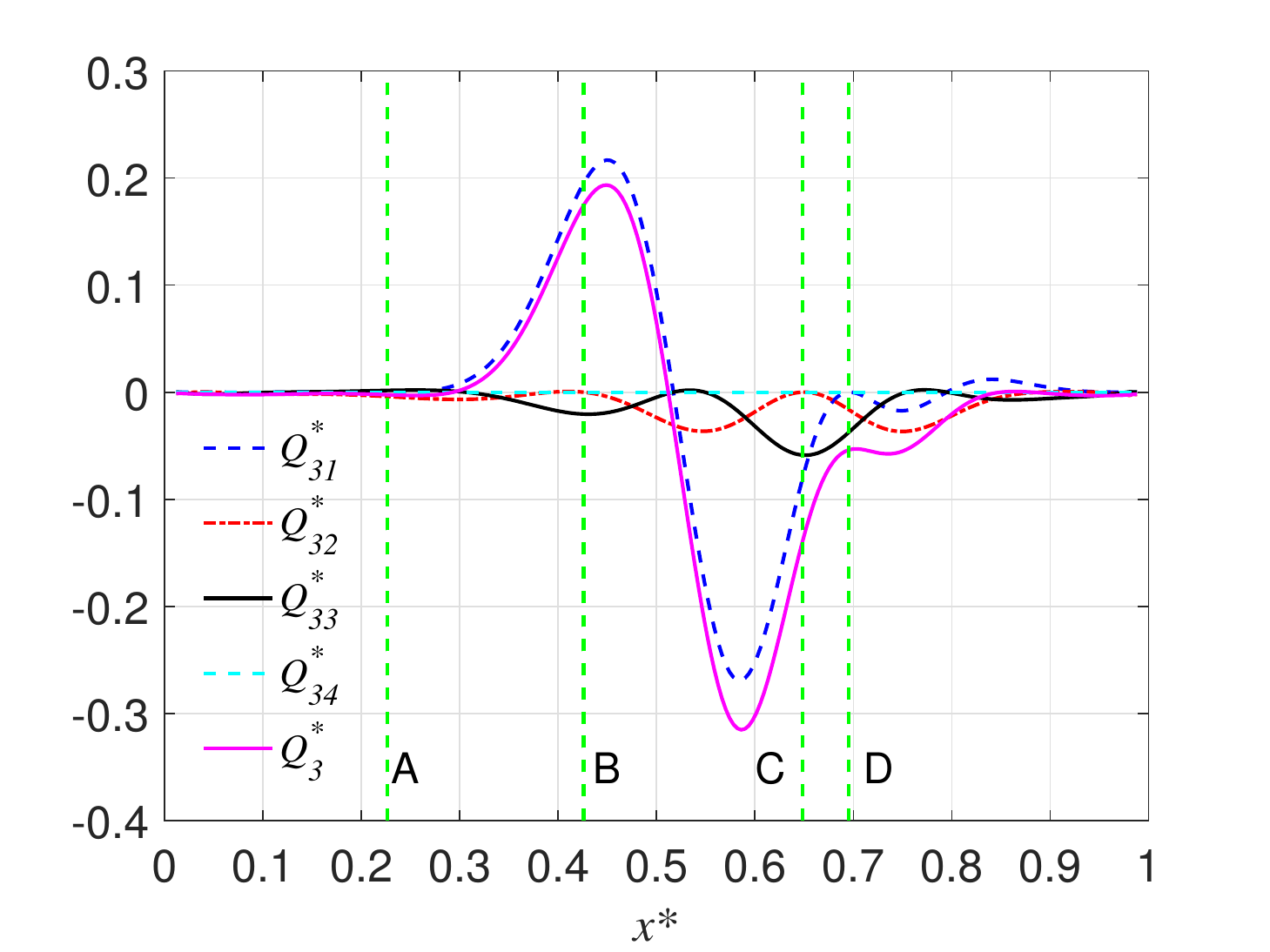}
			\label{cavity_Q3}
		\end{minipage}%
	}%
	
	\subfigure[]{
		\begin{minipage}[t]{0.5\linewidth}
			\centering
			\includegraphics[width=1.0\columnwidth,trim={0.1cm 0.14cm 0.1cm 0.6cm},clip]{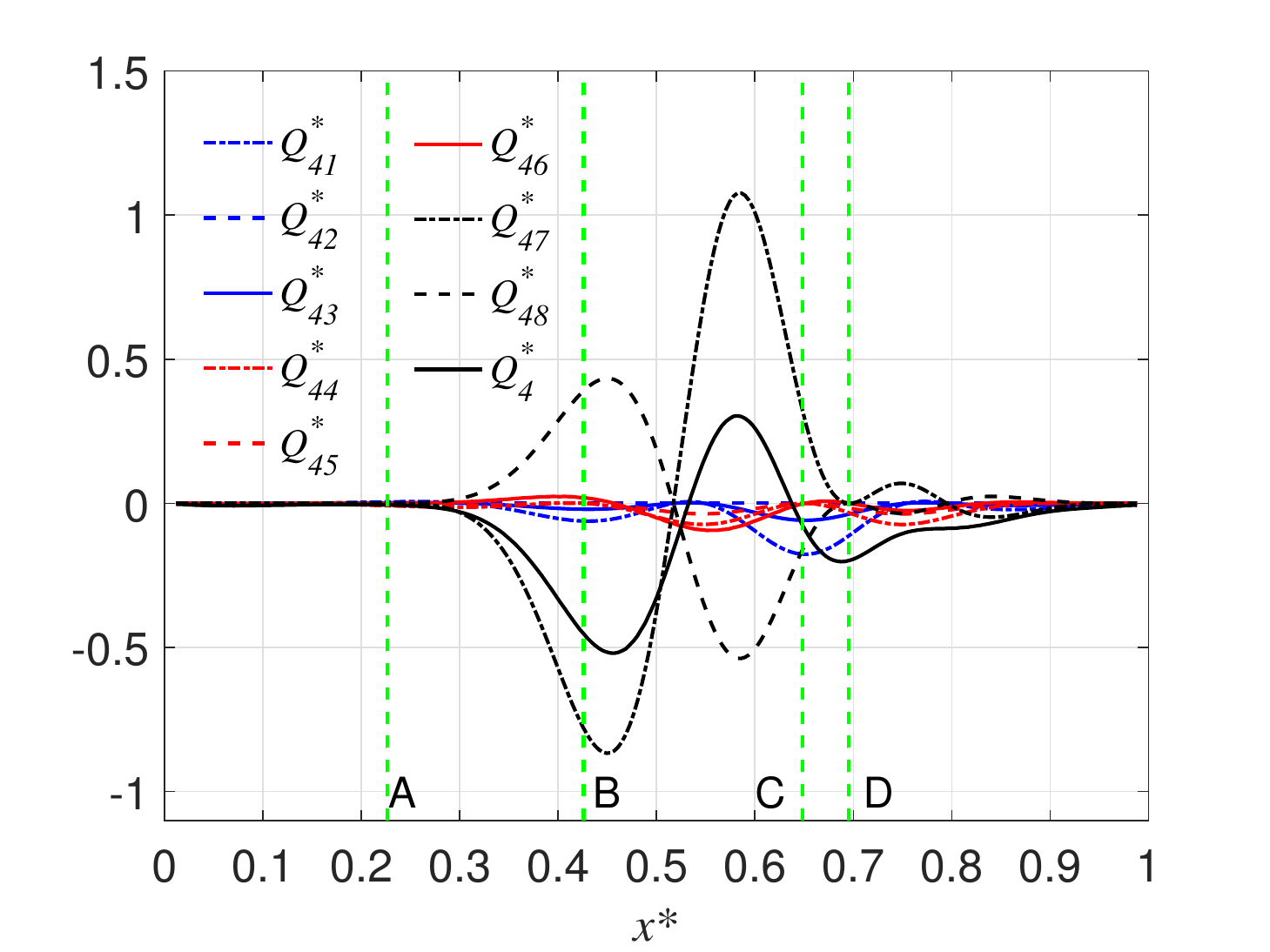}
			\label{cavity_Q4}
		\end{minipage}%
	}%
	\subfigure[]{
		\begin{minipage}[t]{0.5\linewidth}
			\centering
			\includegraphics[width=1.0\columnwidth,trim={0.1cm 0.14cm 0.1cm 0.6cm},clip]{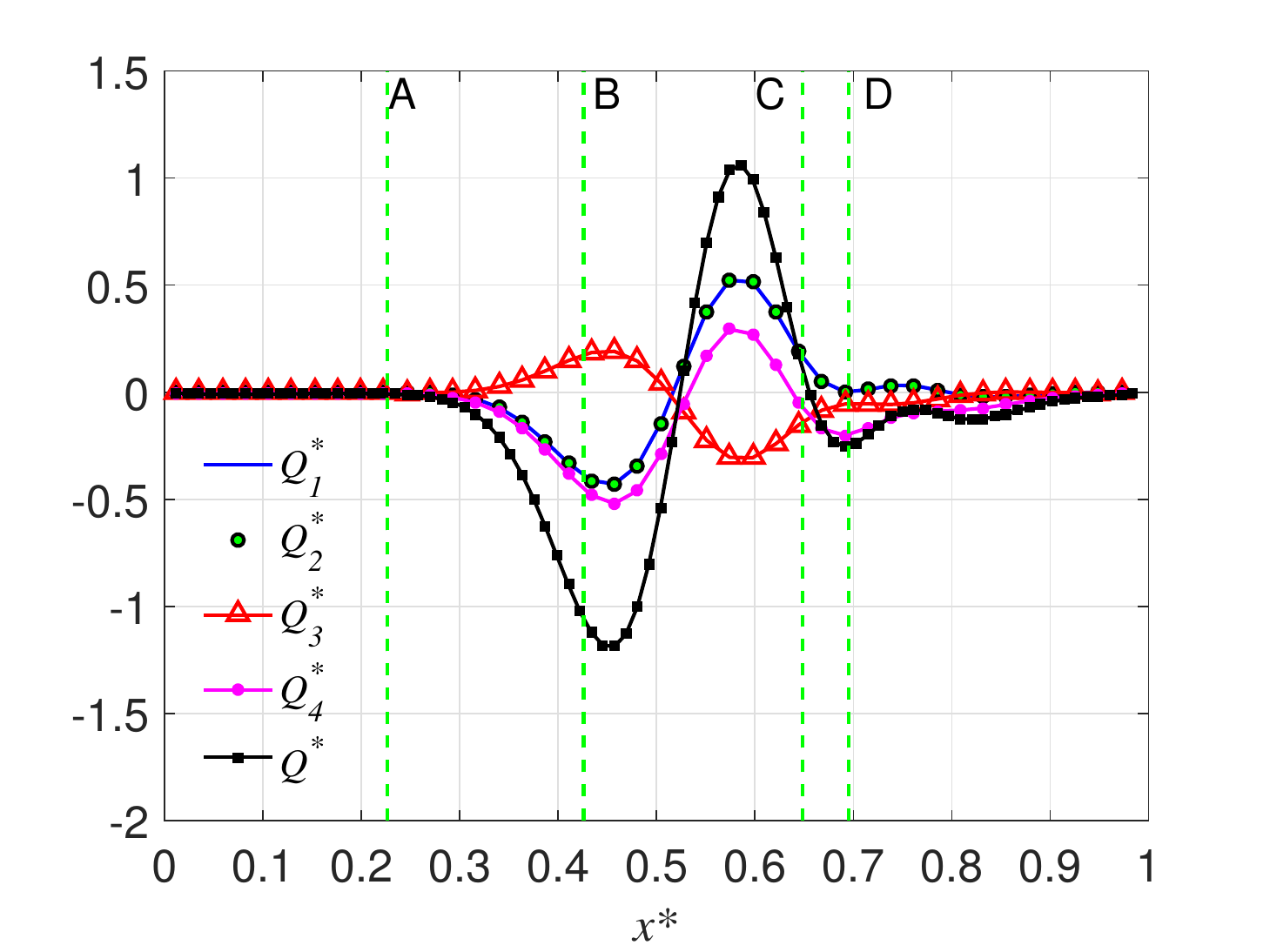}
			\label{cavity_Q}
		\end{minipage}%
	}%
	\caption{Comparison of different contributions to ({\it a}) the wall-normal diffusion of $-2\bm{\nabla L}\bm{:}\bm{\nabla u}^T$ (i.e., $Q_2$), ({\it b}) the wall-normal diffusion of $-\bm{\nabla}h_0\bm{:}\bm{\nabla}\times\bm{\omega}$ (i.e., $Q_{3}$), ({\it c}) the wall-normal
		diffusion of $\bm{\nabla}\bm{\cdot}\bm{q}$ (i.e., $Q_4$) and ({\it d}) the temporal-spatial evolution rate of the WNLDF at the wall (i.e., $Q$) in a 2D lid-driven cavity flow at $Re=1000$. The source terms are normalized by $\rho U^4/L^3$.} 
	\label{xx0}
\end{figure}

Next, the temporal-spatial evolution rate of the WNLDF at the wall is discussed based on Eqs.~\eqref{LF} and~\eqref{LF2}. These expressions can be greatly simplified since the skin friction line is parallel to the surface pressure gradient line at the one-dimensional bottom wall.
Fig.~\ref{cavity_Q2} displays different contributions to the wall-normal diffusion of $Q_{2}$.
Since the global maximum point of the skin friction magnitude (denoted as $F$ where $d\tau_x/dx=0$) lies inside the region $BC$, we have $d\tau_x/dx<0$ and $d\tau_x/dx>0$ in the regions $AF$ and $FD$, respectively. 
Note that the coupling between the skin friction and the boundary enstrophy gradient is 
$Q_{21}=4\bm{\tau}\bm{\cdot}\bm{\nabla}_{\partial B}\Omega_{\partial B}=4\mu^{-2}\tau_{x}^{2}(d\tau_{x}/dx)$ and the skin friction divergence-boundary enstrophy coupling term is $Q_{22}=-4(\bm{\nabla}_{\partial B}\bm{\cdot}\bm{\tau})\Omega_{\partial B}=-2\mu^{-2}\tau_{x}^{2}(d\tau_{x}/dx)$. Therefore, the magnitude of $Q_{21}$ is twice that of $Q_{22}$. The sum of $Q_{21}$ and $Q_{22}$ gives $Q_{2}=2\mu^{-2}\tau_{x}^{2}(d\tau_{x}/dx)$, which has the same magnitude compared to $Q_{22}$ (with different sign) and $Q_{1}$. The variation of $Q_{2}$ mainly concentrates in the region between the two zero skin friction points. $Q_{2}$ is basically less than zero inside $AF$ while is greater than zero inside $FD$.

Fig.~\ref{cavity_Q3} shows different contributions to $Q_3$.
$Q_{31}=-2(\bm{\nabla}_{\partial B}\bm{\cdot}\bm{\tau})\Omega_{\partial B}$ makes the primary contribution to $Q_3$, which is determined by the coupling between
the skin friction divergence $\bm{\nabla}_{\partial B}\bm{\cdot}\bm{\tau}$ and the boundary enstrophy $\Omega_{\partial B}$.
$Q_{32}$ and $Q_{33}$ represent the non-linear coupling between the spatial derivatives of the skin friction
and those of the surface pressure. They demonstrate relatively small contributions compared to that from $Q_{31}$.
Their contributions are mainly concentrated near the pressure maximum point $C$ and the zero skin friction point $D$ due to the relatively
strong sweep motion induced by the primary vortex.  
Similar comparison is shown for $Q_{4}$ in Fig.~\ref{cavity_Q4}.
It is clear that the main contributions come from $Q_{47}=4\bm{\tau}\bm{\cdot}\bm{\nabla}_{\partial B}\Omega_{\partial B}=4\mu^{-2}\tau_{x}^{2}(d\tau_{x}/dx)$ and $Q_{48}=-4(\bm{\nabla}_{\partial B}\bm{\cdot}\bm{\tau})\Omega_{\partial B}=-2\mu^{-2}\tau_{x}^{2}(d\tau_{x}/dx)$. The obvious but relatively small variation of $Q_{41}$ are also observed 
around the pressure maximum point. Contributions from all the other terms are less significant.
Fig.~\ref{cavity_Q} illustrates different contributions to the temporal-spatial evolution rate of 
the WNLDF at the wall (i.e., $Q$). The variations of all the source terms show similar tendency that enhances the 
magnitude of $Q$ within the region $AD$. 

\begin{figure}[h!]
	\centering
	\includegraphics[width=0.5\columnwidth,trim={0.1cm 0.14cm 0.1cm 0.6cm},clip]{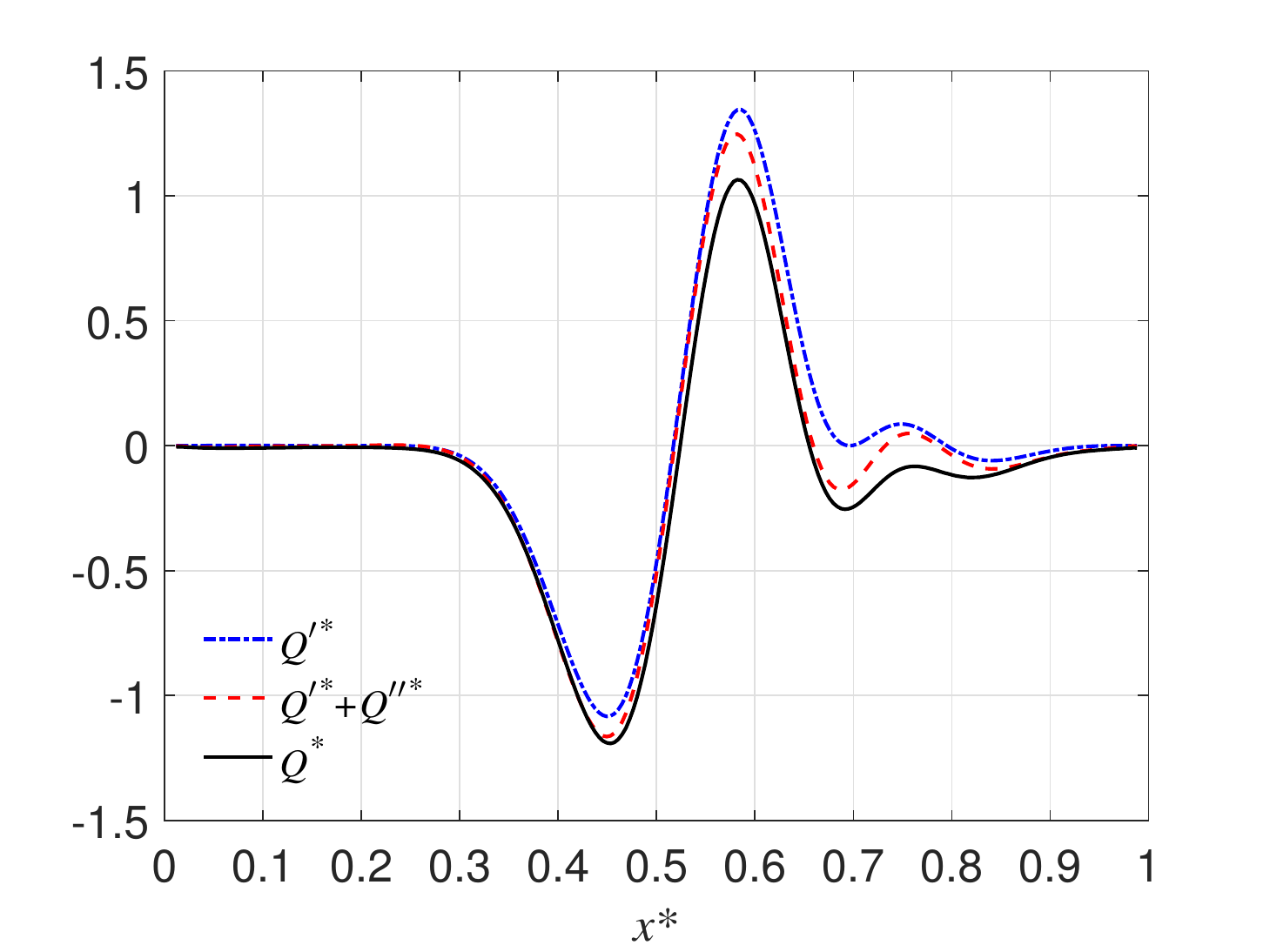}
	\caption{Comparison of $Q^{\prime}$, $Q^{\prime}+Q^{\prime\prime}$ and $Q$ on the bottom wall in a 2D lid-driven cavity flow at $Re=1000$.} 
	\label{cavity_Qmodel}
\end{figure}

Based on the above observation and analysis, the temporal-spatial evolution rate of the WNLDF at the wall (i.e., $Q$) is found to be predominantly contributed by
$Q^{\prime}\equiv10\bm{\tau}\bm{\cdot}\bm{\nabla}_{\partial B}\Omega_{\partial B}-10(\bm{\nabla}_{\partial B}\bm{\cdot}\bm{\tau})\Omega_{\partial B}$ and is less significantly influenced by the coupling effects between the spatial derivatives of skin friction and surface pressure gradient (dominated by $Q_{33}$ and $Q_{41}$, $Q^{\prime\prime}\equiv Q_{33}+Q_{41}$ denotes their sum). Comparison of $Q^{\prime}$, $Q^{\prime}+Q^{\prime\prime}$ and $Q$ on the bottom wall is demonstrated in Fig.~\ref{cavity_Qmodel}. It is clear that $Q^{\prime}$ approximates $Q$ very well and the discrepancy near the peak regions could be further reduced by adding the coupling terms in $Q^{\prime\prime}$.
\subsection{3D turbulent channel flow}
A single-phase incompressible turbulent channel flow is simulated at the frictional Reynolds number $Re_{\tau}=u_{\tau}H/\nu=180$.
$H$ is the half channel height between the two no-slip flat walls.
The domain size is $L_x\times L_y\times L_z=1800\times299\times600$ lattices with $x$, $y$ and $z$ being streamwise, wall-normal and spanwise directions, respectively.
$u_{\tau}=\sqrt{\langle\tau_x\rangle/\rho_{0}}$ is the friction velocity and $y_{\tau}=\nu/u_{\tau}$
is the wall viscous length unit in the viscous sublayer, where $\langle\bm{\cdot}\rangle$ denotes the ensemble average.
After the turbulence evolves into a statistically steady state, the pertubation force is switched off
and the flow is only driven by a constant driving force $g$ in the streamwise direction, which is equivalent to a mean 
pressure gradient $\partial\langle p\rangle/\partial{x}=-\rho_{0}g$.
Three hundred grid points are applied in the wall-normal direction with the first and the last points 
at the channel walls. In the following figures, a physical quantity with a superscript $+$ means the normalization by $u_{\tau}$ and $y_{\tau}$.
\begin{figure}[h]
	\centering
	\includegraphics[width=0.6\columnwidth,trim={0.1cm 1.5cm 0.1cm 1.7cm},clip]{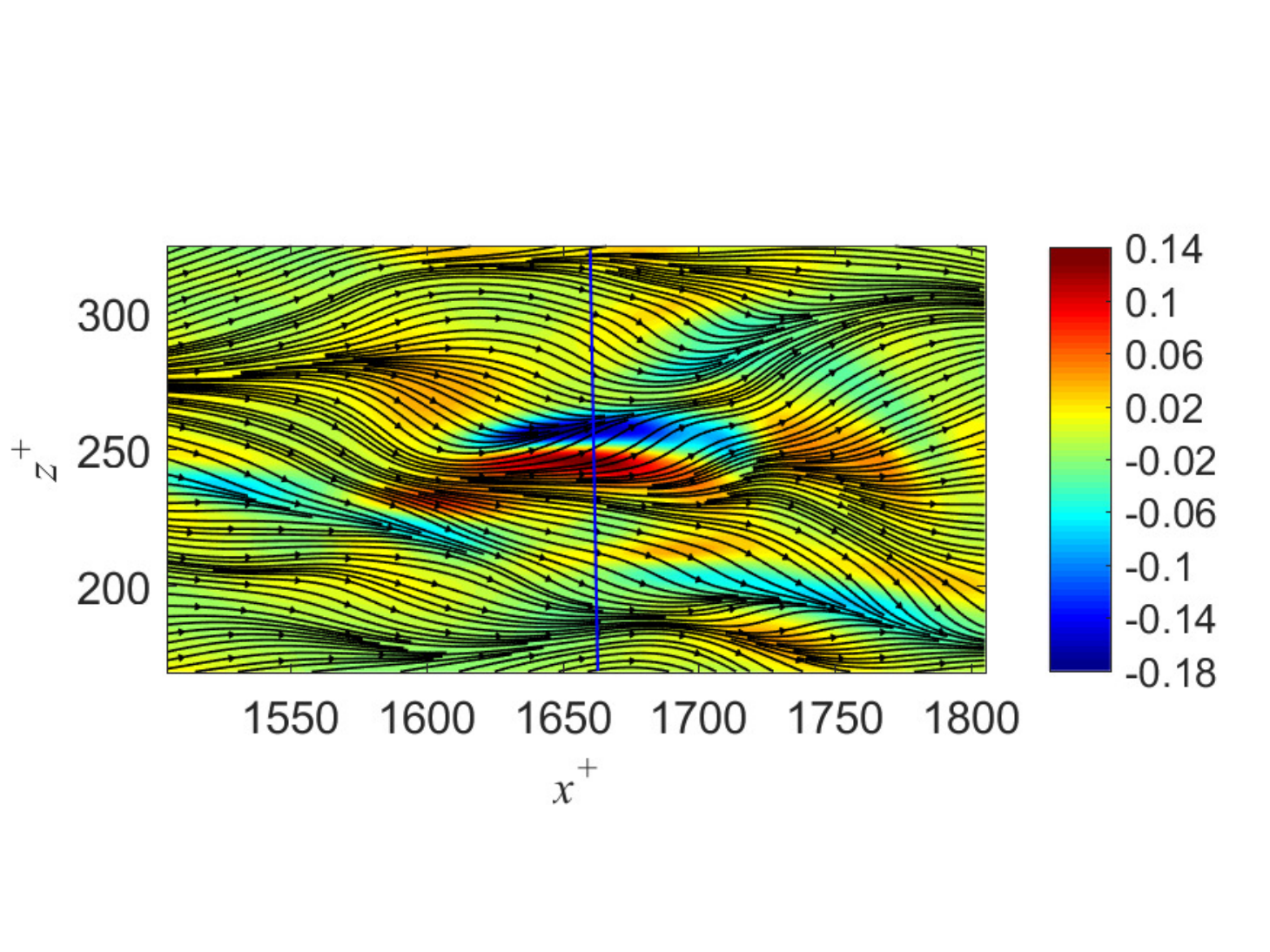}
	\caption{Normalized snapshot of the skin friction divergence $\bm{\nabla}_{\partial B}^{+}\bm{\cdot}\bm{\tau}^{+}$ around the SWNVE where the skin friction lines are superposed. The blue vertical line denotes the path across the SWNVE along the spanwise direction at the streamwise location $x^{+}=1661.54$.} 
	\label{contour_divtauw}
\end{figure}

\begin{figure}[h]
	\centering
	\subfigure[$y^+=0$]{
		\begin{minipage}[t]{0.5\linewidth}
			\centering
			\includegraphics[width=1.0\columnwidth,trim={0.1cm 1.5cm 0.1cm 1.7cm},clip]{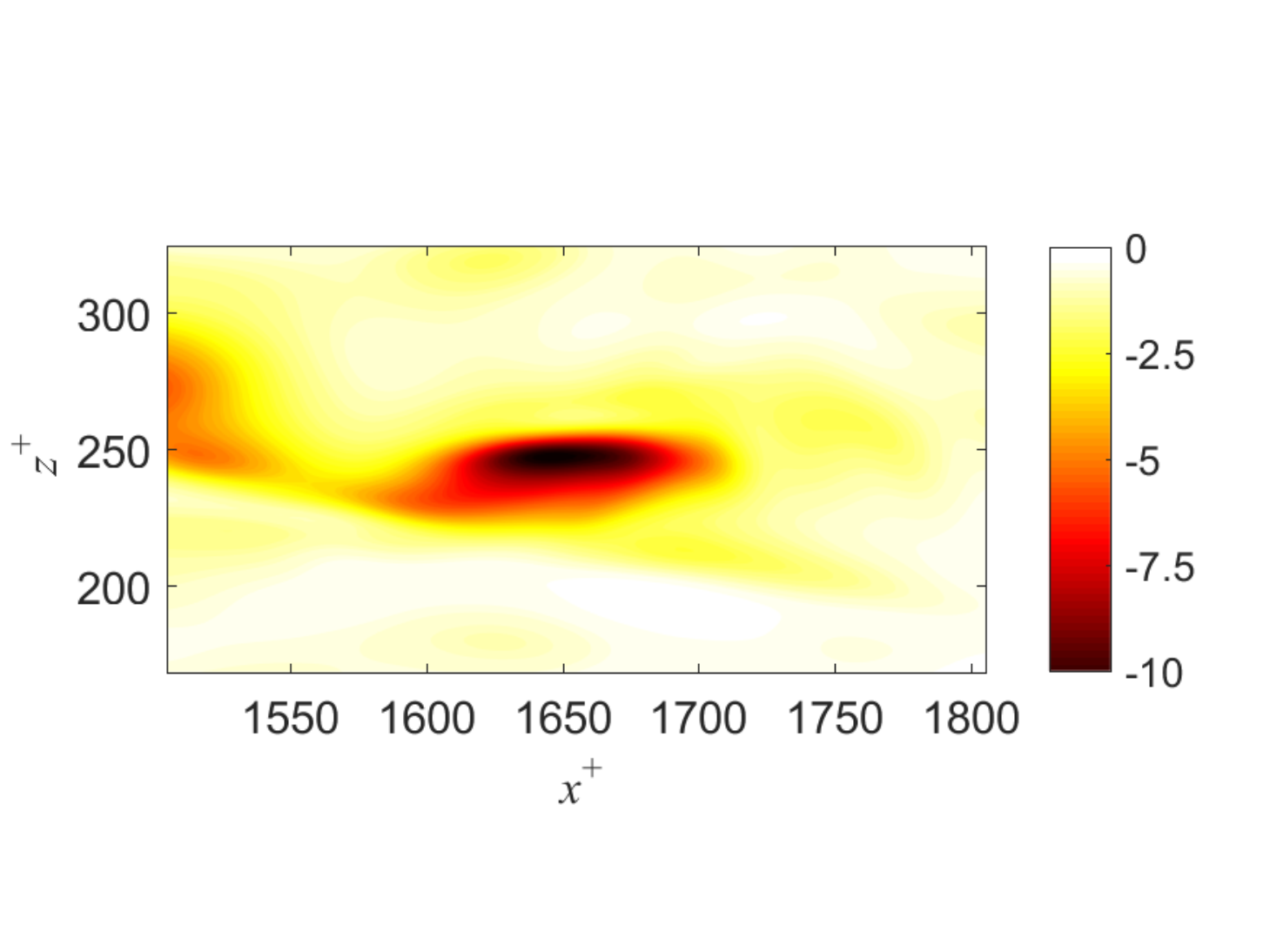}
			\label{contour_divlamb1}
		\end{minipage}%
	}%
	\subfigure[$y^+=1.204$]{
		\begin{minipage}[t]{0.5\linewidth}
			\centering
			\includegraphics[width=1.0\columnwidth,trim={0.1cm 1.5cm 0.1cm 1.7cm},clip]{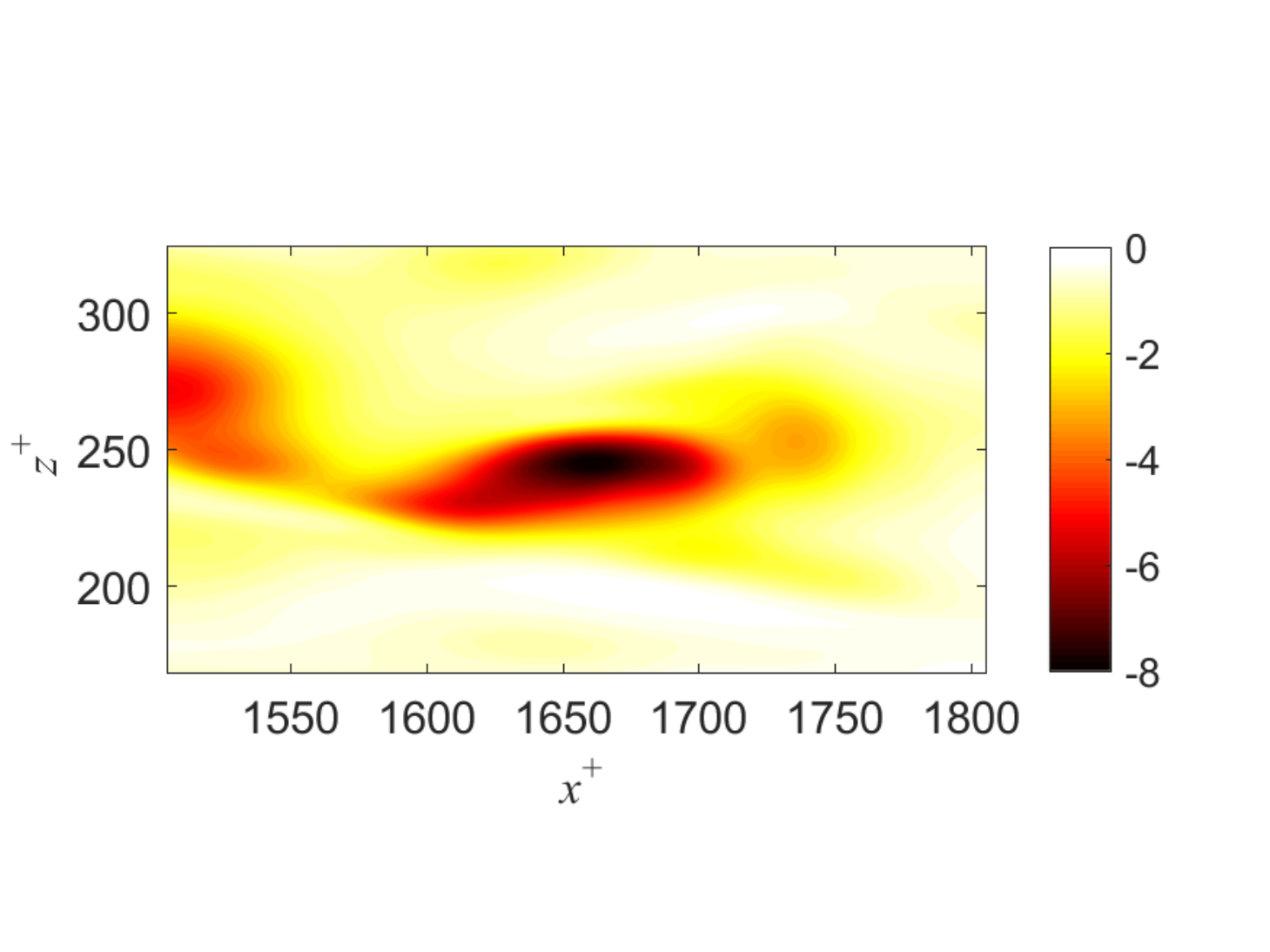}
			\label{contour_divlamb2}
		\end{minipage}%
	}%
	
	\subfigure[$y^+=2.408$]{
		\begin{minipage}[t]{0.5\linewidth}
			\centering
			\includegraphics[width=1.0\columnwidth,trim={0.1cm 1.5cm 0.1cm 1.7cm},clip]{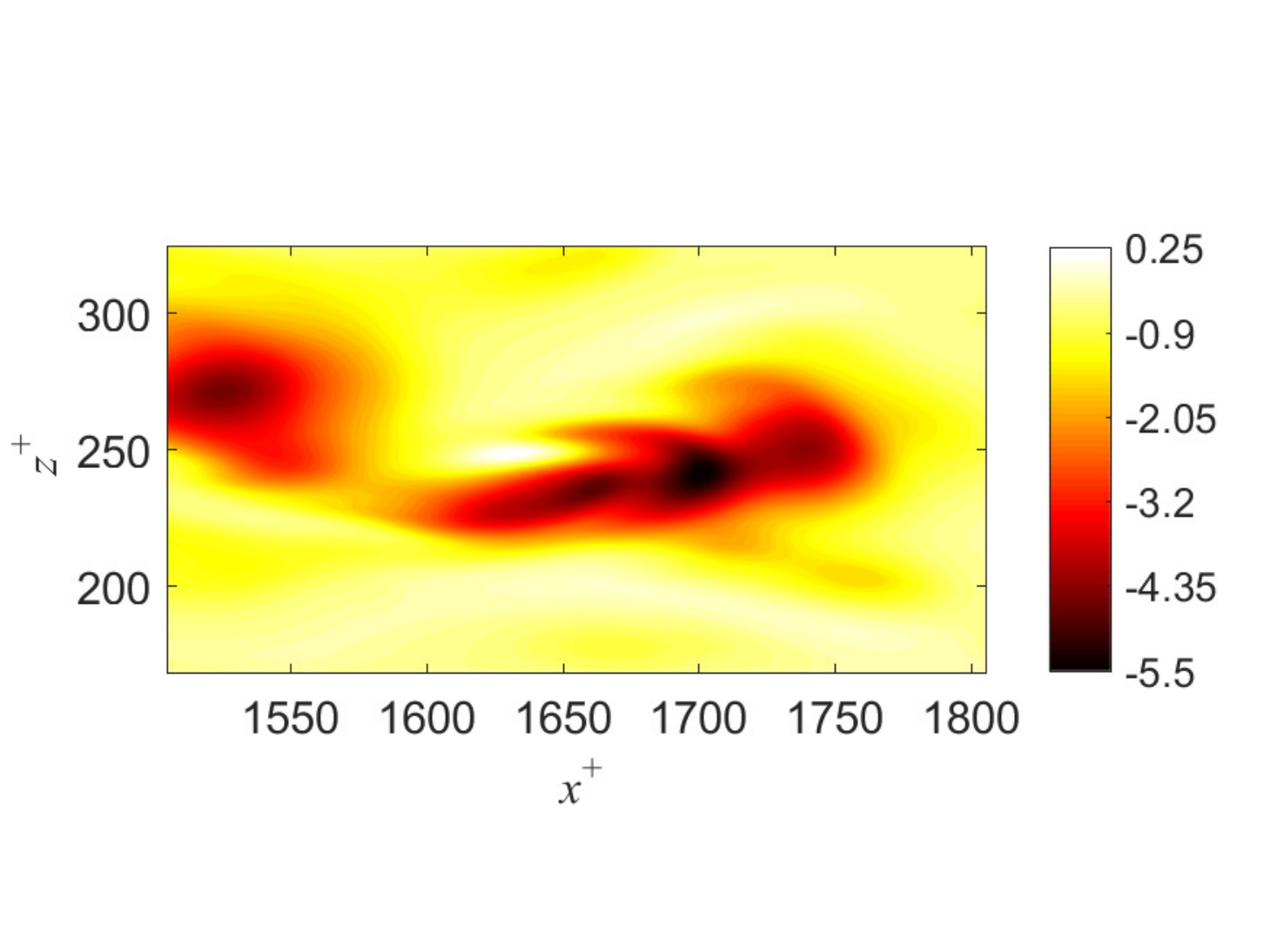}
			\label{contour_divlamb3}
		\end{minipage}%
	}%
	\subfigure[$y^+=3.612$]{
		\begin{minipage}[t]{0.5\linewidth}
			\centering
			\includegraphics[width=1.0\columnwidth,trim={0.1cm 1.5cm 0.1cm 1.7cm},clip]{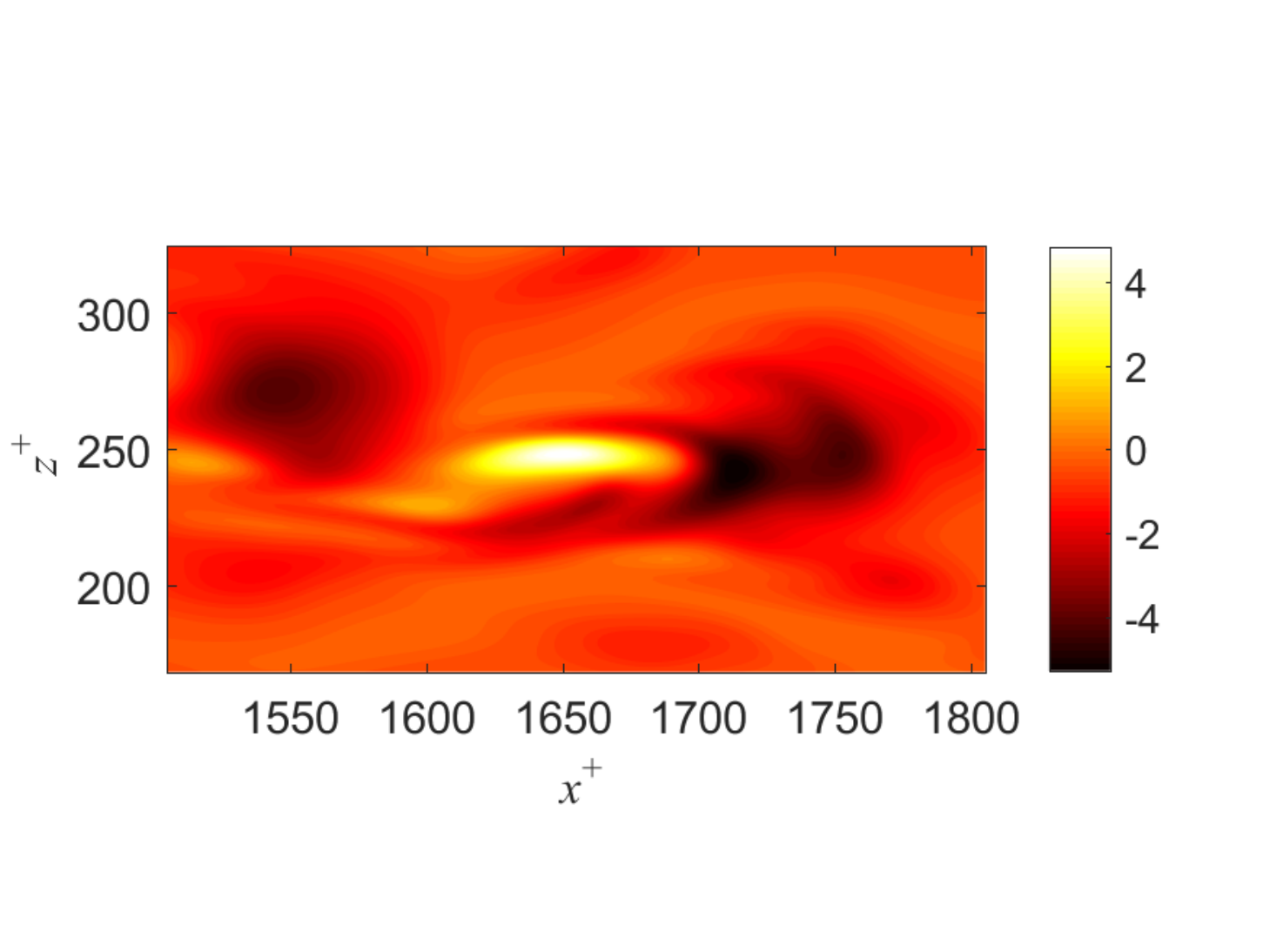}
			\label{contour_divlamb4}
		\end{minipage}%
	}%
	\caption{Normalized snapshots of the Lamb dilatation $\vartheta_L^{+}$ corresponding to the SWNVE at different wall-normal locations inside the viscous sublayer. ({\it a}) $y^+=0$, ({\it b}) $y^+=1.204$, ({\it c}) $y^+=2.408$, and ({\it d}) $y^+=3.612$.} 
	\label{divlamb}
\end{figure}

Previous studies show that strong wall-normal velocity events (SWNVEs) are responsible for high intermittency of the viscous sublayer, which are closely related to quasi-streamwise vortices and associated near-wall self-sustaining momentum transport processes~\cite{ChenTao2021POF,Guerrero2020,Guerrero2022JFM}. The SWNVEs induce strong near-wall sweep and ejection events, which are usually accompanied by high skin friction magnitude and violent surface pressure fluctuation. In Fig.~\ref{contour_divtauw}, the footprint of the quasi-streamwise vortice is identified by the skin friction divergence $\bm{\nabla}_{\partial B}\bm{\cdot}\bm{\tau}$, which is a critical surface quantity to characterize the topological features of the skin friction field~\cite{ChongMS2012,ChenTao2022AIPb,LiuT2021wedge}. According to the near-wall Taylor series expansion solution of the NS equations, the wall-normal velocity component is $u_y=-(2\mu)^{-1}(\bm{\nabla}_{\partial B}\bm{\cdot}\bm{\tau})y^2+O(y^3)$. Therefore, positive skin friction divergence $(\bm{\nabla}_{\partial B}\bm{\cdot}\bm{\tau}>0)$ corresponds to the attachment lines due to sweep motions ($u_y<0$), while negative skin friction divergence $(\bm{\nabla}_{\partial B}\bm{\cdot}\bm{\tau}<0)$ corresponds to the separation lines due to ejection motions ($u_y>0$). 

\begin{figure}[h]
	\centering
	\subfigure[$y^+=0$]{
		\begin{minipage}[t]{0.5\linewidth}
			\centering
			\includegraphics[width=1.0\columnwidth,trim={0.1cm 1.5cm 0.1cm 1.7cm},clip]{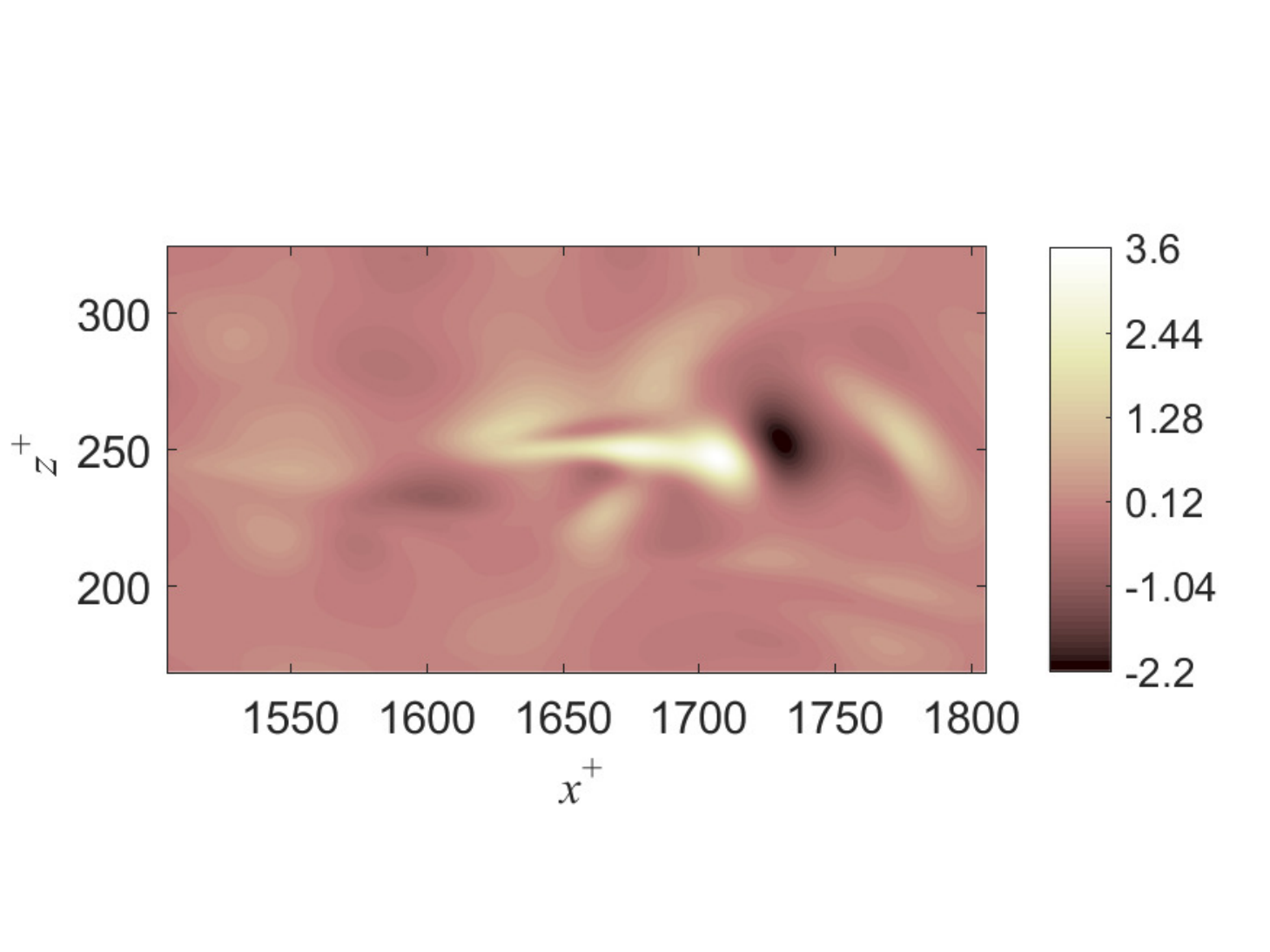}
			\label{contour_fthetaL1}
		\end{minipage}%
	}%
	\subfigure[$y^+=1.204$]{
		\begin{minipage}[t]{0.5\linewidth}
			\centering
			\includegraphics[width=1.0\columnwidth,trim={0.1cm 1.5cm 0.1cm 1.7cm},clip]{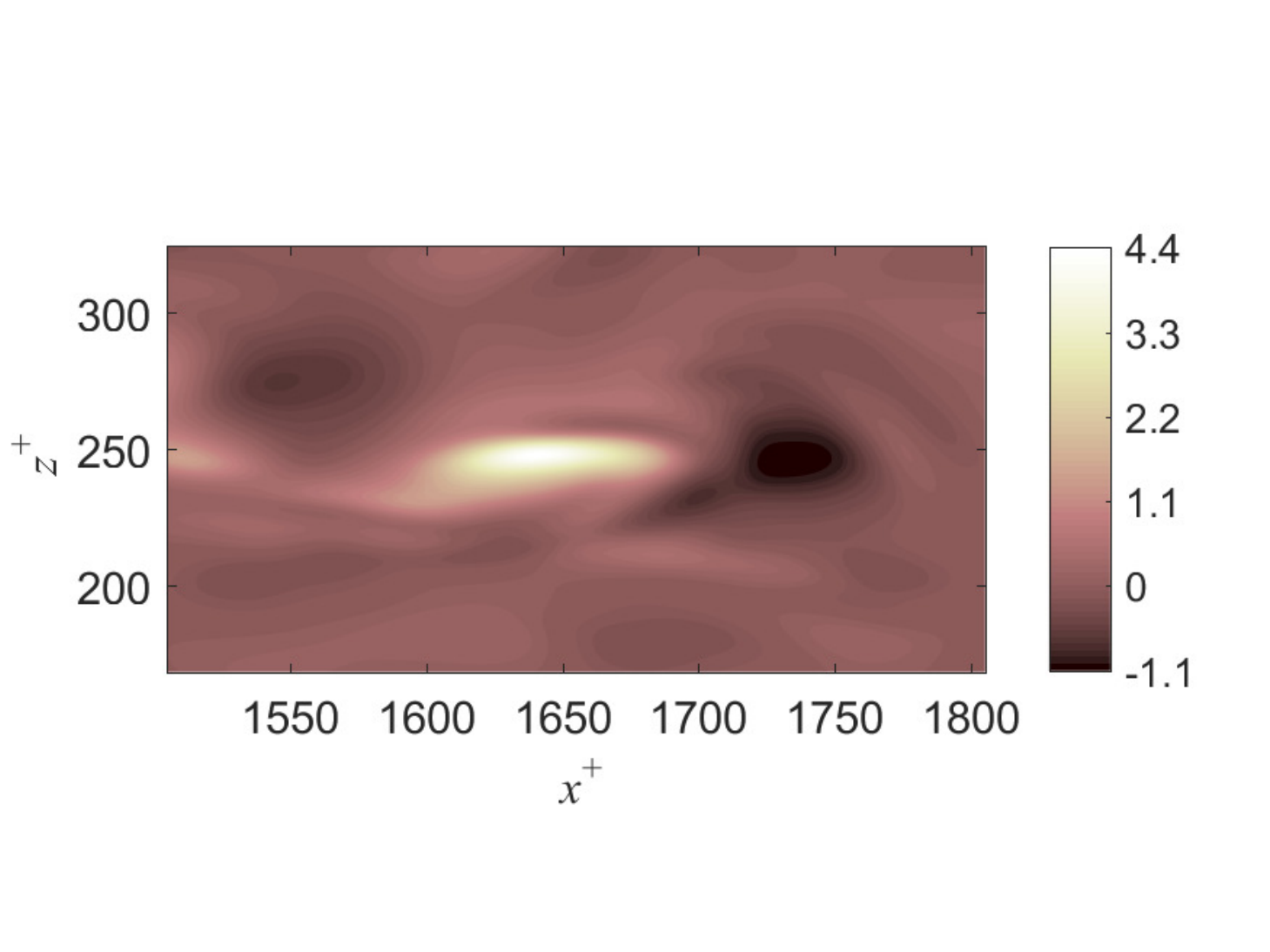}
			\label{contour_fthetaL2}
		\end{minipage}%
	}%
	
	\subfigure[$y^+=2.408$]{
		\begin{minipage}[t]{0.5\linewidth}
			\centering
			\includegraphics[width=1.0\columnwidth,trim={0.1cm 1.5cm 0.1cm 1.7cm},clip]{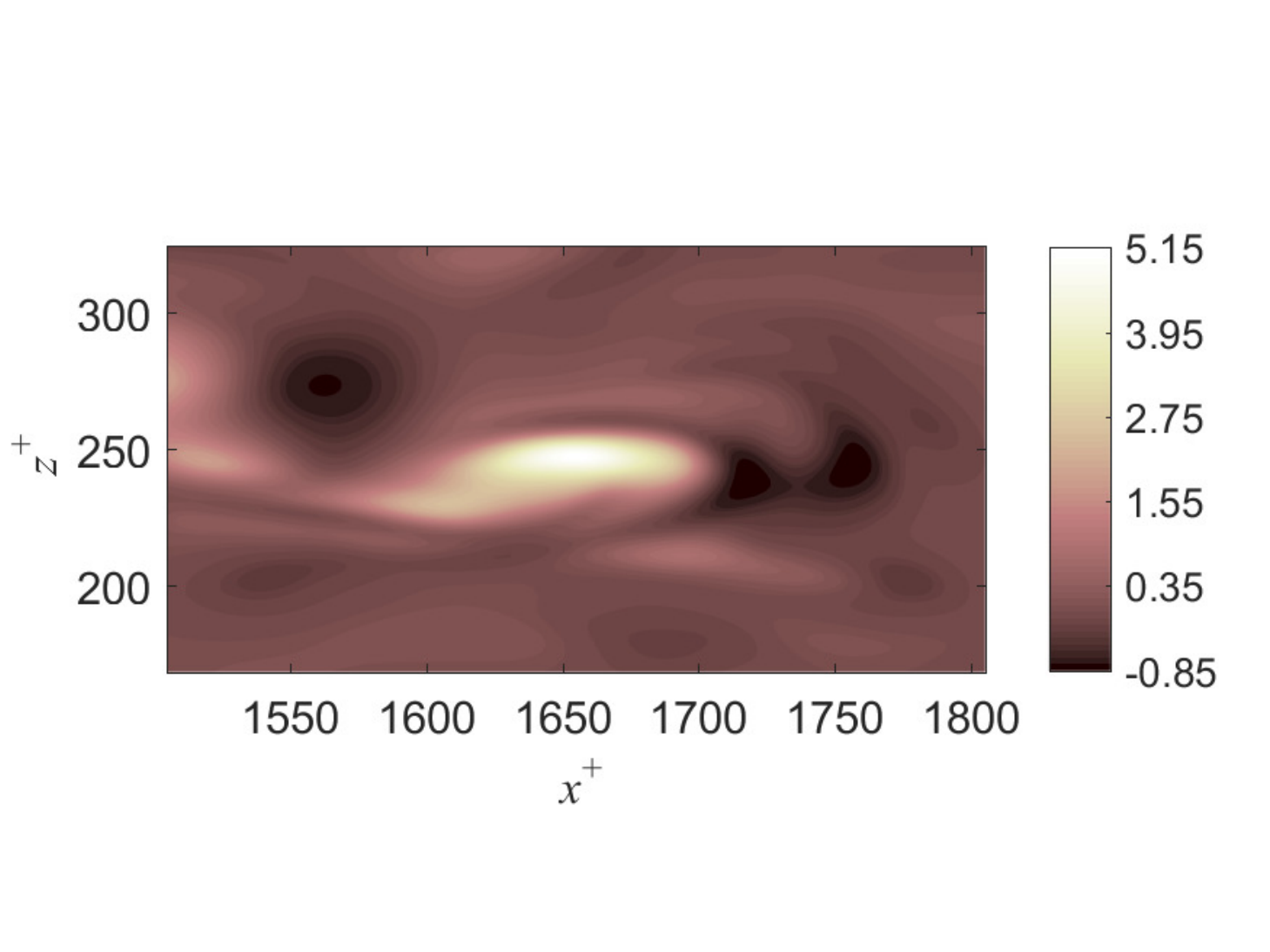}
			\label{contour_fthetaL3}
		\end{minipage}%
	}%
	\subfigure[$y^+=3.612$]{
		\begin{minipage}[t]{0.5\linewidth}
			\centering
			\includegraphics[width=1.0\columnwidth,trim={0.1cm 1.5cm 0.1cm 1.7cm},clip]{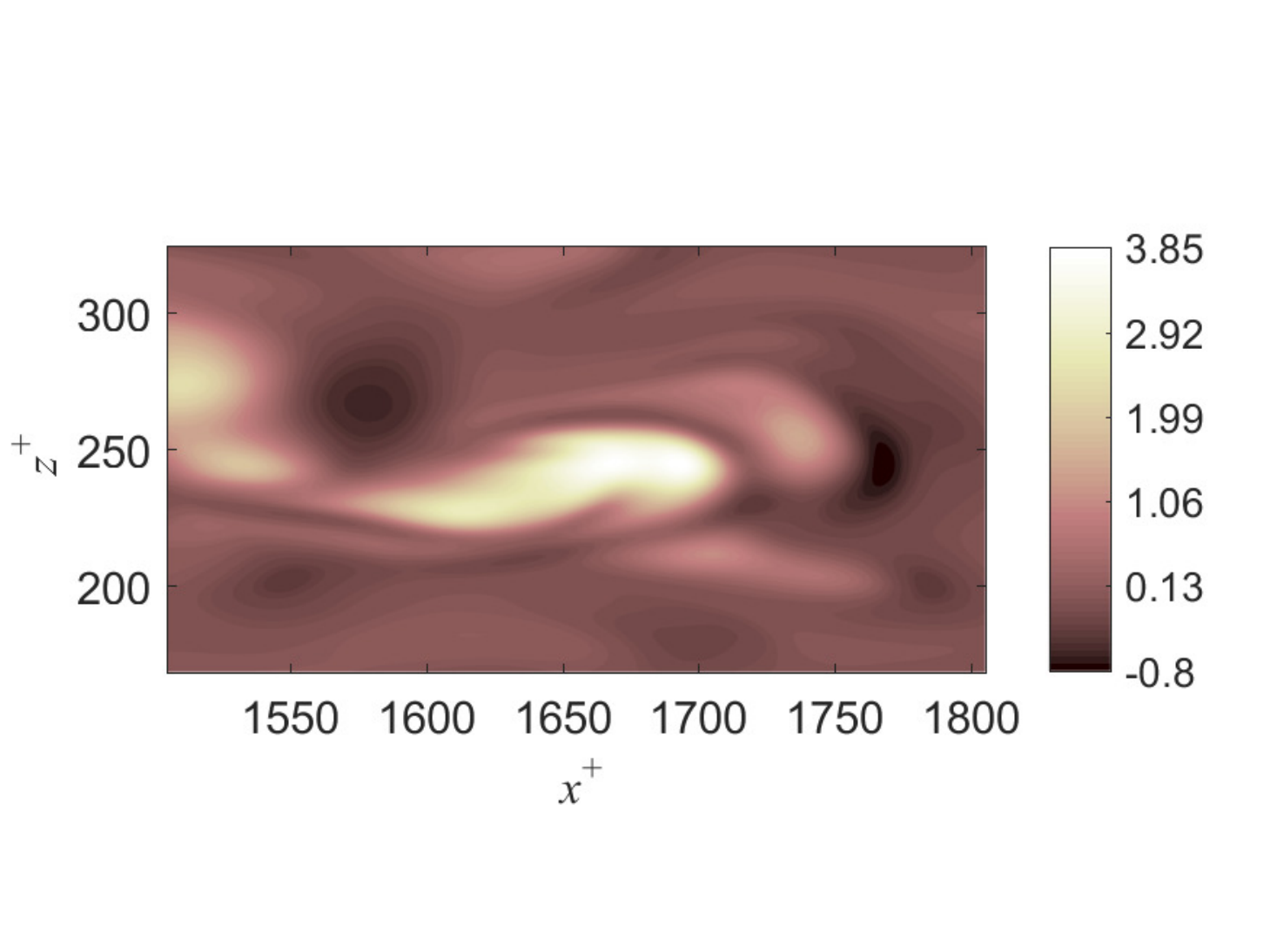}
			\label{contour_fthetaL4}
		\end{minipage}%
	}%
	\caption{Normalized snapshots of the WNLDF $F_{\vartheta_L}^{+}$ corresponding to the SWNVE at dfferent wall-normal locations inside the viscous sublayer. ({\it a}) $y^+=0$ (i.e. the BLDF $f_{\vartheta_L}^{+}$), ({\it b}) $y^+=1.204$, ({\it c}) $y^+=2.408$, and ({\it d}) $y^+=3.612$.} 
	\label{F00}
\end{figure}

The Lamb dilatation captures the temporal-spatial evolution of local high- and low- momentum fluids, which is directly related to the turbulent mixing~\cite{Hamman2008}.
Relevant to the selected SWNVE, Fig.~\ref{divlamb} shows the normalized snapshots of the Lamb dilatation $\vartheta_{L}$ at different wall-normal locations inside the viscous sublayer.
On the wall, the boundary Lamb dilatation is determined by the boundary enstrophy, which is always non-positive and vorticity-dominated. Slightly away from the wall, the Lamb dilatation is determined by the competition between the flexion product $\bm{u}\cdot\bm{\nabla}\times\bm{\omega}$ and the enstrophy $-2\Omega$ such that both positive and negative Lamb dilatation centers can coexist. 
Positive flexion product are usually associated with the enhanced the straining motion and the depletion of the vortical motion.
The flexion product increases rapidly with the increasing wall-normal distance and finally dominates the core pattern of the Lamb dilatation at the outer edge of the viscous sublayer. For example, in Fig.~\ref{contour_divlamb4}, positive and negative Lamb dilatation regions already coexist, among which the interaction and interference will drive the momentum and energy redistribution in the near-wall region.
Normalized snapshots of the WNLDF $F_{\vartheta_{L}}$ are demonstrated in Fig.~\ref{F00}. High-magnitude positive and negative regions of $F_{\vartheta_{L}}$ are observed around the SWNVE. Particularly, the BLDF is plotted in Fig.~\ref{contour_fthetaL1}, which is caused by the non-linear coupling between the skin friction and the surface pressure gradient. It is interesting to note that the pattern of $F_{\vartheta_{L}}$ rapidly changes even within the viscous sublayer, as the wall-normal distance increases. This distinct change should be directly related to sweep and ejection events induced by the quasi-streamwise vortice, and the concentrated wall-normal momentum transfer in this region.

\begin{figure}[h]
	\centering
	\subfigure[]{
		\begin{minipage}[t]{0.5\linewidth}
			\centering
			\includegraphics[width=1.0\columnwidth,trim={0.1cm 0.14cm 0.1cm 0.6cm},clip]{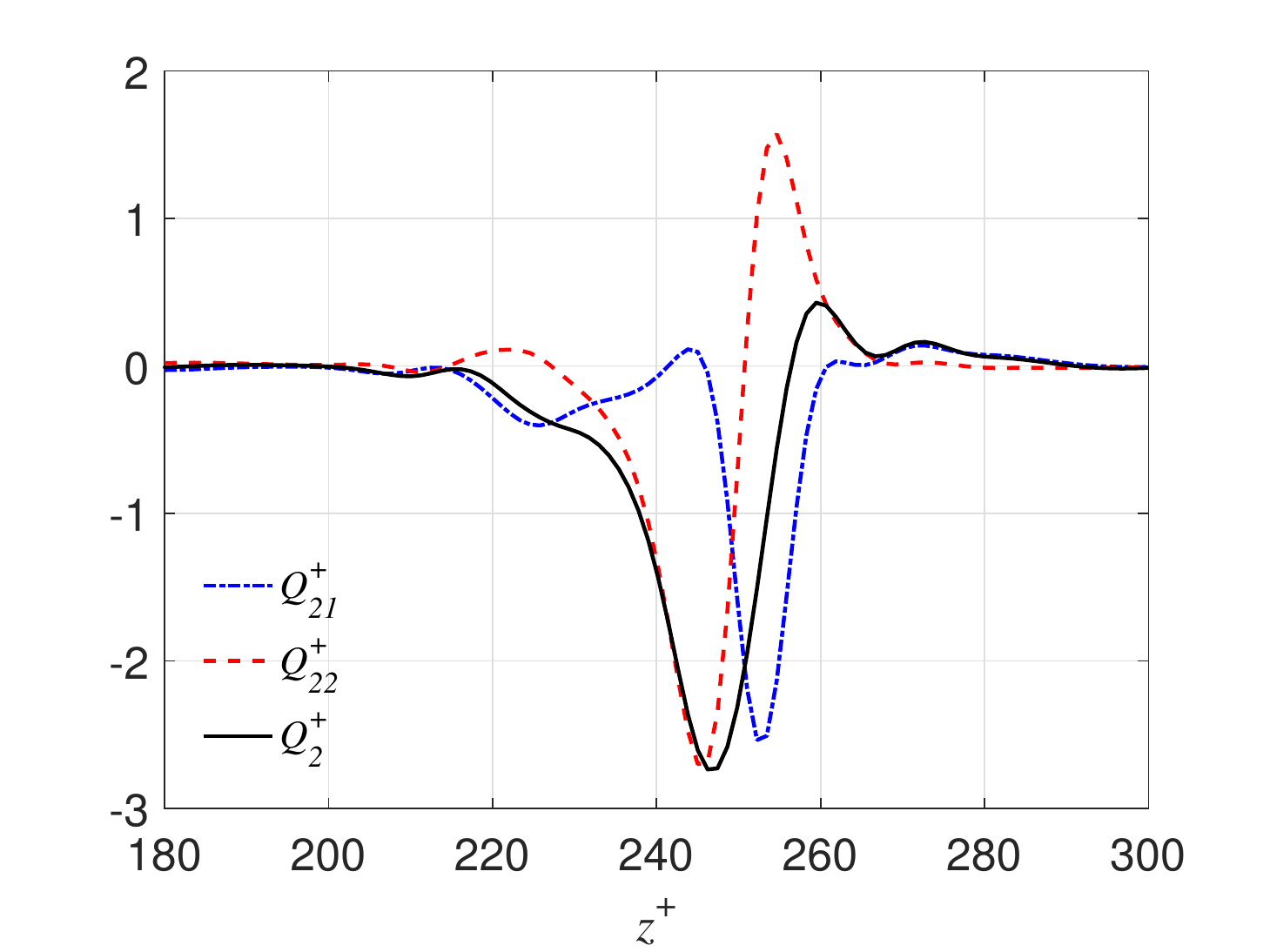}
			\label{Q2compare}
		\end{minipage}%
	}%
	\subfigure[]{
		\begin{minipage}[t]{0.5\linewidth}
			\centering
			\includegraphics[width=1.0\columnwidth,trim={0.1cm 0.14cm 0.1cm 0.6cm},clip]{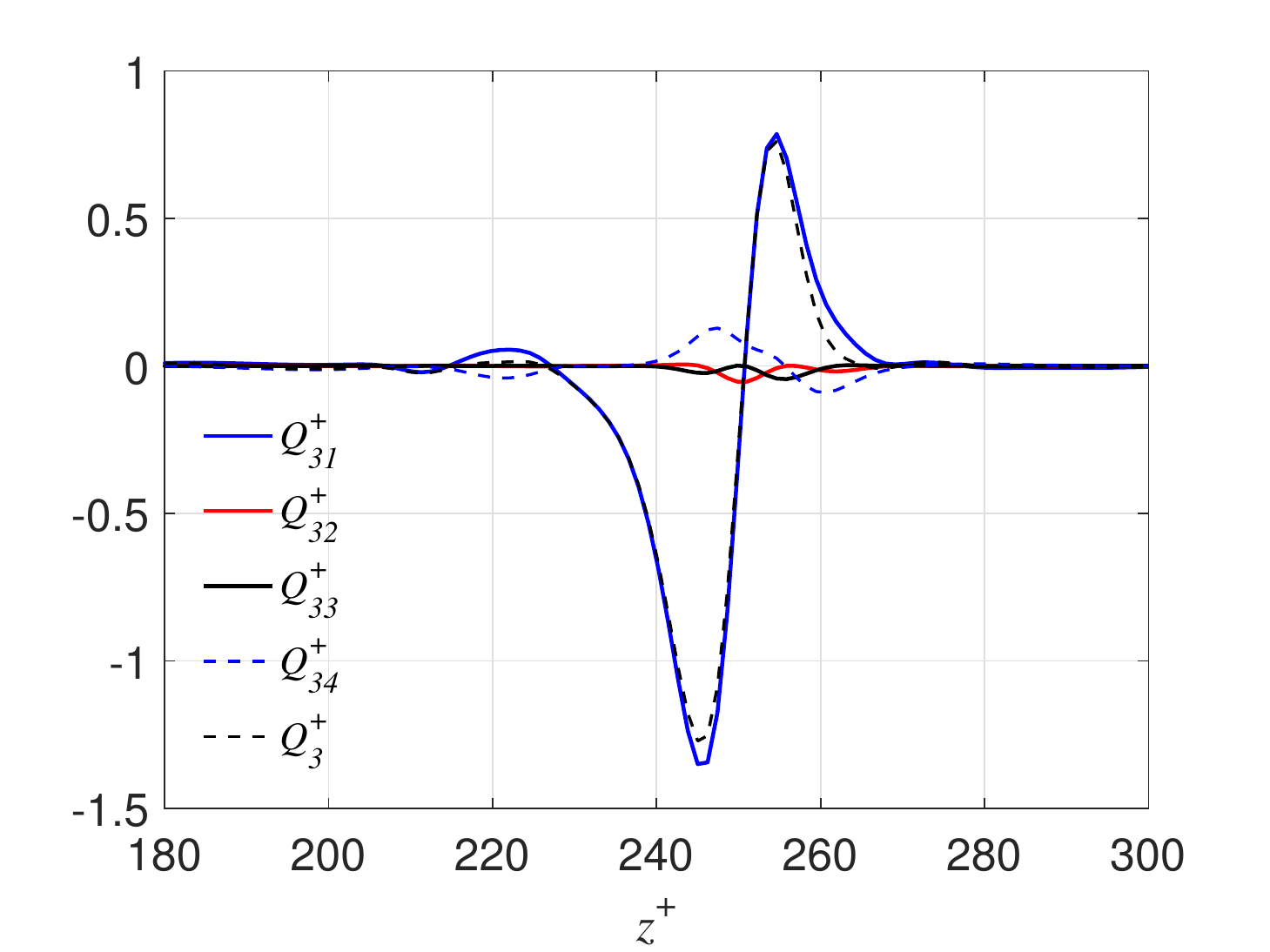}
			\label{Q3compare}
		\end{minipage}%
	}%
	
	\subfigure[]{
		\begin{minipage}[t]{0.5\linewidth}
			\centering
			\includegraphics[width=1.0\columnwidth,trim={0.1cm 0.14cm 0.1cm 0.6cm},clip]{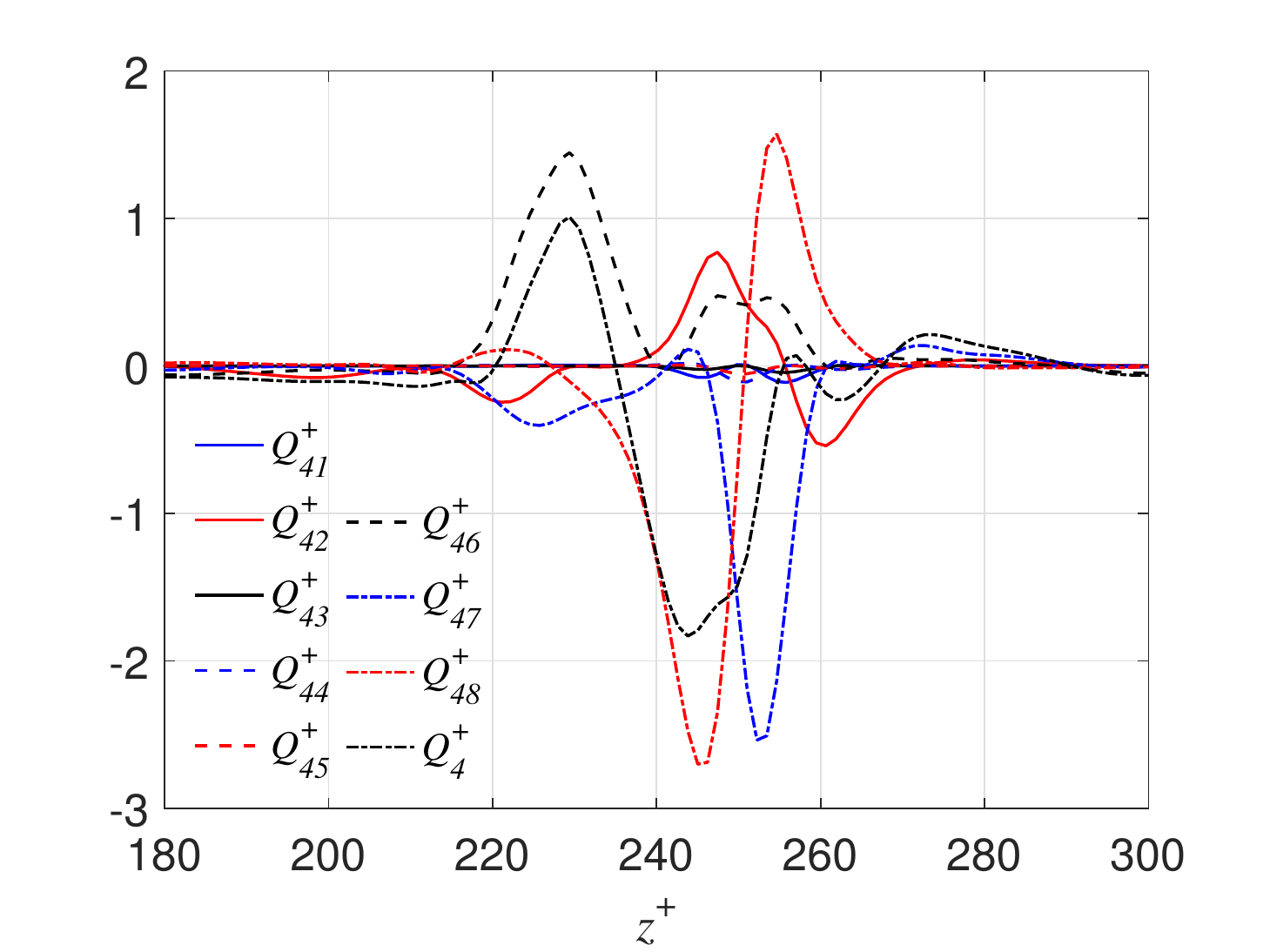}
			\label{Q4compare}
		\end{minipage}%
	}%
	\subfigure[]{
		\begin{minipage}[t]{0.5\linewidth}
			\centering
			\includegraphics[width=1.0\columnwidth,trim={0.1cm 0.14cm 0.1cm 0.6cm},clip]{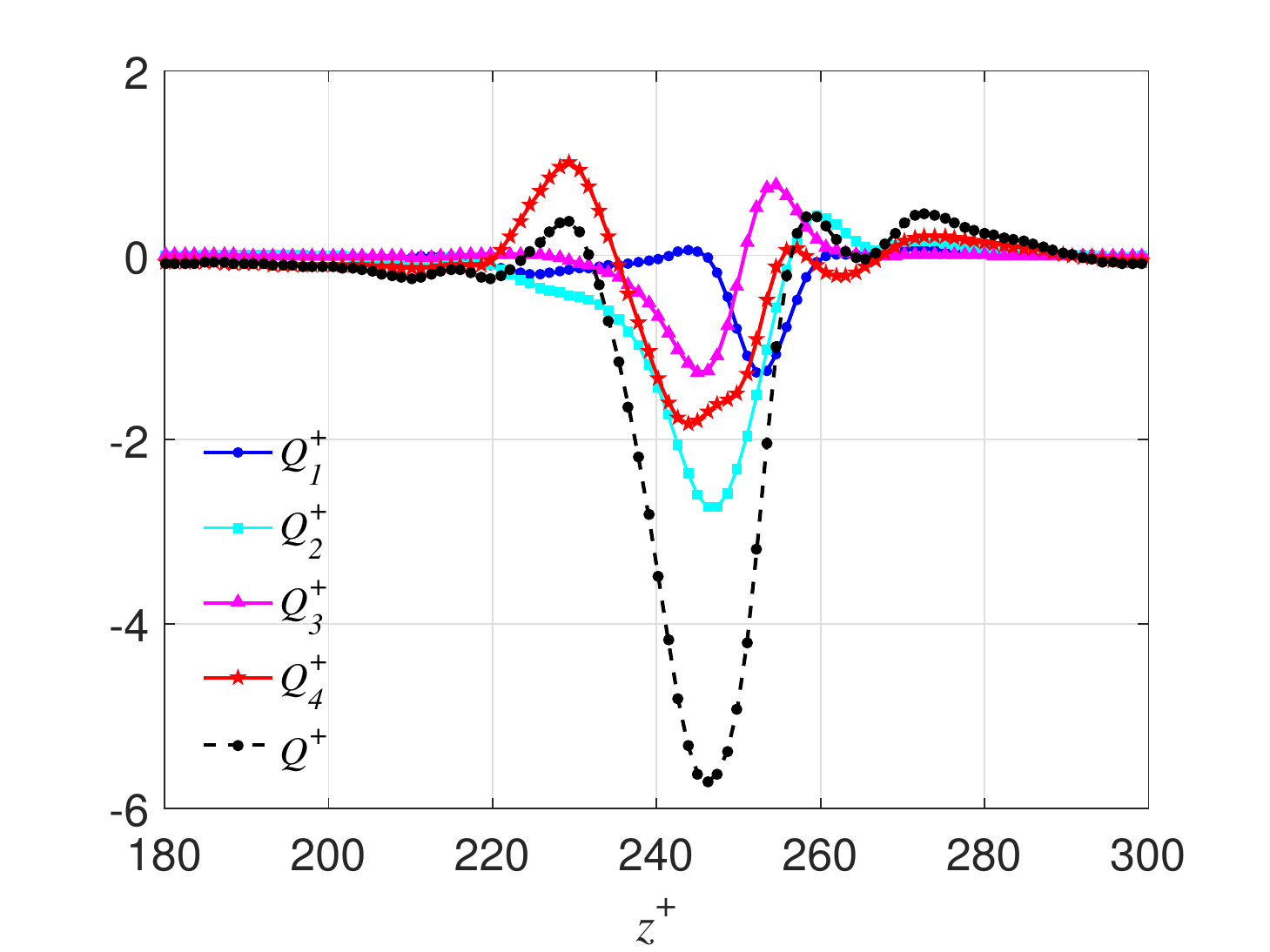}
			\label{Qcompare}
		\end{minipage}%
	}%
	\caption{Comparison of different contributions to ({\it a}) the wall-normal diffusion of $-2\bm{\nabla L}\bm{:}\bm{\nabla u}^T$ (i.e., $Q_2$), ({\it b}) the wall-normal diffusion of $-\bm{\nabla}h_0\bm{:}\bm{\nabla}\times\bm{\omega}$ (i.e., $Q_{3}$), ({\it c}) the wall-normal
		diffusion of $\bm{\nabla}\bm{\cdot}\bm{q}$ (i.e., $Q_4$) and ({\it d}) the temporal-spatial evolution rate of the WNLDF at the wall (i.e., $Q$) in a single-phase turbulent channel flow at $Re_{\tau}=180$.} 
\end{figure}
\begin{figure}[h]
	\centering
	\subfigure[]{
		\begin{minipage}[t]{0.5\linewidth}
			\centering
			\includegraphics[width=1.0\columnwidth,trim={0.1cm 0.12cm 0.1cm 0.6cm},clip]{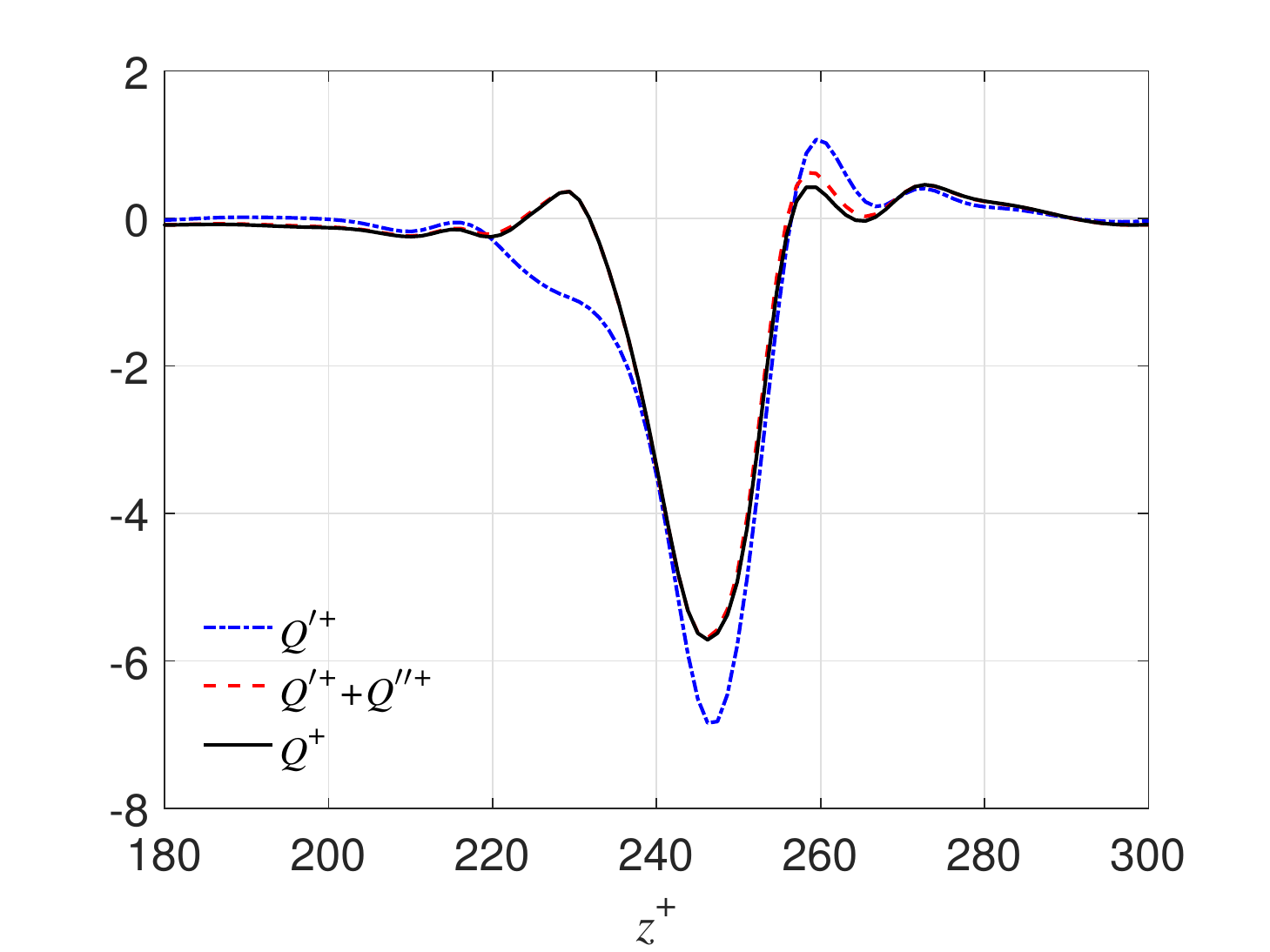}
			\label{Qmodelcompare}
		\end{minipage}%
	}%
	\subfigure[]{
		\begin{minipage}[t]{0.5\linewidth}
			\centering
			\includegraphics[width=1.0\columnwidth,trim={0.1cm 1.75cm 0.1cm 1.8cm},clip]{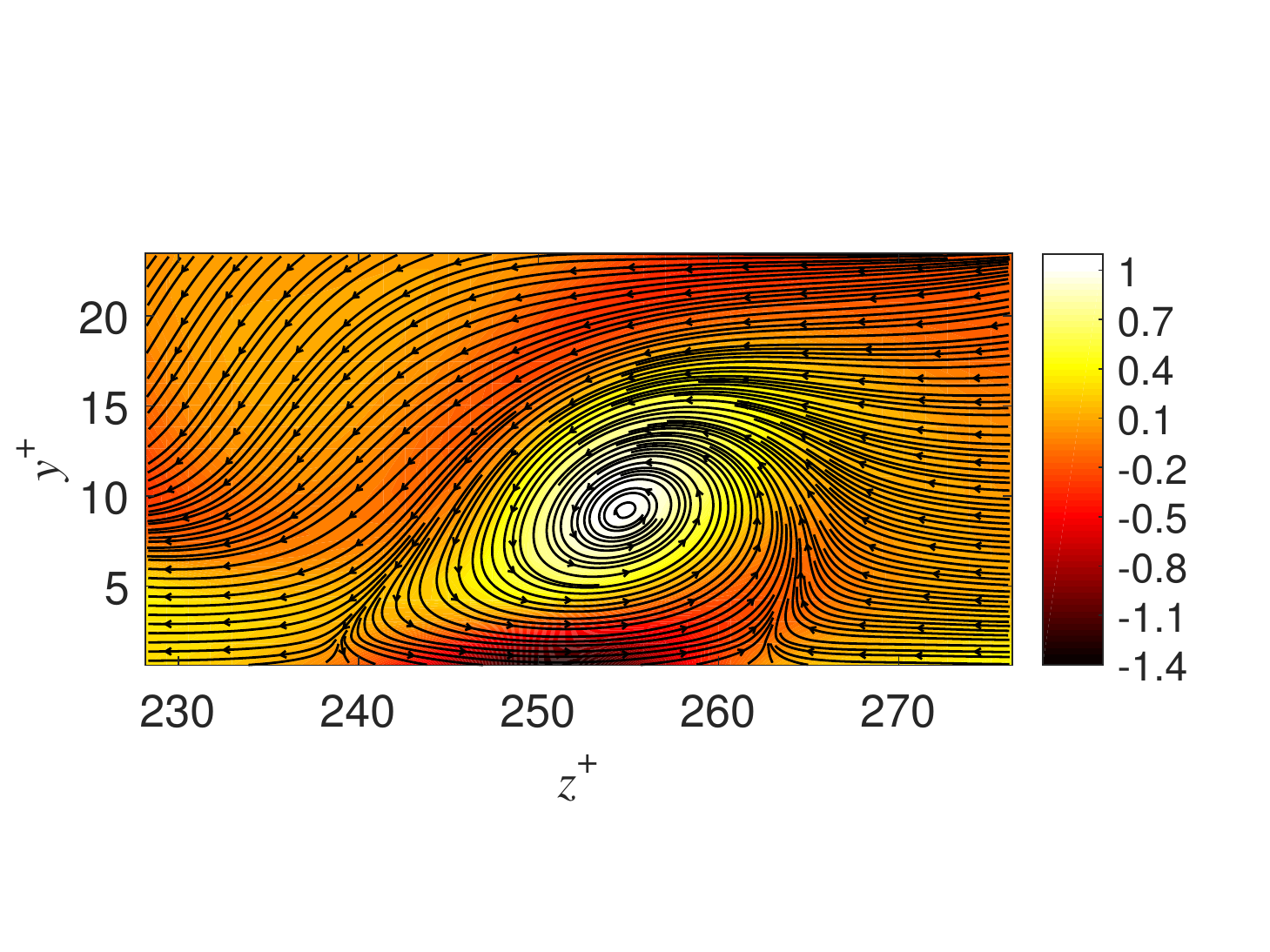}
			\label{near_wall_vortex_slice}
		\end{minipage}%
	}%
	\caption{({\it a}) Comparison of $Q^{\prime}$, $Q^{\prime}+Q^{\prime\prime}$ and $Q$ along the line $x^+=1661.54$ across the SWNVE. ({\it b}) Streamlines in the $y-z$ plane at $x^+=1661.54$ where the normalized streamwise vorticity component is used as the background.} 
\end{figure}

After briefly studying the basic features of near-wall Lamb dilatation and its wall-normal flux, by using Eqs.~\eqref{LF} and~\eqref{LF2}, the temporal-spatial evolution rate of the WNLDF at the bottom channel wall can be quantitatively evaluated with the information of surface quantities.
In order to demonstrate the dominant mechanism to each source term, a line across the SWNVE along the spanwise direction (i.e., $x^+=1661.54$) is selected, as shown in Fig.~\ref{contour_divtauw}.
In Fig.~\ref{Q2compare}, both $Q_{21}=4\bm{\tau}\bm{\cdot}\bm{\nabla}_{\partial B}\Omega_{\partial B}$ and $Q_{22}=-4(\bm{\nabla}_{\partial B}\bm{\cdot}\bm{\tau})\Omega_{\partial B}$ are of primary importance in determining the spatial distribution of $Q_2$.
The positive peak region of $Q_{22}$ is partially counteracted and translated by $Q_{21}$.
As shown in Fig.~\ref{Q3compare}, $Q_3$ is mainly contributed by $Q_{31}=-2(\bm{\nabla}_{\partial B}\bm{\cdot}\bm{\tau})\Omega_{\partial B}$, namely, the coupling
between the skin friction divergence and the boundary enstrophy. Other terms associated with the coupling between the temporal-spatial derivatives of the skin friction and the surface pressure only have negligible
contributions to $Q_{3}$. As displayed in Fig.~\ref{Q4compare},
$Q_4$ is mainly contributed by $Q_{47}=4\bm{\tau}\bm{\cdot}\bm{\nabla}_{\partial B}\Omega_{\partial B}$ and $Q_{48}=-4(\bm{\nabla}_{\partial B}\bm{\cdot}\bm{\tau})\Omega_{\partial B}$. 
Interestingly, $Q_{46}=2\mu^{-1}\bm{\tau}\bm{\cdot}\mathcal{L}_{\partial B}(\bm{\nabla}_{\partial B}p_{\partial B})$ mainly shows positive contribution to $Q_4$ and the positive peak region of $Q_4$ is mainly caused by $Q_{46}$.
In addition, the contribution from $Q_{42}=6\mu^{-1}(\partial\bm{\tau}/\partial t)\bm{\cdot}\bm{\nabla}_{\partial B}p_{\partial B}$ can not be totally neglected
in both the sweep and ejection sides of the SWNVE. In Fig.~\ref{Qcompare}, the superposition of different source
terms ($Q_{1}$, $Q_{2}$, $Q_{3}$ and $Q_{4}$) gives the total temporal-spatial evolution rate $Q$, which shows a high-magnitude negative peak in the region occupied by the SWNVE. 

The preliminary exploration for the SWNVE implies that the temporal-spatial evolution rate of the WNLDF at the wall is mainly contributed by $Q^{\prime}\equiv10\bm{\tau}\bm{\cdot}\bm{\nabla}_{\partial B}\Omega_{\partial B}-10(\bm{\nabla}_{\partial B}\bm{\cdot}\bm{\tau})\Omega_{\partial B}$ and is influenced by the coupling terms including $6\mu^{-1}(\partial\bm{\tau}/\partial t)\bm{\cdot}\bm{\nabla}_{\partial B}p_{\partial B}$ and $2\mu^{-1}\bm{\tau}\bm{\cdot}\mathcal{L}_{\partial B}(\bm{\nabla}_{\partial B}p_{\partial B})$. 
The sum of the latter two terms are denoted as $Q^{\prime\prime}$ for convenience.
Fig.~\ref{Qmodelcompare} plots the distribution of the dominant term $Q^{\prime}$, the sum of the dominant and the coupling terms $Q^{\prime}+Q^{\prime\prime}$ and the DNS result for the total evolution rate $Q$ along $x^+=1661.54$. It is clearly observed that $Q^{\prime}+Q^{\prime\prime}$ provides a very good approximation for $Q$. The discrepancy between the blue and red lines is therefore caused by the contribution from $Q^{\prime\prime}$.
Furthermore, the dominant terms found in this turbulence case is very similar to those in the lid-driven cavity flow (see Section~\ref{cavity_flow_sec}). This similarity can be intuitively understood, because for the two cases considered here, both the flow separation and attachment in the near-wall region are induced by a near-wall vortex interacting with the solid wall, which can be further evidenced from the streamlines in the $y-z$ plane (see Fig.~\ref{near_wall_vortex_slice} for a separation bubble-like structure).
%%%%%%%%%%%%%%%%%%%

\section{Conclusions and discussions}\label{Conclusions}
In this paper, we present a theoretical study of the Lamb dilatation and its hydrodynamic viscous flux for
near-wall incompressible viscous flows. By applying the derived relations to the simulation results,  the underlying physics in instantaneous fluid dynamics near the wall is explored.

Firstly, based on the evolution equation of the Lamb vector obtained by Wu {\it et al.}~\cite{Wu1999POF}, we derive a new form of the transport
equation of the Lamb dilatation. The mathematical equivalence between the present form and that previously reported
by Hamman {\it et al.}~\cite{Hamman2008} is clearly elucidated. 

Secondly, by analogy with the physical concepts of the boundary vorticity flux (BVF), in order to characterize the wall-normal diffusion rate of the Lamb dilatation, we introduce a new physical concept referred to as the wall-normal Lamb dilatation flux (WNLDF, defined as
the wall-normal derivative of the Lamb dilatation multiplied by the dynamic viscosity), whose
value on the wall is termed as the boundary Lamb dilatation flux (BLDF).
These concepts are physically meaningful since the interaction between positive and negative regions of the Lamb dilatation are directly related to the energy redistribution and momentum transfer in near-wall viscous flows~\cite{Hamman2008}. The BLDF is found to be determined by the coupling between skin friction and 
the surface pressure gradient as well as another quadratic term representing the interaction between the 
boundary vorticity and the surface curvature. Therefore, it implies that the wall-normal diffusion of the Lamb dilatation at the wall is directly related to the boundary enstrophy generation. As a direct application, a new aerodynamic force expression is given for a two-dimensional closed body, where the aerodynamic force is contributed by the integral of the wall-normal diffusion of the square root of the Lamb dilatation. Therefore, manipulating the positive and negative regions of the Lamb dilatation by suitable approaches could effectively
control the pressure drag and lift. Compared to the formulation of Hamman {\it et al.}~\cite{Hamman2008},
the aerodynamic force expression presented in this paper provides a clearer physical interpretation for
the pressure drag reduction mechanism.

Thirdly, the temporal-spatial evolution rate of the WNLDF is discussed for a stationary no-slip flat wall. This evolution rate is mainly contributed by four source terms on the wall.
These source terms are expressed using the fundamental surface physical quantities (including skin friction, surface vorticity and surface pressure) and their temporal-spatial derivatives on the wall. These relations provide intrinsic connections between the near-wall flow and the surface physical quantities.
Generally speaking, resolving the real flow physics with high accuracy in such extremely near-wall region is important but difficult for both numerical simulations and experiments. In this regard, these exact relations are useful in understanding and interpreting numerical and experimental results of complex flows. Combined with the existing numerical and experimental studies on near-wall turbulence statistics~\cite{Adrian2007,Jamenez2018,LeeCB2019,Guerrero2020,Guerrero2022JFM}, the present study provides a theoretical foundation to understand the near-wall self-sustaining process and the interaction mechanism between the coherent structures and the solid wall.

Finally, these newly derived relations are applied to two simulated cases: a 2D lid-driven cavity flow (at $Re=1000$) and a turbulent channel flow (at $Re_{\tau}=180$).
For the lid-driven cavity flow, the BLDF at the bottom wall has four zero-crossing points. 
Two of them correspond with the two zero skin friction points (namely, the left separation point and the right attachment point), while the other two points correspond to a pressure maximum point and a pressure minimum point.
Therefore, the bottom wall is basically divided into five regions with different variations of straining and vortical motions in a small vicinity of the wall. 
For both the simulated cases, large spatial variations of the source terms in the evolution equation Eq.~\eqref{LF} are observed in the contact region below the primary vortex, which are directly related to the concentrated Lamb dilatation and its wall-normal viscous flux in the near-wall region.
The dominant mechanism for each source term is evaluated. The wall-normal diffusion of the Lamb dilatation convection term is only contributed by $2\bm{\tau}\bm{\cdot}\bm{\nabla}_{\partial B}\Omega_{\partial B}$, where $\bm{\tau}$ is the skin friction and $\Omega_{\partial B}$ is the boundary enstrophy. The wall-normal diffusion of the coupled Lamb dilatation and velocity gradients is dominated by $4\bm{\tau}\bm{\cdot}\bm{\nabla}_{\partial B}\Omega_{\partial B}$ and $-4(\bm{\nabla}_{\partial B}\bm{\cdot}\bm{\tau})\Omega_{\partial B}$, which also dominates the wall-normal diffusion of the divergence of the viscous source term $\bm{q}$ (see Eqs.~\eqref{qvis} and~\eqref{CRd}).
The wall-normal diffusion of the coupled stagnation enthalpy gradient and the curl of the vorticity is dominated by $-2(\bm{\nabla}_{\partial B}\bm{\cdot}\bm{\tau})\Omega_{\partial B}$. By summing over all these dominant mechanisms, it is found that the total temporal-spatial evolution rate of the WNLDF at the wall is predominantly contributed by $Q^{\prime}\equiv10\bm{\tau}\bm{\cdot}\bm{\nabla}_{\partial B}\Omega_{\partial B}-10(\bm{\nabla}_{\partial B}\bm{\cdot}\bm{\tau})\Omega_{\partial B}$, in addition to the less significant contributions due to the extra coupling effects
related to the temporal-spatial derivatives of the skin friction and the surface pressure. The evolution rate shows 
a highly negative peak associated with the SWNVE in the turbulent channel flow, which reflects the high intermittency feature of the viscous sublayer.

\section*{Acknowledgments }
T. Liu is partially supported by the John O. Hallquist Endowed Professorship and the Presidential Innovation Professorship.   

\section*{Conflict of Interest}
The authors have no conflicts to disclose.
\section*{DATA AVAILABILITY}
The data that support the findings of this study are available
from the corresponding author upon reasonable request.

\appendix
\section{Some derivation details and discussions}\label{Appendix1}
This appendix documents some necessary technical details in deriving Eqs.~\eqref{aa}--~\eqref{dd}.

Using Eq.~(\ref{BLD}), we obtain
\begin{eqnarray}
	Q_1=-\mu\left[\frac{\partial\bm{u}}{\partial n}\right]_{\partial B}\bm{\cdot}\left[\bm{\nabla}\vartheta_{L}\right]_{\partial B}
	=2\bm{\tau}\bm{\cdot}\bm{\nabla}_{\partial B}\Omega_{\partial B}.
\end{eqnarray}
Therefore, Eq.~\eqref{aa} is proved.

On the wall $\partial B$, the wall-normal derivative of the Lamb vector gradient is
\begin{eqnarray}
	&&\left[\frac{\partial\bm{\nabla L}}{\partial n}\right]_{\partial B}
	=\left[\bm{\nabla}\frac{\partial\bm{L}}{\partial n}\right]_{\partial B}
	=\bm{\nabla}_{\partial B}\left[\frac{\partial\bm{L}}{\partial n}\right]_{\partial B}
	+\bm{n}\left[\frac{\partial^2\bm{L}}{\partial n^2}\right]_{\partial B}\nonumber\\
	& &=-2\left(\bm{\nabla}_{\partial B}\Omega_{\partial B}\right)\bm{n}
	+\bm{n}\left[-2(\bm{\nabla}_{\partial B}\bm{\cdot}\bm{\omega}_{\partial B})\bm{\omega}_{\partial B}
	-\frac{1}{\mu^2}(\bm{\nabla}_{\partial B}\bm{\cdot}\bm{\tau})\bm{\tau}-\frac{3}{\mu}f_{\Omega}\bm{n}\right],
\end{eqnarray}
where the BEF $f_{\Omega}$ is given in Eq.~\eqref{BEF}. Therefore, we have
\begin{eqnarray}\label{d1}
	\left[\frac{\partial\bm{\nabla L}}{\partial n}\bm{:}
	\bm{\nabla u}^T\right]_{\partial B}=-\frac{2}{\mu}\bm{\tau}\bm{\cdot}\bm{\nabla}_{\partial B}\Omega_{\partial B}.
\end{eqnarray}

By applying Eq.~\eqref{NS2} on the wall $\partial B$ and using the relation $[\partial p/\partial n]_{\partial B}=-\bm{\nabla}_{\partial B}\bm{\cdot}\bm{\tau}$, we obtain
\begin{eqnarray}\label{s1}
	\left[\mu\nabla^2\bm{u}\right]_{\partial B}=-\left[\mu\bm{\nabla}\times\bm{\omega}\right]_{\partial B}=\bm{\nabla}_{\partial B}p_{\partial B}-(\bm{\nabla}_{\partial B}\bm{\cdot}\bm{\tau})\bm{n},
\end{eqnarray}
Therefore, direct evaluation gives
\begin{eqnarray}\label{s2}
	\left[\frac{\partial\bm{\nabla u}}{\partial n}\right]_{\partial B}
	=\frac{1}{\mu}\bm{\nabla}_{\partial B}\bm{\tau}+
	\frac{1}{\mu}\bm{n}\left[\bm{\nabla}_{\partial B}p_{\partial B}-
	(\bm{\nabla}_{\partial B}\bm{\cdot}\bm{\tau})\bm{n}\right],
\end{eqnarray}
Then, by noticing that $[\bm{\nabla L}]_{\partial B}=-2\Omega_{\partial B}\bm{nn}$ and using Eq.~\eqref{s2}, we have
\begin{eqnarray}\label{d2}
	\left[\bm{\nabla L}\bm{:}\frac{\partial\bm{\nabla u}^T}{\partial n}\right]_{\partial B}
	=\frac{2}{\mu}(\bm{\nabla}_{\partial B}\bm{\cdot}\bm{\tau})\Omega_{\partial B}.
\end{eqnarray}
Using Eqs.~\eqref{d1} and~\eqref{d2}, we have
\begin{eqnarray}
	Q_2&=&-2\mu\left[\frac{\partial\bm{\nabla L}}{\partial n}\bm{:}
	\bm{\nabla u}^T\right]_{\partial B}-2\mu\left[\bm{\nabla L}\bm{:}\frac{\partial\bm{\nabla u}^T}{\partial n}\right]_{\partial B}\nonumber\\
	&=&4\bm{\tau}\bm{\cdot}\bm{\nabla}_{\partial B}\Omega_{\partial B}-4(\bm{\nabla}_{\partial B}\bm{\cdot}\bm{\tau})\Omega_{\partial B}.
\end{eqnarray}
Therefore, Eq.~\eqref{bb} is proved.

The term $Q_3$ on the wall is decomposed as
\begin{eqnarray}\label{ff3}
	Q_3=-\mu\left[\bm{\nabla}h_0\right]_{\partial B}\bm{\cdot}\left[\frac{\partial\bm{\nabla}\times\bm{\omega}}{\partial n}\right]_{\partial B}
	-\mu\left[\frac{\partial\bm{\nabla}h_0}{\partial n}\right]_{\partial B}\bm{\cdot}\left[\bm{\nabla}\times\bm{\omega}\right]_{\partial B}.
\end{eqnarray}

On the wall, the stagnation enthalpy gradient is 
\begin{eqnarray}\label{ff1}
	\left[\bm{\nabla}h_0\right]_{\partial B}=\frac{1}{\rho}\bm{\nabla}_{\partial B}p_{\partial B}-\frac{1}{\rho}\left(\bm{\nabla}_{\partial B}\bm{\cdot}\bm{\tau}\right)\bm{n},
\end{eqnarray}
and the wall-normal derivative of the stagnation enthalpy gradient is
\begin{eqnarray}\label{ff2}
	\left[\frac{\partial\bm{\nabla}h_0}{\partial n}\right]_{\partial B}
	=-\frac{1}{\rho}\bm{\nabla}_{\partial B}\left(\bm{\nabla}_{\partial B}\bm{\cdot}\bm{\tau}\right)
	-\bm{n}\left(\frac{1}{\rho}\nabla_{\partial B}^{2}p_{\partial B}-2\Omega_{\partial B}\right).
\end{eqnarray}

The wall-normal derivative of $\bm{\nabla}\times\bm{\omega}$ on the wall is evaluated as
\begin{eqnarray}\label{m0}
	\left[\frac{\partial\bm{\nabla}\times\bm{\omega}}{\partial n}\right]_{\partial B}=\frac{1}{\mu}\bm{\nabla}_{\partial B}\times\bm{\sigma}+\bm{n}\times\left[\frac{\partial^2\bm{\omega}}{\partial n^2}\right]_{\partial B}.
\end{eqnarray}
From Eq.~\eqref{BVF}, we have
\begin{eqnarray}\label{m1}
	\bm{\nabla}_{\partial B}\times\bm{\sigma}
	=\left({\nabla}_{\partial B}^{2}p_{\partial B}\right)\bm{n}+\mu\bm{n}\times\bm{\nabla}_{\partial B}\left(\bm{\nabla}_{\partial B}\bm{\cdot}\bm{\omega}_{\partial B}\right).
\end{eqnarray}
Interestingly, it can be shown that
\begin{eqnarray}\label{m2}
	\bm{n}\times\bm{\nabla}_{\partial B}\left(\bm{\nabla}_{\partial B}\bm{\cdot}\bm{\omega}_{\partial B}\right)
	=\frac{1}{\mu}\bm{\nabla}_{\partial B}
	(\bm{\nabla}_{\partial B}\bm{\cdot}\bm{\tau})
	-\frac{1}{\mu}\nabla_{\partial B}^2\bm{\tau},
\end{eqnarray}
which results in
\begin{eqnarray}\label{m3}
	\bm{\nabla}_{\partial B}\times\bm{\sigma}
	=\left(\bm{\nabla}_{\partial B}^{2}p_{\partial B}\right)\bm{n}+\bm{\nabla}_{\partial B}
	(\bm{\nabla}_{\partial B}\bm{\cdot}\bm{\tau})-\nabla_{\partial B}^2\bm{\tau}.
\end{eqnarray}

The vorticity transport equation can be expressed as
\begin{eqnarray}\label{m4}
	\frac{\partial\bm{\omega}}{\partial t}
	+\bm{u}\bm{\cdot}\bm{\nabla\omega}
	=\bm{\omega}\bm{\cdot}\bm{\nabla u}+\nu\nabla^2\bm{\omega},
\end{eqnarray}
from which it readily follows that
\begin{eqnarray}\label{d2omegadn2}
	\left[\frac{\partial^2\bm{\omega}}{\partial n^2}\right]_{\partial B}=\frac{1}{\nu}\left(\frac{\partial}{\partial t}-\nu\nabla_{\partial B}^2\right)\bm{\omega}_{\partial B}
	=\frac{1}{\nu}\mathcal{L}_{\partial B}\bm{\omega}_{\partial B}.
\end{eqnarray}
Eq.~\eqref{d2omegadn2} implies that for any point on a stationary flat wall, the wall-normal diffusion of the vorticity is fully determined by its unsteady and tangential diffusion effects. Combining Eqs.~(\ref{m1}),~(\ref{m3}) and~(\ref{d2omegadn2}) yields
\begin{eqnarray}\label{s5}
	\left[\frac{\partial\bm{\nabla}\times\bm{\omega}}{\partial n}\right]_{\partial B}
	=-\frac{1}{\mu\nu}\mathcal{L}_{\partial B}\bm{\tau}+\frac{1}{\mu}\bm{\nabla}_{\partial B}(\bm{\nabla}_{\partial B}\bm{\cdot}\bm{\tau})-\frac{1}{\mu}\nabla_{\partial B}^{2}\bm{\tau}+\frac{1}{\mu}
	(\nabla_{\partial B}^{2}p_{\partial B})\bm{n}.
\end{eqnarray}

By using Eqs.~\eqref{ff1} and~\eqref{s5}, the first term in the right hand side of Eq.~\eqref{ff3} can be evaluated as
\begin{eqnarray}\label{ff3a}
	-\mu\left[\bm{\nabla}h_0\right]_{\partial B}\bm{\cdot}\left[\frac{\partial\bm{\nabla}\times\bm{\omega}}{\partial n}\right]_{\partial B}
	&=&\frac{1}{\mu}\mathcal{L}_{\partial B}\bm{\tau}\bm{\cdot}\bm{\nabla}_{\partial B}p_{\partial B}-\frac{1}{\rho}\bm{\nabla}_{\partial B}(\bm{\nabla}_{\partial B}\bm{\cdot}\bm{\tau})\cdot\bm{\nabla}_{\partial B}p_{\partial B}\nonumber\\
	& &+\frac{1}{\rho}\nabla_{\partial B}^{2}\bm{\tau}\bm{\cdot}\bm{\nabla}_{\partial B}p_{\partial B}+\frac{1}{\rho}\left(\bm{\nabla}_{\partial B}\bm{\cdot}\bm{\tau}\right)\left(\nabla_{\partial B}^{2}p_{\partial B}\right).
\end{eqnarray}
By using Eqs.~\eqref{s1} and~\eqref{ff2}, the second term in the right hand side of Eq.~\eqref{ff3} can be evaluated as
\begin{eqnarray}\label{ff3b}
	-\mu\left[\frac{\partial\bm{\nabla}h_0}{\partial n}\right]_{\partial B}\bm{\cdot}\left[\bm{\nabla}\times\bm{\omega}\right]_{\partial B}
	&=&-\frac{1}{\rho}\bm{\nabla}_{\partial B}(\bm{\nabla}_{\partial B}\bm{\cdot}\bm{\tau})\bm{\cdot}\bm{\nabla}_{\partial B}p_{\partial B}\nonumber\\
	& &+\frac{1}{\rho}\left(\bm{\nabla}_{\partial B}\bm{\cdot}\bm{\tau}\right)\left(\nabla_{\partial B}^{2}p_{\partial B}\right)-2\left(\bm{\nabla}_{\partial B}\bm{\cdot}\bm{\tau}\right)\Omega_{\partial B}.
\end{eqnarray}
From Eqs.~\eqref{ff3},~\eqref{ff3a} and~\eqref{ff3b}, Eq.~\eqref{cc} is proved.

By employing the vector identity (for any two vectors $\bm{a}$ and $\bm{b}$)
\begin{eqnarray}\label{q0}
	\bm{\nabla}\bm{\cdot}\left(\bm{a}\times\bm{b}\right)=\bm{b}\bm{\cdot}(\bm{\nabla}\times\bm{a})-\bm{a}\bm{\cdot}(\bm{\nabla}\times\bm{b}),
\end{eqnarray}
the divergence of the source term $\bm{q}$ can be written as
\begin{eqnarray}\label{q1}
	\bm{\nabla}\bm{\cdot}\bm{q}=-2\nu\bm{\nabla}\bm{\cdot}\left(\frac{\partial\bm{\omega}}{\partial x_j}\times
	\frac{\partial\bm{u}}{\partial x_j}\right)=-2\nu\left[\bm{\nabla}\bm{u}\bm{:}\bm{\nabla}\left(\bm{\nabla}\times\bm{\omega}\right)-\bm{\nabla\omega}\bm{:}\bm{\nabla\omega}\right].
\end{eqnarray}
Then, taking the wall-normal derivatives of both sides of Eq.~(\ref{q1}) gives
\begin{eqnarray}\label{q2}
	\frac{\partial(\bm{\nabla}\bm{\cdot}\bm{q})}{\partial n}=-2\nu\left[\frac{\partial\bm{\nabla u}}{\partial n}\bm{:}\bm{\nabla}\left(\bm{\nabla}\times\bm{\omega}\right)
	+\bm{\nabla u}\bm{:}\frac{\partial\bm{\nabla}\left(\bm{\nabla}\times\bm{\omega}\right)}{\partial n}
	-\frac{\partial\bm{\nabla\omega}\bm{:}\bm{\nabla\omega}}{\partial n}
	\right].
\end{eqnarray}
The three terms in the right hand side of Eq.~(\ref{q2}) are evaluated at the wall as follows.

On the wall, $\bm{\nabla}\left(\bm{\nabla}\times\bm{\omega}\right)$ can be decomposed as
\begin{eqnarray}\label{s3}
	\left[\bm{\nabla}\left(\bm{\nabla}\times\bm{\omega}\right)\right]_{\partial B}=\bm{\nabla}_{\partial B}\left[\bm{\nabla}\times\bm{\omega}\right]_{\partial B}+\bm{n}\left[\frac{\partial\bm{\nabla}\times\bm{\omega}}{\partial n}\right]_{\partial B},
\end{eqnarray}
By virtue of Eq.~(\ref{s1}), the first term in the right hand side of Eq.~(\ref{s3}) can be evaluated as
\begin{eqnarray}\label{s4}
	\bm{\nabla}_{\partial B}\left[\bm{\nabla}\times\bm{\omega}\right]_{\partial B}=-\frac{1}{\mu}\bm{\nabla}_{\partial B}\bm{\nabla}_{\partial B}p_{\partial B}
	+\frac{1}{\mu}\bm{\nabla}_{\partial B}\left(\bm{\nabla}_{\partial B}\bm{\cdot}\bm{\tau}\right)\bm{n}.
\end{eqnarray}

Using Eqs.~(\ref{s2}),~(\ref{s5}),~(\ref{s3}) and~(\ref{s4}), we obtain the first term in the right hand side of Eq.~(\ref{q2}):
\begin{eqnarray}\label{w1}
	\left[\frac{\partial\bm{\nabla u}}{\partial n}\bm{:}\bm{\nabla}\left(\bm{\nabla}\times\bm{\omega}\right)\right]_{\partial B}
	&=&-\frac{1}{\mu^2}\bm{\nabla}_{\partial B}\bm{\tau}\bm{:}\bm{\nabla}_{\partial B}\bm{\nabla}_{\partial B}p_{\partial B}\nonumber\\
	& &-\frac{1}{\mu^2\nu}\mathcal{L}_{\partial B}\bm{\tau}\bm{\cdot}\bm{\nabla}_{\partial B}p_{\partial B}+\frac{1}{\mu^2}\bm{\nabla}_{\partial B}(\bm{\nabla}_{\partial B}\bm{\cdot}\bm{\tau})\bm{\cdot}\bm{\nabla}_{\partial B}p_{\partial B}\nonumber\\
	& &-\frac{1}{\mu^2}{\nabla}_{\partial B}^2\bm{\tau}\bm{\cdot}\bm{\nabla}_{\partial B}p_{\partial B}-\frac{1}{\mu^2}(\nabla_{\partial B}^2p_{\partial B})(\bm{\nabla}_{\partial B}\bm{\cdot}\bm{\tau}).
\end{eqnarray}

On $\partial B$, the second term in the right hand side of Eq.~(\ref{q2}) can be evaluated as
\begin{eqnarray}\label{ppw7}
	\left[\bm{\nabla u}\bm{:}\frac{\partial\bm{\nabla}\left(\bm{\nabla}\times\bm{\omega}\right)}{\partial n}\right]_{\partial B}
	&=&\frac{1}{\mu}\bm{\tau}\bm{\cdot}\left[\bm{\nabla}\times\frac{\partial^2\bm{\omega}}{\partial n^2}\right]_{\partial B}\nonumber\\
	&=&\frac{1}{\mu}\bm{\tau}\bm{\cdot}\bm{\nabla}_{\partial B}\times\left[\frac{\partial^2\bm{\omega}}{\partial n^2}\right]_{\partial B}+\frac{1}{\mu}\bm{\tau}\bm{\cdot}\bm{n}\times\left[\frac{\partial^3\bm{\omega}}{\partial n^3}\right]_{\partial B}\nonumber\\
	&=&\frac{1}{\mu}\bm{\tau}\bm{\cdot}\bm{\nabla}_{\partial B}\times\frac{1}{\nu}\mathcal{L}_{\partial B}\bm{\omega}_{\partial B}+\frac{1}{\mu}\bm{\tau}\bm{\cdot}\bm{n}\times\left[\frac{\partial^3\bm{\omega}}{\partial n^3}\right]_{\partial B}\nonumber\\
	&=&\frac{1}{\mu}\bm{\tau}\bm{\cdot}\bm{n}\times\left[\frac{\partial^3\bm{\omega}}{\partial n^3}\right]_{\partial B}
	=-\frac{1}{\mu}\bm{n}\times\bm{\tau}\bm{\cdot}\left[\frac{\partial^3\bm{\omega}}{\partial n^3}\right]_{\partial B}\nonumber\\
	&=&-\bm{\omega}_{\partial B}\bm{\cdot}\left[\frac{\partial^3\bm{\omega}}{\partial n^3}\right]_{\partial B}.
\end{eqnarray}

The following task is to evaluate the third-order wall-normal derivative $[\partial^3\bm{\omega}/\partial n^3]_{\partial B}$. To this end, acting $\mu\partial/\partial n$ on both sides of Eq.~(\ref{m4}) gives
\begin{eqnarray}\label{ppw}
	\frac{\partial}{\partial t}\left(\mu\frac{\partial\bm{\omega}}{\partial n}\right)
	+\mu\frac{\partial(\bm{u}\bm{\cdot}\bm{\nabla\omega})}{\partial n}
	=\mu\frac{\partial(\bm{\omega}\bm{\cdot}\bm{S})}{\partial n}+\nu\nabla^2\left(\mu\frac{\partial\bm{\omega}}{\partial n}\right).
\end{eqnarray}
On the wall, the first term in Eq.~(\ref{ppw}) is equal to the temporal partial derivative of the BVF, namely,
\begin{eqnarray}\label{ppw1}
	\left[\frac{\partial}{\partial t}\left(\mu\frac{\partial\bm{\omega}}{\partial n}\right)\right]_{\partial B}=
	\frac{\partial\bm{\sigma}}{\partial t},
\end{eqnarray}
where the BVF $\bm{\sigma}$ is given in Eq.~\eqref{BVF}.
The second term in Eq.~\eqref{ppw} is simplified as
\begin{eqnarray}\label{ppw2}
	\left[\mu\frac{\partial(\bm{u}\bm{\cdot}\bm{\nabla\omega})}{\partial n}\right]_{\partial B}
	=\mu\left[\frac{\partial\bm{u}}{\partial n}\right]_{\partial B}\bm{\cdot}\left[\bm{\nabla\omega}\right]_{\partial B}
	=\bm{\tau}\bm{\cdot}\bm{\nabla}_{\partial B}\bm{\omega}_{\partial B}.
\end{eqnarray}
From Eq.~\eqref{s2}, we have
\begin{eqnarray}\label{psomega}
	\left[\frac{\partial\bm{S}}{\partial n}\right]_{\partial B}\bm{\cdot}\bm{\omega}_{\partial B}
	&=&\frac{1}{2\mu}\bm{\nabla}_{\partial B}\bm{\tau}\bm{\cdot}\bm{\omega}_{\partial B}+\frac{1}{2\mu}\bm{\omega}_{\partial B}\bm{\cdot}\bm{\nabla}_{\partial B}\bm{\tau}+\frac{1}{2\mu}\left(\bm{\omega}_{\partial B}\bm{\cdot}\bm{\nabla}_{\partial B}p_{\partial B}\right)\bm{n}\nonumber\\
	&=&-\frac{1}{2\mu}\left(\bm{\nabla}_{\partial B}\bm{\tau}^{T}-\bm{\nabla}_{\partial B}\bm{\tau}\right)\bm{\cdot}\bm{\omega}_{\partial B}\nonumber\\
	& &+\frac{1}{\mu}\bm{\omega}_{\partial B}\bm{\cdot}\bm{\nabla}_{\partial B}\bm{\tau}+\frac{1}{2\mu}\left(\bm{\omega}_{\partial B}\bm{\cdot}\bm{\nabla}_{\partial B}p_{\partial B}\right)\bm{n}.
\end{eqnarray}
The first term in Eq.~\eqref{psomega} can be further simplified as
\begin{eqnarray}\label{psomega2}
	-\frac{1}{2\mu}\left(\bm{\nabla}_{\partial B}\bm{\tau}^{T}-\bm{\nabla}_{\partial B}\bm{\tau}\right)\bm{\cdot}\bm{\omega}_{\partial B}=\frac{1}{2\mu}\bm{\omega}_{\partial B}\times\left(\bm{\nabla}_{\partial B}\times\bm{\tau}\right)
	=-\frac{1}{2\mu}\bm{\tau}\left(\bm{\nabla}_{\partial B}\bm{\cdot}\bm{\omega}_{\partial B}\right).
\end{eqnarray}
Combining Eqs.~\eqref{psomega} and~\eqref{psomega2} gives
\begin{eqnarray}\label{psomega3}
	\mu\left[\frac{\partial\bm{S}}{\partial n}\right]_{\partial B}\bm{\cdot}\bm{\omega}_{\partial B}
	=-\frac{1}{2}\bm{\tau}\left(\bm{\nabla}_{\partial B}\bm{\cdot}\bm{\omega}_{\partial B}\right)+\bm{\omega}_{\partial B}\bm{\cdot}\bm{\nabla}_{\partial B}\bm{\tau}+\frac{1}{2}\left(\bm{\omega}_{\partial B}\bm{\cdot}\bm{\nabla}_{\partial B}p_{\partial B}\right)\bm{n}.
\end{eqnarray}
In addition, we have
\begin{eqnarray}\label{psomega4}
	\bm{\sigma}\bm{\cdot}\bm{S}_{\partial B}=-\frac{1}{2}\bm{\tau}\left(\bm{\nabla}_{\partial B}\bm{\cdot}\bm{\omega}_{\partial B}\right)-\frac{1}{2}\left(\bm{\omega}_{\partial B}\bm{\cdot}\bm{\nabla}_{\partial B}p_{\partial B}\right)\bm{n}.
\end{eqnarray}
Using Eqs.~\eqref{psomega3} and \eqref{psomega4}, the coupling between the surface vorticity $\bm{\omega}_{\partial B}$ and surface pressure gradient $\bm{\nabla}_{\partial B}p_{\partial B}$ cancel each other so that the third term in Eq.~(\ref{ppw}) reduces to
\begin{eqnarray}\label{ppw3}
	\left[\mu\frac{\partial(\bm{\omega}\bm{\cdot}\bm{S})}{\partial n}\right]_{\partial B}=-\bm{\tau}(\bm{\nabla}_{\partial B}\bm{\cdot}\bm{\omega}_{\partial B})+\bm{\omega}_{\partial B}\bm{\cdot}\bm{\nabla}_{\partial B}\bm{\tau}.
\end{eqnarray}
Note that half of the first term in Eq.~\eqref{ppw3} comes from Eq.~\eqref{psomega3} while the other half comes from Eq.~\eqref{psomega4}.
The last term in Eq.~(\ref{ppw}) reduces to
\begin{eqnarray}\label{ppw4}
	\left[\nu\nabla^2\left(\mu\frac{\partial\bm{\omega}}{\partial n}\right)\right]_{\partial B}=
	\nu\nabla_{\partial B}^{2}\bm{\sigma}+\nu\mu\left[\frac{\partial^3\bm{\omega}}{\partial n^3}\right]_{\partial B}.
\end{eqnarray}

Combination of Eqs.~(\ref{ppw1})--(\ref{ppw4}) gives
\begin{eqnarray}\label{ppw5}
	\nu\mu\left[\frac{\partial^3\bm{\omega}}{\partial n^3}\right]_{\partial B}
	=\mathcal{L}_{\partial B}\bm{\sigma}+\bm{\tau}(\bm{\nabla}_{\partial B}\bm{\cdot}\bm{\omega}_{\partial B})+\bm{\tau}\bm{\cdot}\bm{\nabla}_{\partial B}\bm{\omega}_{\partial B}
	-\bm{\omega}_{\partial B}\bm{\cdot}\bm{\nabla}_{\partial B}\bm{\tau}.
\end{eqnarray}

By virtue of the identity 
\begin{eqnarray}\label{ppw6}
	&&\bm{\tau}\bm{\cdot}\bm{\nabla}_{\partial B}\bm{\omega}_{\partial B}
	-\bm{\omega}_{\partial B}\bm{\cdot}\bm{\nabla}_{\partial B}\bm{\tau}\nonumber\\
	&&=2\mu\bm{n}\times\bm{\nabla}_{\partial B}\Omega_{\partial B}
	+(\bm{\nabla}_{\partial B}\bm{\cdot}\bm{\omega}_{\partial B})\bm{\tau}
	-(\bm{\nabla}_{\partial B}\bm{\cdot}\bm{\tau})\bm{\omega}_{\partial B},
\end{eqnarray}
Eq.~(\ref{ppw5}) can be equivalently expressed as
\begin{eqnarray}\label{ppw8}
	\left[\frac{\partial^3\bm{\omega}}{\partial n^3}\right]_{\partial B}
	&=&\frac{1}{\nu\mu}\mathcal{L}_{\partial B}\bm{\sigma}+\frac{1}{\nu\mu}\bm{\tau}(\bm{\nabla}_{\partial B}\bm{\cdot}\bm{\omega}_{\partial B})\nonumber\\
	&&+2\frac{1}{\nu}\bm{n}\times\bm{\nabla}_{\partial B}\Omega_{\partial B}
	+\frac{1}{\nu\mu}(\bm{\nabla}_{\partial B}\bm{\cdot}\bm{\omega}_{\partial B})\bm{\tau}
	-\frac{1}{\nu\mu}(\bm{\nabla}_{\partial B}\bm{\cdot}\bm{\tau})\bm{\omega}_{\partial B}.
\end{eqnarray}
Substituting Eq.~(\ref{ppw8}) into Eq.~(\ref{ppw7}), we obtain
\begin{eqnarray}\label{ppw10}
	\left[\bm{\nabla u}\bm{:}\frac{\partial\bm{\nabla}\left(\bm{\nabla}\times\bm{\omega}\right)}{\partial n}\right]_{\partial B}
	&=&-\frac{1}{\nu\mu}\bm{\omega}_{\partial B}\bm{\cdot}\mathcal{L}_{\partial B}\bm{\sigma}
	-\frac{2}{\nu\mu}\bm{\tau}\bm{\cdot}\bm{\nabla}_{\partial B}\Omega_{\partial B}\nonumber\\
	& &+\frac{2}{\nu\mu}(\bm{\nabla}_{\partial B}\bm{\cdot}\bm{\tau})\Omega_{\partial B}.
\end{eqnarray}
It is noted that $\bm{\omega}_{\partial B}\bm{\cdot}\mathcal{L}_{\partial B}\bm{\sigma}$ in Eq.~\eqref{ppw10} can be further simplified as
\begin{eqnarray}
	\bm{\omega}_{\partial B}\bm{\cdot}\mathcal{L}_{\partial B}\bm{\sigma}=\bm{\omega}_{\partial B}\bm{\cdot}\bm{n}\times\mathcal{L}_{\partial B}\bm{\nabla}_{\partial B}p_{\partial B}
	=\frac{1}{\mu}\bm{\tau}\bm{\cdot}\mathcal{L}_{\partial B}\bm{\nabla}_{\partial B}p_{\partial B},
\end{eqnarray}
which is determined by the coupling between the skin friction and the temporal-spatial evolution of the surface pressure gradient. Therefore, 
\begin{eqnarray}\label{ppwss}
	\left[\bm{\nabla u}\bm{:}\frac{\partial\bm{\nabla}\left(\bm{\nabla}\times\bm{\omega}\right)}{\partial n}\right]_{\partial B}
	&=&-\frac{1}{\mu^2\nu}\bm{\tau}\bm{\cdot}\mathcal{L}_{\partial B}\bm{\nabla}_{\partial B}p_{\partial B}
	-\frac{2}{\nu\mu}\bm{\tau}\bm{\cdot}\bm{\nabla}_{\partial B}\Omega_{\partial B}\nonumber\\
	& &+\frac{2}{\nu\mu}(\bm{\nabla}_{\partial B}\bm{\cdot}\bm{\tau})\Omega_{\partial B}.
\end{eqnarray}

Now we consider the last term in Eq.~\eqref{q2}.
It is easy to verify that
\begin{eqnarray}\label{u1}
	\left[\mu\frac{\partial}{\partial n}\left(\nu\bm{\nabla\omega}\bm{:}\bm{\bm{\nabla\omega}}\right)\right]_{\partial B}
	=2\nu\bm{\nabla}_{\partial B}\bm{\sigma}\bm{:}\bm{\nabla}_{\partial B}\bm{\omega}_{\partial B}
	+2\nu\left[\frac{\partial^2\bm{\omega}}{\partial n^2}\right]_{\partial B}\bm{\cdot}\bm{\sigma}.
\end{eqnarray}
By using the vector identity (for any four vectors $\bm{a}$, $\bm{b}$, $\bm{c}$ and $\bm{d}$)
\begin{eqnarray}
	(\bm{a}\times\bm{b})\cdot(\bm{c}\times\bm{d})
	=(\bm{a}\cdot\bm{c})(\bm{b}\cdot\bm{d})-
	(\bm{a}\cdot\bm{d})(\bm{b}\cdot\bm{c}),
\end{eqnarray}
the first term in the right hand side of Eq.~\eqref{u1} is
\begin{eqnarray}\label{u2}
	2\nu\bm{\nabla}_{\partial B}\bm{\sigma}\bm{:}\bm{\nabla}_{\partial B}\bm{\omega}_{\partial B}
	=\frac{2}{\rho}\bm{\nabla}_{\partial B}\bm{\tau}\bm{:}\bm{\nabla}_{\partial B}\bm{\nabla}_{\partial B}p_{\partial B}.
\end{eqnarray}
The second term in the right hand side of Eq.~\eqref{u1} is
\begin{eqnarray}\label{u3}
	2\nu\left[\frac{\partial^2\bm{\omega}}{\partial n^2}\right]_{\partial B}\bm{\cdot}\bm{\sigma}=2\mathcal{L}_{\partial B}\bm{\omega}_{\partial B}\bm{\cdot}\bm{\sigma}
	=\frac{2}{\mu}\mathcal{L}_{\partial B}\bm{\tau}\bm{\cdot}\bm{\nabla}_{\partial B}p_{\partial B}.
\end{eqnarray}
Combining Eqs.~\eqref{u1},~\eqref{u2} and~\eqref{u3} gives the following expression
\begin{eqnarray}\label{ppw9}
	\left[-\frac{\partial\bm{\nabla\omega}\bm{:}\bm{\nabla\omega}}{\partial n}\right]_{\partial B}
	=-\frac{2}{\mu^2}\bm{\nabla}_{\partial B}\bm{\tau}\bm{:}\bm{\nabla}_{\partial B}\bm{\nabla}_{\partial B}p_{\partial B}
	-\frac{2}{\mu^2\nu}\mathcal{L}_{\partial B}\bm{\tau}\bm{\cdot}\bm{\nabla}_{\partial B}p_{\partial B}.
\end{eqnarray}
Combination of Eqs.~\eqref{q2},~\eqref{w1},~\eqref{ppwss} and~\eqref{ppw9} gives Eq.~\eqref{dd}.
\bibliography{Chenetalref}% Produces the bibliography via BibTeX.

\end{document}